\documentclass{jfm}
\usepackage{graphicx,epsf,subfigure}
\usepackage{natbib}
\usepackage{epsfig} 
\usepackage{epstopdf}
\usepackage{amsmath}
\usepackage{textcomp}
\usepackage{multirow}

\let\realverbatim=\verbatim
\let\realendverbatim=\endverbatim
\renewcommand\verbatim{\par\addvspace{6pt plus 2pt minus 1pt}\realverbatim}
\renewcommand\endverbatim{\realendverbatim\addvspace{6pt plus 2pt minus 1pt}}
\makeatletter
\newcommand\verbsize{\@setfontsize\verbsize{10}\@xiipt}
\renewcommand\verbatim@font{\verbsize\normalfont\ttfamily}
\makeatother

\ifCUPmtlplainloaded \else
  \checkfont{eurm10}
  \iffontfound
    \IfFileExists{upmath.sty}
      {\typeout{^^JFound AMS Euler Roman fonts on the system,
                   using the 'upmath' package.^^J}%
       \usepackage{upmath}}
      {\typeout{^^JFound AMS Euler Roman fonts on the system, but you
                   don't seem to have the}%
       \typeout{'upmath' package installed. jfm.cls can take advantage
                 of these fonts,^^Jif you use 'upmath' package.^^J}%
      }
  \else
  \fi
\fi

\ifCUPmtlplainloaded \else
  \checkfont{msam10}
  \iffontfound
    \IfFileExists{amssymb.sty}
      {\typeout{^^JFound AMS Symbol fonts on the system, using the
                'amssymb' package.^^J}%
       \usepackage{amssymb}%
       \let\le=\leqslant  \let\leq=\leqslant
       \let\ge=\geqslant  \let\geq=\geqslant
      }{}
  \fi
\fi

\ifCUPmtlplainloaded \else
  \IfFileExists{amsbsy.sty}
    {\typeout{^^JFound the 'amsbsy' package on the system, using it.^^J}%
     \usepackage{amsbsy}}
    {\providecommand\boldsymbol[1]{\mbox{\boldmath $##1$}}}
\fi

\def \be{\begin{equation}}
\def \ee{\end{equation}}
\def \ba{\begin{eqnarray}}
\def \ea{\end{eqnarray}}
\def \dd{\mbox{d}}
\def \pd{\partial }
\def \bs{\boldsymbol}

\newsavebox{\astrutbox}
\sbox{\astrutbox}{\rule[-5pt]{0pt}{20pt}}

\title[Libration driven multipolar instabilities]{Libration driven multipolar instabilities}
\author[D. C\'ebron, S. Vantieghem and W. Herreman]{
D.\ns C\ls \'E\ls B\ls R\ls O\ls N$^1$\footnote{Email adress for correspondance: david.cebron@erdw.ethz.ch}, \ns S.\ns V\ls A\ls N\ls T\ls I\ls E\ls G\ls H\ls E\ls M$^1$ \ns \and \ns W.\ns H\ls E\ls R\ls R\ls E\ls M\ls A\ls N$^2$}
\affiliation{$^1$Institut f\"ur Geophysik, Sonneggstrasse 5, ETH Z\"urich, Z\"urich, CH-8092, Switzerland. \\
$^2$ Universit\'e de Paris-Sud (LIMSI-CNRS), BP 133, F-91403 Orsay, Cedex, France}

\date{2013}
\pagerange{\pageref{firstpage}--\pageref{lastpage}}
\doi{S002211200100456X}

\begin{document}

\label{firstpage}
\maketitle

\begin{abstract}

We consider rotating flows in non-axisymmetric enclosures that are driven by libration, i.e. by a small periodic modulation of the rotation rate. Thanks to its simplicity, this model is relevant to various contexts, from industrial containers (with small oscillations of the rotation rate) to fluid layers of terrestial planets (with length-of-day variations). Assuming a multipolar $n$-fold boundary deformation, we first obtain the two-dimensional basic flow. We then perform a short-wavelength local stability analysis of the basic flow, showing that an instability may occur in three dimensions. We christen it the Libration Driven Multipolar Instability (LDMI). The growth rates of the LDMI are computed by a Floquet analysis in a systematic way, and compared to analytical expressions obtained by perturbation methods.

We then focus on the simplest geometry allowing the LDMI, a librating deformed cylinder. To take into account viscous and confinement effects, we perform a global stability analysis, which shows that the LDMI results from a parametric resonance of inertial modes. Performing numerical simulations of this librating cylinder, we confirm that the basic flow is indeed established and report the first numerical evidence of the LDMI. Numerical results, in excellent agreement with the stability results, are used to explore the non-linear regime of the instability (amplitude and viscous dissipation of the driven flow). We finally provide an example of LDMI in a deformed spherical container to show that the instability mechanism is generic. Our results show that the previously studied libration driven elliptical instability simply corresponds to the particular case $n=2$ of a wider class of instabilities. Summarizing, this work shows that any oscillating non-axisymmetric container in rotation may excite intermittent, space-filling LDMI flows, and this instability should thus be easy to observe experimentally.

\end{abstract}


\section{Introduction}

It is basic planetary and astrophysical knowledge that celestial objects are rapidly rotating and orbit around each other. This combination of rapid rotation and mutual gravitational interaction forces these objects to synchronize, get phase locked, precess, librate and be tidally deformed. Since the pioneering work of \cite{Poincare_precession}, we are interested in modeling how the liquid parts of such rotating objects can respond to this precession, libration and tidal deformations. In this work, we combine both effects of longitudinal libration and deformations by considering the ideal case of a deformed rotating rigid container which undergoes a periodic modulation of its rotation rate. The simplicity of this ideal model makes it relevant to various contexts, from industrial rotating fluid containers to planetary fluid layers. To understand the particularity of this study, we give below a short review on previous relevant studies on both ingredients, i.e. librations and deformed containers.

\subsection{Libration driven flows}

Librations can be decomposed in two classes: longitudinal and latitudinal libration. In the first case, the rotation speed of the object oscillates around some mean value, but the planet continues to rotate about the same axis. The case of latitudinal libration is more complex as it is the figure axes of the object that oscillate about some mean direction.  In most cases, persistent librations are due to gravitational coupling between an astrophysical body and its main gravitational partner around which it orbits \cite[][]{comstock2003solar}, but they can also be due to exchange of angular momentum between the solid mantle of a planet and its atmosphere, as on Titan. On the Earth, the change in polar ice sheets, between ice and non-ice ages, modifies the inertia tensor of the Earth, which leads to oscillations of the mantle rotation rate \cite[see e.g.][where the consequences on the Earth magnetic field are discussed]{PRLlibration}. Note that these oscillations are called length-of-day (LOD) variations, rather than libration, for non-synchronized planets like the Earth. Finally, strong enough meteorite impacts may be responsible for the occurrence of transient decaying libration movements \cite[][]{wieczorek2009did,lebarsNature}. The analysis of the librations of a planet allows to define constraints on its internal structure \cite[e.g.][explain Mercury's longitudinal libration by the presence of a liquid core]{margot2007}. The main problem of these models is that they presuppose that the fluid rotates without being disturbed by the libration of the solid shell, except in a thin viscous boundary layer, the Ekman layer, at the solid liquid interface. This is not a valid approximation in many cases and the non-rigid response of the fluid in the liquid layer of the planet has thus to be characterized.

Because of astrophysical applications of libration driven flows, a number of studies has been devoted to librating axisymmetric containers in order to investigate the role of the viscous coupling. It has been shown that longitudinal libration in axisymmetric containers can drive inertial waves in the bulk of the fluid as well as boundary layer centrifugal instabilities in the form of Taylor-G\"ortler rolls \cite[][]{aldridge1967experimental, aldridge1969axisymmetric,aldridge75,tilgner1999driven,noir2009experimental,calkins2010axisymmetric,sauret2012fluid}. In addition, laboratory and numerical studies have corroborated the analytically predicted generation of a mainly retrograde axisymmetric and stationary zonal flow in the bulk, based upon non-linear interactions within the Ekman boundary layers \cite[][]{wang1970cylindrical,busse2010mean,busse2010zonal,calkins2010axisymmetric,sauret2010experimental,noir2010experimental,Noir2012,sauret2012fluid,Sauret2013}.

Although practical to isolate the effect of viscous coupling, the spherical approximation of the core-mantle or ice shell-subsurface ocean boundaries, is not fully accurate from a planetary point of view and very restrictive from a fluid dynamics standpoint. Indeed, due to the rotation of the planet and the gravitational interactions with companion bodies, the general shape of the core-mantle boundary can significantly differ from that of a sphere. In this article, we precisely study the impact of the combination of libration and boundary deformations. 

\subsection{Elliptical and multipolar instabilities}

The effect of boundary deformations on rotating flows has been carefully studied in the case of tidal deformations. It is recognized that tides generate flows in the mantle that may dissipate enough energy to lead to a synchronisation. Since Malkus' experiments \cite[][]{ malkus1989experimental}, the impact of tides on liquid planetary interiors has been ongoing and it has been quite actively studied in recent years. Most previous studies consider the case of an elliptically deformed container, with a constant, non-zero differential rotation between the fluid and the elliptical distortion. In a geophysical context, it corresponds to a non-synchronized body with a constant spin rate $\Omega_0$, subject to dynamical tides rotating at the constant orbital rotation rate $\Omega_{orb}$ (i.e. the axes of the Core-Mantle Boundary (CMB) elliptical deformation rotates at $\Omega_{orb}$). In this case, the elliptical streamlines of the two-dimensional (2D) basic flow can be destabilized into a fully three-dimensional (3D) flow by an elliptical instability, the so-called tidally-driven elliptical instability or TDEI \cite[see e.g.][]{kerswell2002elliptical}. 

Generally speaking, the elliptical instability can be seen as the inherent local instability due to the non-zero strain of elliptical streamlines \cite[][]
{bayly1986three,waleffe1990three}, or as the parametric resonance between two free inertial waves (resp. modes) of the rotating
unbounded (resp. bounded) fluid and an elliptical strain, which is not an inertial wave or mode  \cite[][]
{moore1975instability,tsai1976stability}. Such a resonance mechanism,
confirmed by numerous works in elliptically deformed cylinders \cite[e.g.][]
{eloy2000experimental,eloy2001stability,eloy2003elliptic,lavorel2010experimental,guimbard2010elliptic} and ellipsoids
\cite[][]{lacaze2004elliptical,lacaze2005elliptical,le2007coriolis,le2010tidal,Cebron2010PEPI,cebron2010tilt,cebron2010tidal}, is not limited to elliptical deformation but also operates for a general $n$-fold deformation \cite[][]{le1999short}. The elliptical instability is thus a particular case of a wider class of instability, the multipolar instability, which can also be seen as the inherent local instability of the multipolar streamlines \cite[][]{le1999short,le2000three}, or as the parametric resonance between two free inertial waves (resp. modes) of the rotating unbounded (resp. bounded) fluid and an $n$-fold strain \cite[][]{eloy2001stability,eloy2003elliptic}.

\subsection{Libration driven multipolar instabilities (LDMI)}

According to \cite{le2000three}, who considers the case of a constant non-zero differential rotation between the fluid and the multipolar distortion, the multipolar instability vanishes in the case of synchronous rotation ($\Omega_0 = \Omega_{orb}$). However, in the very particular case of elliptical deformation, \cite{Cebron2012MHD,Cebron_Pof} have recently numerically and experimentally confirmed that oscillations around this synchronous state is sufficient to excite elliptical instability, the so-called Libration Driven Elliptical Instability (LDEI), as previously suggested by previous local stability studies of unbounded inviscid flows \cite[][]{kerswell1998tidal,herreman2009effects,cebronAA}. The results obtained by \cite{wu2013on} in a spheroid are in agreement with these studies: by considering a particular class of perturbations which satisfy the non-penetration boundary conditions and volume conservation, they show that an instability is possible, and they confirm this result with simulations. This could be of fundamental importance in planetary liquid cores and subsurface oceans of synchronized bodies, where librations are generically present \cite[e.g.][]{lebarsNature,Noir2012}. 

In this work, we aim at showing that this results holds for a general multipolar deformation, which leads to consider the LDEI as a particular case of a wider class of instabilities, the Libration Driven Multipolar Instability (LDMI). To do so, this work is organized as follows. In section \ref{sec:def_problem}, we define the problem and the considered basic flow. Then, we perform in section \ref{sec:stability} various theoretical stability analysis of the basic flow: considering unbounded inviscid flows, we first generalize previous local analysis to arbitrary multipolar flow (section \ref{sec:local_stability}), and we then take into account viscosity and confinement effects in a cylinder by developing the first eigenmodes global stability analysis of libration driven flows (section \ref{sec:global_stability}). In section \ref{res}, we finally compare our theoretical predictions with three-dimensional non-linear viscous simulations and explore the non-linear regime using a home-made massively parallel finite-volume code.

\begin{figure}                   
  \begin{center}
        \setlength{\epsfysize}{6.0cm}
	\epsfbox{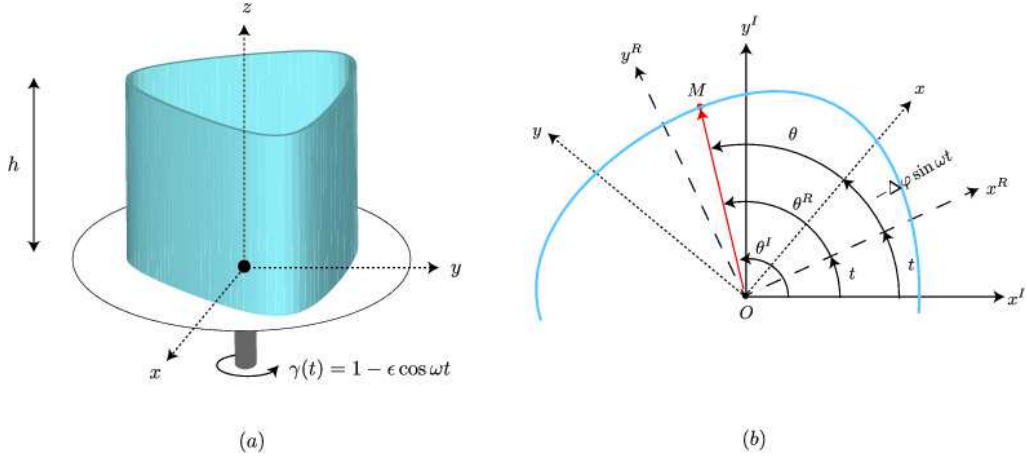}
  \end{center}
  \caption{(a) Sketch of the studied set-up. A deformed cylinder is put on a turntable that rotates at the dimensionless rotation speed $\gamma (t)= 1- \varepsilon \cos \omega t$. The combination of libration and deformation drives a 2D basic flow, which can be destabilized into 3D flows through the Libration Driven Multipolar Instability (LDMI). (b) Different frames of reference are used: the librating frame ($x$, $y$, $\theta$, no superscript) attached to the turntable, the inertial (or lab) frame ($x^I$, $y^I$, $\theta^I$) and the frame rotating at constant speed $1$ ($x^R$, $y^R$, $\theta^R$).} \label{fig:schema}
  \end{figure}
  
\section{Problem definition} \label{sec:def_problem}

\subsection{Dimensionless equations}

In this article, we consider a fluid contained in a librating and deformed rigid cylinder as sketched in figure \ref{fig:schema}(a). In dimensional form, the librating frame of reference $(x,y,z)$ rotates at variable speed $ \Omega_0\,  \gamma(t) \, \boldsymbol{e}_z$ with respect to the inertial frame of reference (suffix $I$, see figure \ref{fig:schema}b), where $\Omega_0$ is the time-averaged rotation rate. Using $1/\Omega_0$ as the timescale, we write the dimensionless rotation speed $\gamma(t)$ as (see fig. \ref{fig:schema})
\be \label{eq:gammadef}
\gamma (t) =  \frac{\dd}{\dd t} \left (  t - \Delta \varphi \sin \omega t    \right ) =  1 - \varepsilon \cos \omega t \, ,
\ee 
and call $\omega$ the libration frequency, $\Delta \varphi$ the libration angle, $\varepsilon = \omega \Delta \varphi$ the libration amplitude. The fluid container is immobile in the librating frame (which rotates at $\gamma (t)$). It takes the form of deformed cylinder limited by two horizontal boundaries at $z=0$, $z=h$, and a lateral wall located at, in cylindrical coordinates $(r, \theta, z)$,
\be 
\zeta (r,\theta) = C + \frac{1}{2} = 0 \, ,
\ee
where
\be \label{eq:Cdef}
C=  - \frac{r^2}{2} + p \, \frac{r^n}{n} \cos (n \theta)
\ee 
defines a multipolar deformation \cite[e.g.][]{le1999short}. Note that, in the limit of small deformations $p \ll 1$, an explicit solution of $\zeta (r,\theta)=0$, i.e. of the lateral wall location, is given by
\be \label{eq:explicitstream}
r=1+\frac{p}{n} \cos n \theta + \mathcal{O}(p^2) \, .
\ee
The coordinates $r, z$ and the container height $h$ are scaled in units of $R$, the radius of the undeformed cylindrical container. We will denote $p$ the amplitude of the multipolar deformation. The integer $n$ sets the type (or order) of the multipolar deformation of the boundary: $n=2$ for elliptical deformation, $n=3$ for tripolar deformation, $n \in \mathbb{N}$ for a general $n$-fold deformation. 

\begin{figure}                   
  \begin{center}
    \begin{tabular}{ccc}
      \setlength{\epsfysize}{5.0cm}
      \subfigure[]{\epsfbox{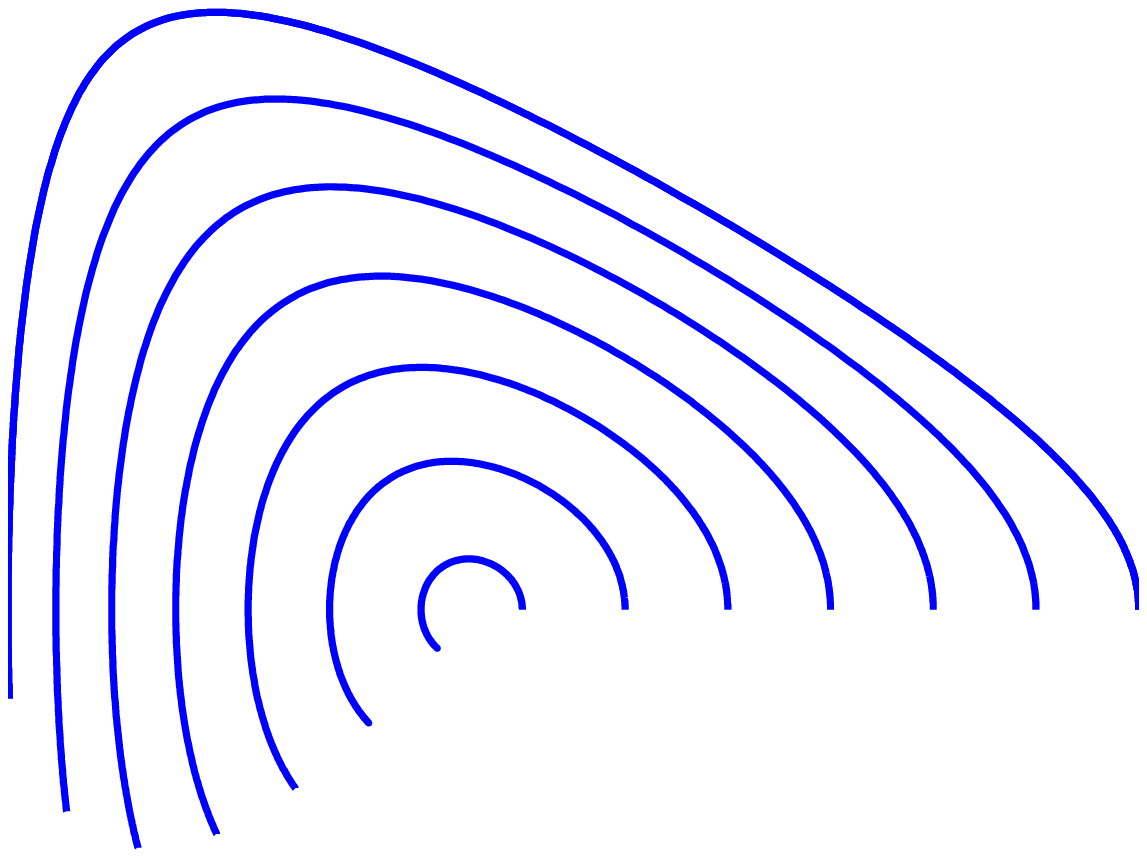}} &
      \setlength{\epsfysize}{5.0cm}
      \subfigure[]{\epsfbox{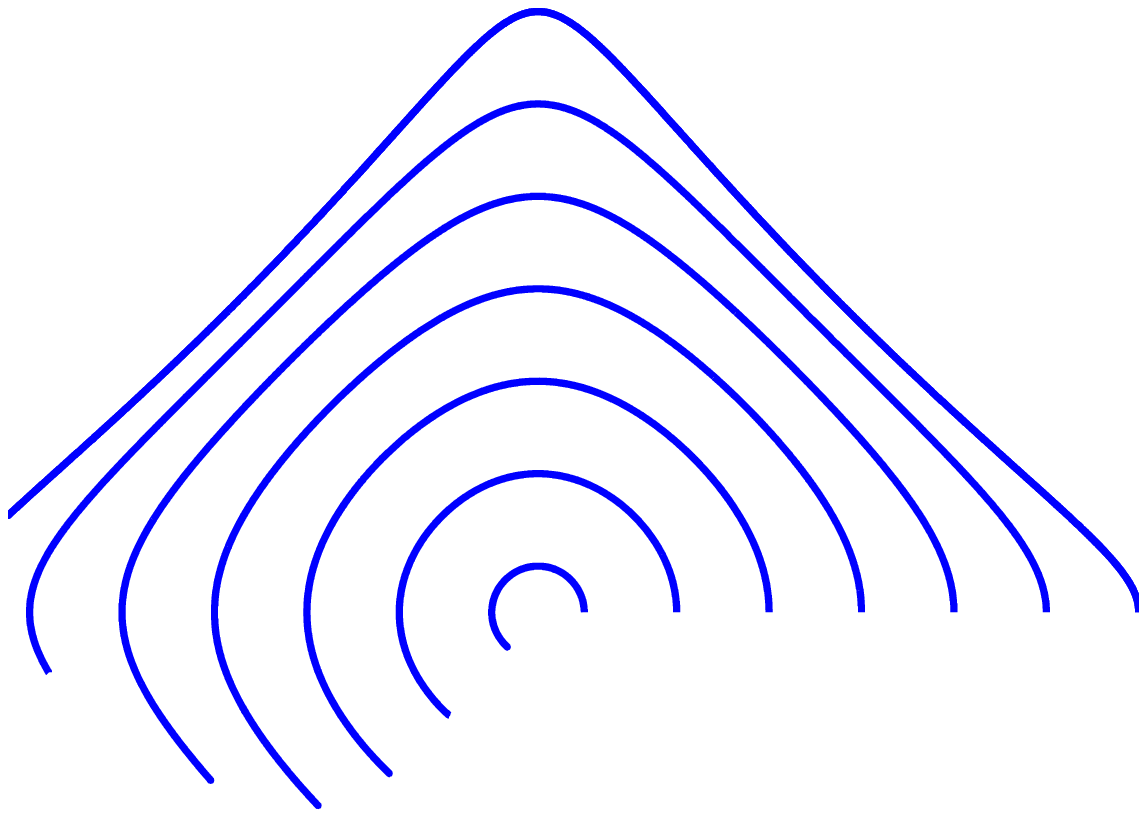}} 
    \end{tabular}
    \caption{Pathlines of the multipolar flow (\ref{eq:basicflow3}) for a libration angle $\Delta \varphi =2$, a deformation $p=0.45$, with initial positions given by $\theta_{(t=0)}=0$ and an initial radius from $r_{(t=0)}= 0.1$ to $r_{(t=0)}=1.3$ by steps of $0.2$. (a) $n=3$. Here, $(C,\beta_3)$ varies between $(C,\beta_3) = (-0.0049, 0.0768)$, reached for $r_{(t=0)}= 0.1$, and $(C,\beta_3)=(-0.5155,0.7914)$, reached for $r_{(t=0)}= 0.8$. (b) $n=4$. Here, $(C,\beta_4)$ varies between $(C,\beta_4)=(-0.005,0.009)$, reached for $r_{(t=0)}= 0.1$, and $(C,\beta_4)=(-0.2739,0.9426)$, reached for $r_{(t=0)}= 0.8$.}
    \label{fig:basic flow}            
  \end{center}
\end{figure}

Considering a Newtonian incompressible fluid with constant and homogeneous material properties, the combination of libration and deformation induces flows governed by the Navier-Stokes and mass conservation equations. Using $1/\Omega_0$ as the timescale, $\Omega_0 R$ for velocity scale, these equations write in the librating frame of reference
 \begin{eqnarray}\label{syst:eqNS_sl}
  \frac{\partial \boldsymbol{u}}{\partial t}+ (\boldsymbol{u}\cdot\boldsymbol{\nabla})\boldsymbol{u}  & = & -\boldsymbol{\nabla}\Pi+E \boldsymbol{\nabla} ^2\boldsymbol{u} -2 \gamma \,\boldsymbol{e}_z \times \boldsymbol{u} - \frac{\mathrm{d} \gamma}{\mathrm{d} t} \boldsymbol{e}_z  \times \boldsymbol{x}\, , \label{syst:eqNS_s1l} \\    
\label{syst:eqNS_s2l} \nabla \cdot \boldsymbol{u}  & = &   0\, ,
\end{eqnarray}
where $E=\nu/(\Omega_0 R^2)$ is the Ekman number, with $\nu$ the fluid kinematic viscosity, $\boldsymbol{u}$ the velocity of the fluid in the librating frame and $\Pi$ the reduced pressure taking the centrifugal force into account. On the lateral boundary, we consider no-slip boundary conditions, i.e. $ \bs{u}(\zeta = 0 ) = \bs{0}$, and the top and bottom are periodic boundaries, i.e. 
\begin{eqnarray}
 \bs{u} |_{z = 0 } = \bs{u} |_{z = h}\, ,\, \textrm{and}\, \left. \frac{\partial \bs{u}}{\partial z} \right|_{z=0} = \left. \frac{\partial \bs{u}}{\partial z} \right|_{z=h}.
\end{eqnarray}
We thus allow for a mean axial flow to exist.  

\subsection{Two-dimensional inviscid basic flow}

In the inviscid limit $E= 0$, one can find a 2D basic flow $\bs{u} = \boldsymbol{U}$ that exactly satisfies the previous equations (\ref{syst:eqNS_s1l})-(\ref{syst:eqNS_s2l}): 
\be\label{eq:basicflow3}
\boldsymbol{U} = \nabla \times (\Psi \boldsymbol{e}_z) \quad , \quad \Psi  = C(r,\theta)\, \varepsilon \cos \omega t = \left (  - \frac{r^2}{2} + p \frac{r^n}{n} \cos n \theta \right ) \varepsilon \cos \omega t \, .
\ee 
Taking the axial component of the curl of (\ref{syst:eqNS_s1l}), the equation for axial vorticity reduces to
\begin{eqnarray}
\frac{\textrm{d} ( \nabla^2 \Psi )}{\textrm{d} t}  =  2\,  \frac{\textrm{d} \gamma}{\textrm{d} t}  ,
\end{eqnarray}
which is exactly satisfied. We also see that the basic flow is always parallel to that boundary, so that the inviscid boundary condition $ \boldsymbol{U} \cdot \bs{e}_n = \bs{0} |_{\zeta = 0 } $ is also satisfied (where $\bs{e}_n$ is the outward boundary normal unit vector). Moreover, due to the separation of the space and time variables in the streamfunction (\ref{eq:basicflow3}) of our 2D basic flow, each fluid particle, as seen from the librating frame, oscillates back and forth along a part of a streamline $C=cst$. Thus, any streamline $C=cst$ does not change with time, and the pathlines of any particle located on this streamline at $t=0$ is thus a part of this streamline. This behavior is illustrated in figure \ref{fig:basic flow}, which shows several pathlines in the librating frame, for $n=3,4$. The particles follow a track $C=cst$ and join their initial position after completing a periodic cycle. Following \cite{le1999short}, we introduce the parameter
\begin{eqnarray}
\beta_n=p \left(\frac{2n|C|}{n-2} \right)^{n/2-1} \label{eq:beta_n}
\end{eqnarray}
that measures the local asymmetry of the streamlines (and pathlines) and varies in $\beta_n \in [0;1]$. For $n=2$, the flow is a uniform elliptical flow and all the pathlines have the same $\beta_2 = p$ that can be identified with the ellipticity ($\beta_2=|1-\Upsilon^2|/(1+\Upsilon^2)$ with $\Upsilon$ the axes ratio). For $n \geq3$, $\beta_n$ varies with $C$ and the flow is thus not uniform. As $\beta_n$ is increased, the streamlines become more and more angular, exhibiting $n$ singular points (corners) for $\beta_n=1$. For $\beta_n >1$, the streamlines of an unbounded flow are no longer closed and our analysis will thus be restricted to the range $\beta_n \in [0;1]$. Note that the maximum value $p_{max}$ for $p$, reached when $\beta_n=1$ on the boundary streamline $\zeta (r,\theta)=0$, i.e. $C=-1/2$, decreases with $n$, tending towards an asymptotic constant value $p_{max}=\textrm{e}^{-1}$ when $n$ becomes infinite (with $\textrm{e}=\exp 1$).

\begin{table}
  \begin{center}
\def~{\hphantom{0}}
  \begin{tabular}{lccc}
      Parameters  & Name   & Range for the simulations  \\[3pt]
       $\omega$   & Libration frequency & $2 - 4.5$ \\
       $\varepsilon$   & Libration amplitude & $0.8 - 1.5$ \\
       $\gamma(t) \boldsymbol{e}_z=[1-\varepsilon \cos \omega t] \boldsymbol{e}_z$  & Rotation vector &   $-$ \\
       $p$  & Multipolar deformation amplitude & $0.2 - 0.5$ \\
       $n$  & Multipolar deformation order & $3$ \\
       $h$  & Height of the cylinder & $2$ \\
       $E=\nu\ /(\Omega_0 R^2)$  & Ekman number & $10^{-4} - 1.5 \cdot 10^{-3}$ \\
       $C$  & Streamfunction spatial dependancy & $-$ \\
       $\beta_n=p \left(\frac{2n|C|}{n-2} \right)^{n/2-1}$  & Local pathline/streamline deformation & $0 - \sqrt{3}/2$\\
  \end{tabular}
  \caption{Dimensionless control parameters involved in the problem definition. }
  \label{tab:kd}
  \end{center}
\end{table}

\subsection{Libration as seen from other frames}

The librating frame is best adapted for numerical simulations, since the boundary has a fixed shape, but the local stability theory will be formulated in the inertial frame (superscript I) to avoid any inertial fictitious force, and the global theory in the frame rotating at constant speed $\Omega_0 \boldsymbol{e}_z$ (superscript R). The different reference frames are illustrated in figure \ref{fig:schema}(b). We can relate azimuthal angles $\theta, \theta^I,\theta^R$ in the three frames by  
\be
\theta^I = \theta^R + t  = \theta + t - \Delta \varphi \sin \omega t \,  .
\ee
Flows in different frames are related as
\ba
\boldsymbol{u}^I &=&  \boldsymbol{u} (r,\theta^I - t + \Delta \varphi \sin \omega t ,z,t ) + (1 - \varepsilon \cos \omega t)  \, \boldsymbol{e}_z \times \boldsymbol{x}^I ,\nonumber \\
\boldsymbol{u}^R  &=&  \boldsymbol{u} (r,\theta^R + \Delta \varphi \sin \omega t ,z,t ) \quad\quad\quad \ \ - \varepsilon \cos \omega t \, \boldsymbol{e}_z \times \boldsymbol{x}^R \, .
\ea
It is instructive to see that the basic flow takes a particularly simple, potential form in the rotating frame of reference, 
\be
 \boldsymbol{U}^R =  \nabla \left [ - \varepsilon \, p\, \frac{r^n}{n}  \, \cos \omega t \,  \sin n (\theta^R +   \Delta \varphi \sin \omega t  )  \right ]  \, . \label{eq:Urotdef}
\ee
This clearly shows that the boundary deformations
\be
\zeta^R (r,\theta^R,t) = \frac{1}{2}- \frac{r^2}{2} + p \frac{r^n}{n} \cos n (\theta^R  +   \Delta \varphi \sin \omega t ) = 0 \label{eq:zetarotdef}
\ee
induces a $\mathcal{O}(\varepsilon p)$ potential flow, stretching some directions, but compressing others. The transverse stretching allows the basic flow to destabilise inertial modes with a horizontal vorticity aligned with the stretched axis, exactly as in the case of elliptical instability \cite[][]{waleffe1990three}.  The particularity of the librational forcing is related to the broader frequency content of the boundary deformation and basic flow. This can be seen by expanding the functions appearing in rotating frame expressions of the basic flow and the boundary deformation: 
\ba
\cos \omega t \, \cos n (\theta + \Delta \varphi \sin \omega t )  &=& \mathrm{e}^{\mathrm{i} n \theta} f(t) + c.c.  \, , \label{eq:freqcontent} \\
\cos n (\theta + \Delta \varphi \sin \omega t )  &=&  \mathrm{e}^{\mathrm{i} n \theta} g(t) + c.c. \, ,
\ea
where $c.c.$ is as usual the complex conjuguate, and 
\ba
f(t) &=&  \frac{1}{4} \sum_{j \in\mathbb{Z}} \Big [ \mathrm{J}_{j-1} (n \Delta \varphi ) + \mathrm{J}_{j+1} (n \Delta \varphi ) \big ]  \,  \mathrm{e}^{\mathrm{i} j \omega t}    \label{eq:fdef}\\
g(t) &= &\frac{1}{2}  \sum_{j \in\mathbb{Z}} \, \mathrm{J}_{j} (n \Delta \varphi )  \,  \mathrm{e}^{\mathrm{i} j \omega t}   \label{eq:gdef}
\ea
with $\mathrm{J}_{j}$ are Bessel functions of the first kind \cite[see formula (9.1.42) \& (9.1.45) in][]{abramowitz1964handbook}. For finite angles $\Delta \varphi$, librational forcing involves more then one frequency. Any of these different frequency components, will be able to couple different inertial modes and this is the basic particularity of Libration Driven Multipolar Instabilities (LDMI) in comparison with previous studies \cite[][]{le1999short,le2007coriolis}.  

We note $f_j$ and $(f^\dagger)_j$ the coefficient in front of the $\exp (\textrm{i}\, j \omega t )$ component of $f(t)$ and its complex conjugate $f^\dagger(t)$ and similar for $g_j$ and $(g^\dagger)_j$. In the limit of small libration angles $\Delta \varphi \rightarrow 0$, there is one dominant frequency in the boundary deformation, as $\mathrm{J}_{0}(0)=1$ and $\mathrm{J}_{j}(0)=0$ for $j \neq 0$: we have $f(t) \rightarrow   (\cos \omega t )/ 2$ and $g(t) \rightarrow 1/2$, but as $\Delta \varphi \rightarrow 0$, the libration amplitude $\varepsilon \rightarrow 0$ for a fixed $\omega$.

\section{Linear stability analysis} \label{sec:stability}

In this section, we are concerned with the linear instability of the flow (\ref{eq:basicflow3}). We will perturb the basic flow $\boldsymbol{U}$ with a small 3D flow $\boldsymbol{u}$ and search to identify under which conditions $\boldsymbol{u}$ can grow in time. In a local stability analysis, this problem is reduced to a stability study of each basic flow particle trajectory separately. This leads to general formula that are broadly applicable. In a global analysis, we analyze the system (fluid + container) as a whole. This leads to precise information on the kind of modes that can be destabilized in a particular fluid domain. Comparing both approaches will lead to useful insights here. Our purpose is to perform a rigorous study on libration driven multipolar instabilities that completes previous work on multipolar instability \cite[][]{le1999short,eloy2001stability,eloy2003elliptic} and libration driven elliptical instability \cite[][]{kerswell1998tidal,herreman2009effects,Cebron2012MHD,cebronAA,Cebron_Pof}.

\begin{table}
  \begin{center}
\def~{\hphantom{0}}
  \begin{tabular}{l|c|c|c|c}
       $ $  &  Unbounded & Inviscid & Short--wavelength & Small forcing   \\[3pt]
       $ $  & $ $  & ($E=0$) & ($\vartheta \ll 1$) & ($\varepsilon p \ll 1$)   \\
       Problem definition: section \ref{sec:def_problem} & $-$ & $-$ & $-$  & $ - $ \\
       Local analysis: section \ref{sec:local_stability}  & $\surd$ & $\surd$ & $\surd$  & $ - $ \\
       Local analytical analysis: section \ref{sec:asympt}  & $\surd$ & $\surd$ & $\surd$  & $\surd$  \\
       Global analysis: section \ref{sec:global_stability} &   $- $ &   $- $ & $-$ & $\surd$ \\
       Simulations: section \ref{res} & $-$ & $-$ & $-$  & $ - $ \\
  \end{tabular}
  \caption{Assumptions used in the different sections of this work. }
  \label{tab:assumptions}
  \end{center}
\end{table}

\subsection{Local Stability analysis: unbounded flows} \label{sec:local_stability}

In this section, we investigate the inviscid stability of the pathlines of the basic flow $\boldsymbol{U}$ in a fluid domain assumed to be unbounded (see table \ref{tab:assumptions}). To do so, we consider a perturbed solution of the equations of motion under the form of localized plane waves along the pathlines of the basic flow, and we assume that the plane waves characteristic wavelength $\vartheta$ is very small (short--wavelength hypothesis).

 \subsubsection{Short--wavelength Lagrangian stability analysis}

The approach we follow here is based on the short--wavelength Lagrangian theory, used by \cite{bayly1986three}, \cite{craik1986evolution}, and then generalized in \cite{friedlander1991instability} and \cite{lifschitz1991local,lifschitz1993localized,lifschitz1994instability} where the whole theory is thoroughly explained. This theory is now rather classical in stability studies of flows \cite[e.g.][]{bayly1996three,lebovitz1996short,leblanc1997three}, and we thus only remind below some basic elements of the stability analysis in following  the approach of \cite{le1999short}. We found it simplest to work in the inertial frame of reference, but the superscript $I$ will be omitted in what follows. The perturbation velocity $\boldsymbol{u}$ is written in the geometrical optics, or WKB (Wentzel-Kramers-Brillouin) form:
\begin{eqnarray}
\boldsymbol{u}(\boldsymbol{x},t)=\boldsymbol{a}(\boldsymbol{x},t)\, \mathrm{e}^{\textrm{i} \chi(\boldsymbol{x},t)/ \vartheta} \, . \label{eq:pert}
\end{eqnarray}
Here, the amplitude $\boldsymbol{a}(\boldsymbol{x},t)$ and phase $\chi(\boldsymbol{x},t)$ are real functions dependent on space $\boldsymbol{x}$ and time $t$. The characteristic wavelength $\vartheta \ll 1$ is the small paramater used for the asymptotic (WKB) expansion. In the inviscid limit, the evolution of (\ref{eq:pert})  is governed by the linearized Euler equations. Along the pathlines of $\boldsymbol{U}$, the leading order problem can be written in Lagrangian form as a system of ordinary differential equations  \cite[][]{lifschitz1994instability}:
\begin{eqnarray}
\frac{\textrm{d} \boldsymbol{X}}{\textrm{d} t}&=&\boldsymbol{U}(\boldsymbol{X},t) \, , \label{eq:lifshitz1} \\ 
\frac{\textrm{d} \boldsymbol{\mathcal{K}}}{\textrm{d} t}&=&-(\boldsymbol{\nabla} \boldsymbol{U})^{\textrm{T}}(\boldsymbol{X},t)\, \boldsymbol{\mathcal{K}} \, , \label{eq:lifshitz2}\\
\frac{\textrm{d} \boldsymbol{a}}{\textrm{d} t}&=& \left( \frac{2 \boldsymbol{\mathcal{K}} \boldsymbol{\mathcal{K}}^{\textrm{T}}}{|\boldsymbol{\mathcal{K}}|^2}-\boldsymbol{I} \right) \boldsymbol{\nabla} \boldsymbol{U}(\boldsymbol{X},t)\,  \boldsymbol{a} \, , \label{eq:lifshitz3} 
\end{eqnarray}
with constraint 
\be
\boldsymbol{\mathcal{K}}  \cdot \boldsymbol{a} =0 \, . \label{eq:lifshitz4}
\ee
Here $\textrm{d}/\textrm{d}t=\partial_t + \boldsymbol{U} \cdot \boldsymbol{\nabla}$ are Lagrangian derivatives, $\boldsymbol{I}$ is the identity matrix, $\boldsymbol{\mathcal{K}}=\boldsymbol{\nabla} \chi$ is the (local) wavevector along the Lagrangian trajectory $\boldsymbol{X}$. The incompressibility condition (\ref{eq:lifshitz4}) is always fulfilled if the initial condition $(\boldsymbol{X}_0,\boldsymbol{\mathcal{K}_0},\boldsymbol{a}_0)$ satisfies $\boldsymbol{\mathcal{K}_0} \cdot \boldsymbol{a}_0=0 $ \cite[][]{le2000three}. As shown by \cite{lifschitz1991local}, the existence of an unbounded solution for $\boldsymbol{a}$ provides a sufficient condition of instability. Assuming closed pathlines, stability is naturally analysed over one turnover period $T$ along the pathline. Note that this system of equations can be seen as an extension of Rapid Distorsion Theory (RDT) to nonhomogeneous flows \cite[][]{cambon1985etude,cambon1994stability,sipp1998elliptic}. 

In practice, the equation (\ref{eq:lifshitz1}) has to be solved as a first step to know the trajectory $\boldsymbol{X}$ emerging out of initial position $\boldsymbol{X}_0$.  Knowing $\boldsymbol{X}$, one can solve the wavevector equation (\ref{eq:lifshitz2}) for an initial vector $\boldsymbol{\mathcal{K}}_0$.  As the magnitude or sign of $\boldsymbol{\mathcal{K}}_0$ cannot influence the growth of $\boldsymbol{a}$ and due to $\boldsymbol{\mathcal{K}}_0 \cdot \boldsymbol{a}_0 = 0$, in the particular case of a two-dimensional flow with closed streamlines in the $x-y$ plane, we can consider the role of different $\boldsymbol{\mathcal{K}}_0$ with a single angle 
\be
\xi =  \arccos \left( \frac{\boldsymbol{\mathcal{K}_{0}} \cdot \boldsymbol{e}_z}{|| \boldsymbol{\mathcal{K}_{0}} ||} \right)
\ee 
that varies in the interval $\xi \in [ 0^\circ, 90^\circ]$ \cite[e.g.][]{le1999short}. Knowledge of $\boldsymbol{X}$ and $\boldsymbol{\mathcal{K}}$ finally allows to solve equation (\ref{eq:lifshitz3}) for the amplitudes $\boldsymbol{a}$ and to look for growing solutions.
As shown by \cite{lifschitz1991local}, the existence of an unbounded time-evolution of $\boldsymbol{a}$ provides a sufficient condition of instability. The result holds for viscous flows \cite[][]{landman1987three,lifschitz1991local} if the characteristic wavelength $\vartheta$ is larger than $\sqrt{E/\sigma}$, where $\sigma$ is the maximum inviscid growth rate of $||\boldsymbol{a}(\boldsymbol{x},t)|| \sim \exp ( \sigma t) $.  Finally, to close this brief description of short--wavelength Lagrangian theory, it is worth mentioning that viscous effects on the perturbations can be easily taken into account by adding to the inviscid growth rate $\sigma$ the viscous damping rate $-\mathcal{K}^2\, E$ \cite[e.g.][]{craik1986evolution,landman1987three,le2000three}.

 \subsubsection{Numerical results: Floquet analysis}

\begin{figure}                   
  \begin{center}
    \begin{tabular}{ccc}
      \setlength{\epsfysize}{5.0cm}
      \subfigure[]{\epsfbox{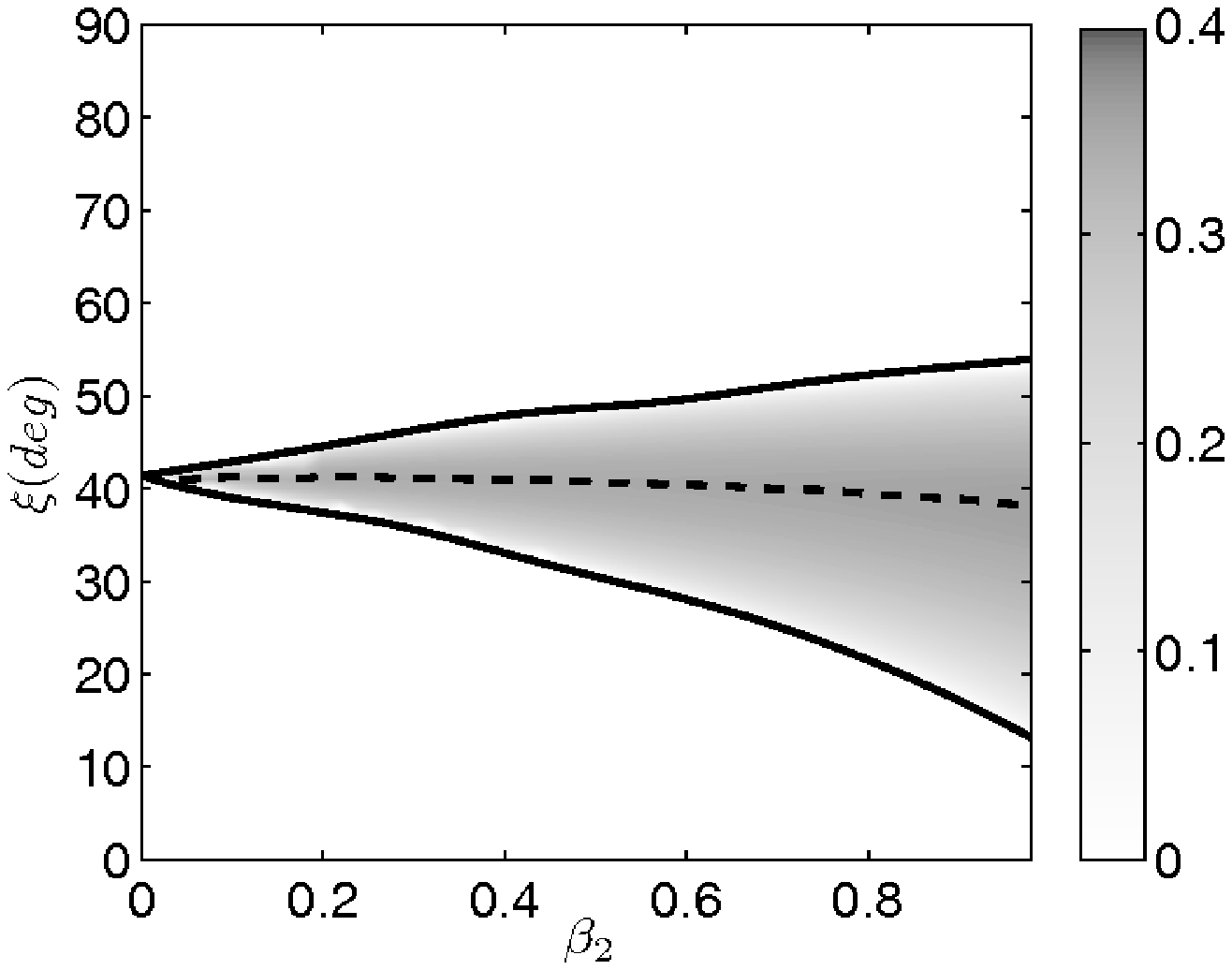}} &
      \setlength{\epsfysize}{5.0cm}
      \subfigure[]{\epsfbox{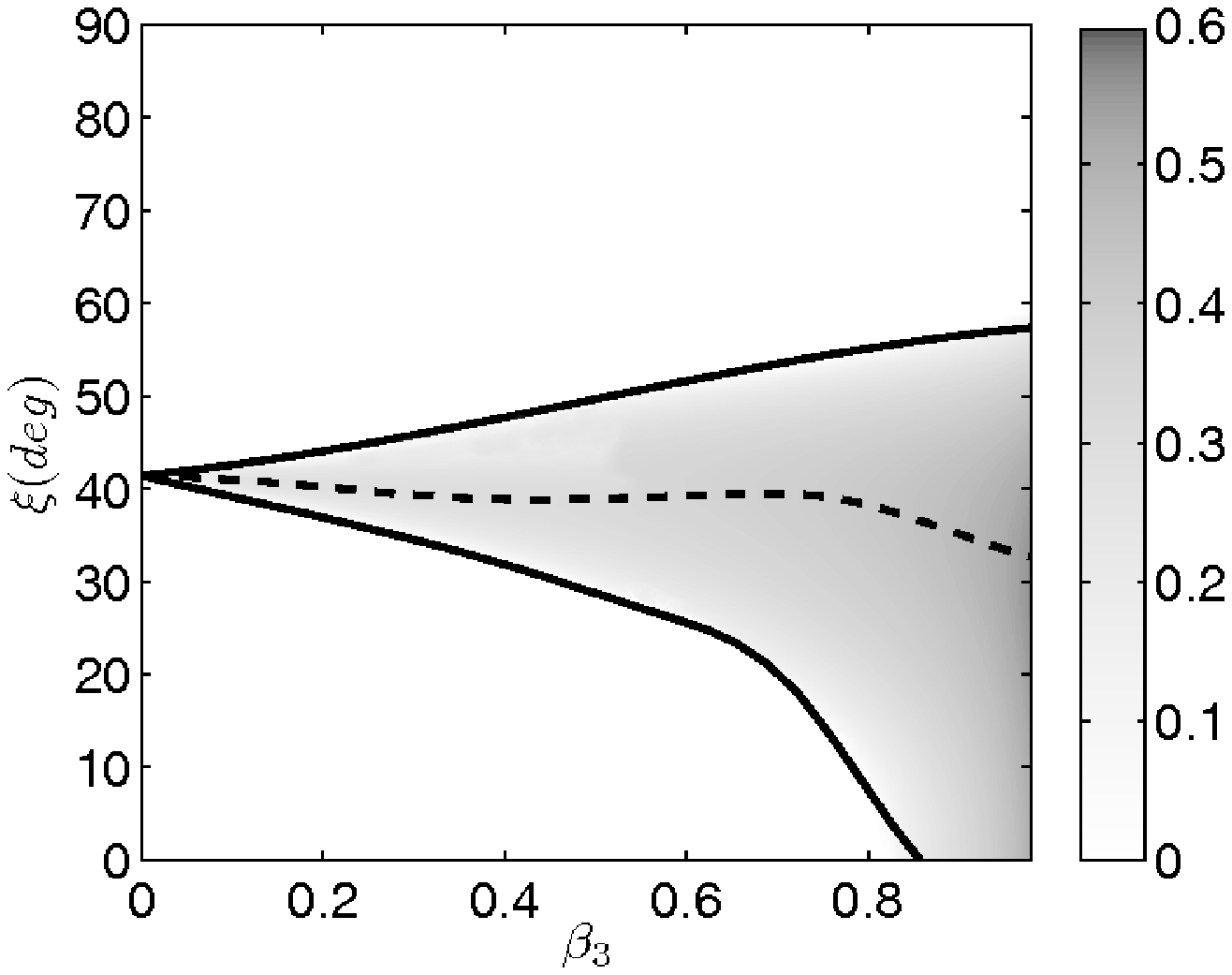}} 
    \end{tabular}
    \caption{Growth rates $\sigma^*$ in the $(\beta_n, \xi)$-plane, obtained from the Floquet analysis, for $\varepsilon =1.5$ and $\omega=3$ (as shown in section \ref{sec:asympt} in the limit $\varepsilon p \ll 1$, only one tongue exists for $\omega>2$). The colorbar represents $\sigma^*$, solid lines are the boundaries of the tongue (corresponding to a non-zero $\sigma^*$), and the dashed line represents the values of $\xi$ which maximise $\sigma^*$. (a) $n=2$ (b) $n=3$.}
    \label{fig:tongues}       
  \end{center}
\end{figure}

We now solve the previous problem in the inertial frame using the basic flow (\ref{eq:basicflow3}). As particles always come back to their initial position after a time that can be noted $T$, we have periodic trajectories $\boldsymbol{X}$, which also results in periodic functions $\boldsymbol{\mathcal{K}}$ whatever the chosen $\boldsymbol{\mathcal{K}}_0$. With $\boldsymbol{\mathcal{K}}(t)$ periodic, (\ref{eq:lifshitz3}) can be analyzed in terms of Floquet theory  \cite[e.g.][]{bender1978advanced}.  Starting from three canonical initial conditions, in matrix form $\boldsymbol{a}(0)=\boldsymbol{I}$, we integrate equation (\ref{eq:lifshitz3}) over exactly one period $t\in [0,T]$ to obtain the monodromy matrix $\boldsymbol{a}(T)$. The Floquet exponents are the three eigenvalues $\varpi_1$, $\varpi_2$, and $\varpi_3$ of this matrix and represent the multiplicative gain of the associated Floquet eigenvector, over one period $T$. As noted by \cite{kerswell1993instability}, $\textrm{det}\ \boldsymbol{a}(T)=1$, and $\boldsymbol{\mathcal{K}}_0 = \boldsymbol{\mathcal{K}}(0)=\boldsymbol{\mathcal{K}}(T)$ is a left eigenvector of $\boldsymbol{a}(T)$, with an eigenvalue $\varpi_1=1$. The two other eigenvalues are then either complex conjugates on the unit circle, indicating stability, or a real, reciprocal pair, one of which lying outside the unit circle. In this last case, an instability is present, with growth rate given by
\begin{eqnarray}
\sigma(n,\varepsilon,\beta_n,\xi)= \frac{1}{T}\, \textrm{ln} |\varpi(n,\varepsilon,\beta_n,\xi)| \, . \label{eq:sigFloquet}
\end{eqnarray}
Figure (\ref{fig:tongues}) show results from some calculations of $\sigma$ in the $\beta_n$-$\xi$ plane, for fixed $\varepsilon=1.5$ and $\omega =3$. We clearly see resonance tongues emerging from a well defined angle, and we will show in section \ref{sec:asympt} that its value is given by $\xi = \arccos (\omega / 4) $, here $\xi=41.41^\circ$. As $\beta_n$ increases, a broadening band of angles gets destabilized.  If we only focus on the largest growth rate, obtained by maximising over all angles $\xi$ and rescale it with respect to the local strain rate
\be
\sigma^*= \max_{\xi} \frac{\sigma}{\varepsilon p (n-1) |2C|^{n/2-1}} \,  , \label{eq:defsigstar}
\ee
we get the curves of figure \ref{fig:compWKB}. For $n=2$, the rescaled growth rate varies increases monotonically with the libration magnitude $\varepsilon$ and $\beta_n$. In the elliptical case $n=2$, the impact of considering large deformation $\beta_2$ remains very small, but in the triangular case $n=3$, $\sigma^*$ may double in magnitude and increases very sharply when $\beta_3 \rightarrow 1$.

\subsubsection{Asymptotic analysis for small forcings $\varepsilon p \ll 1$} \label{sec:asympt}

\begin{figure}                   
  \begin{center}
    \begin{tabular}{ccc}
      \setlength{\epsfysize}{5.0cm}
      \subfigure[]{\epsfbox{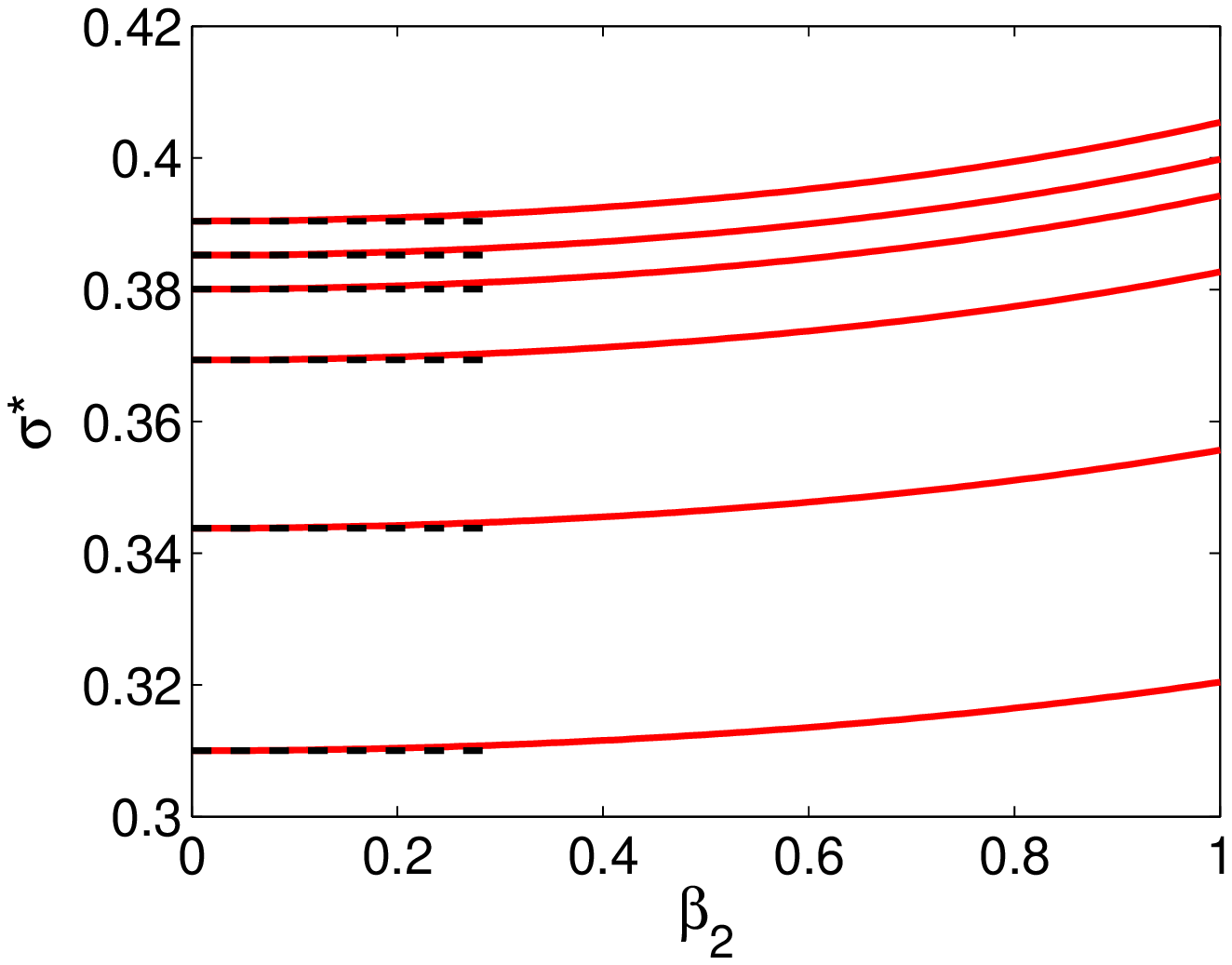}} &
      \setlength{\epsfysize}{5.0cm}
      \subfigure[]{\epsfbox{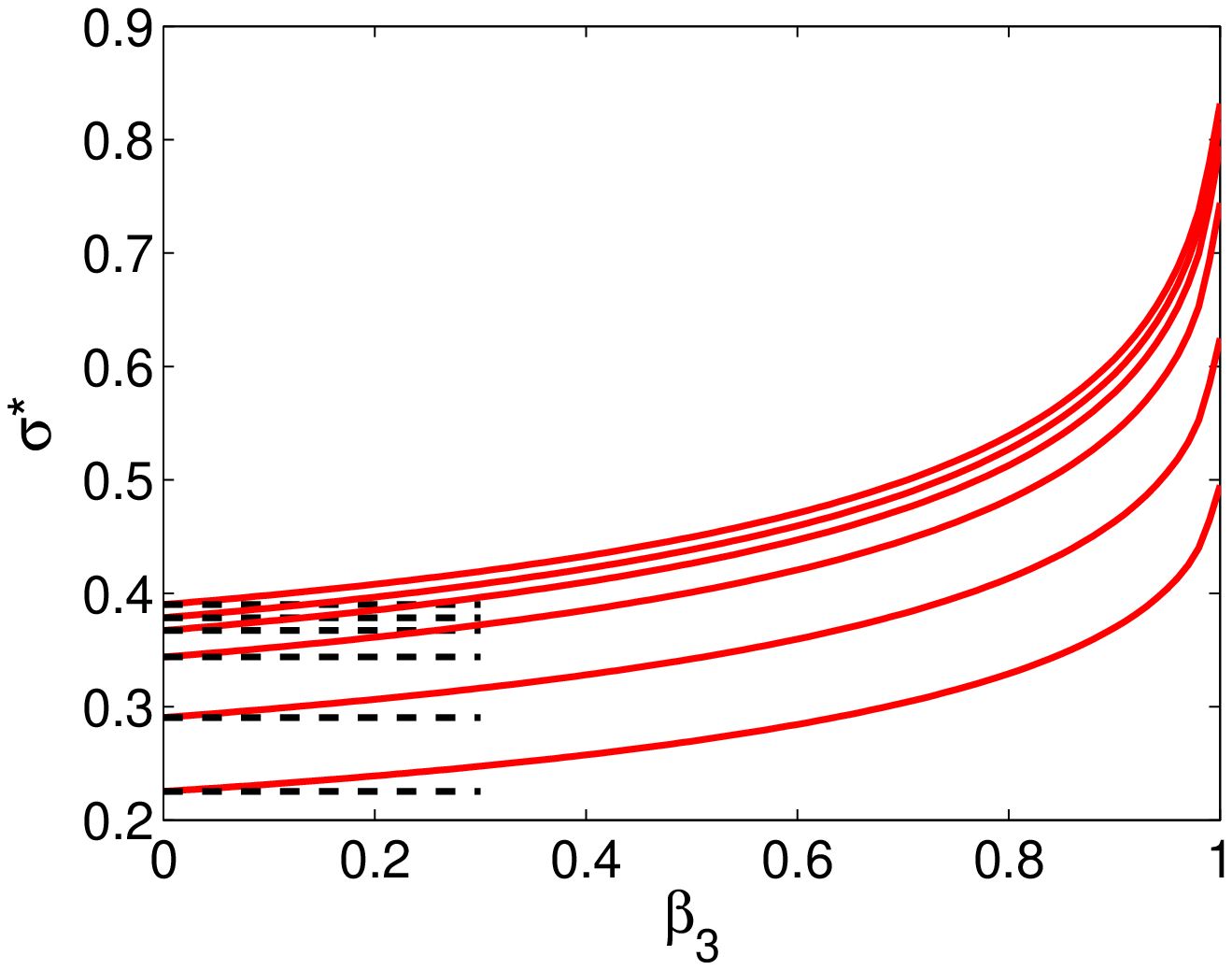}} \\
      \setlength{\epsfysize}{5.0cm}
      \subfigure[]{\epsfbox{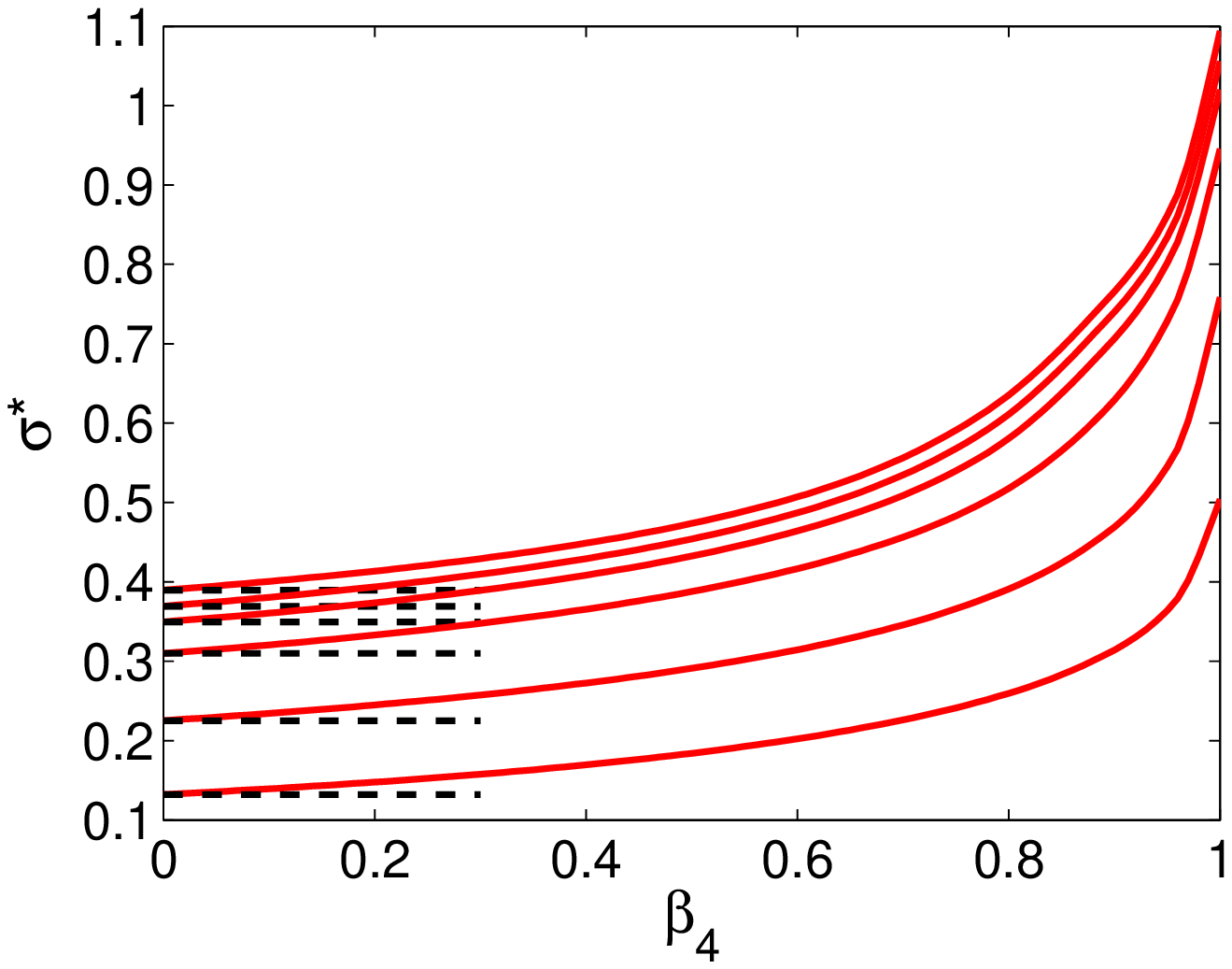}} &
      \setlength{\epsfysize}{5.0cm}
      \subfigure[]{\epsfbox{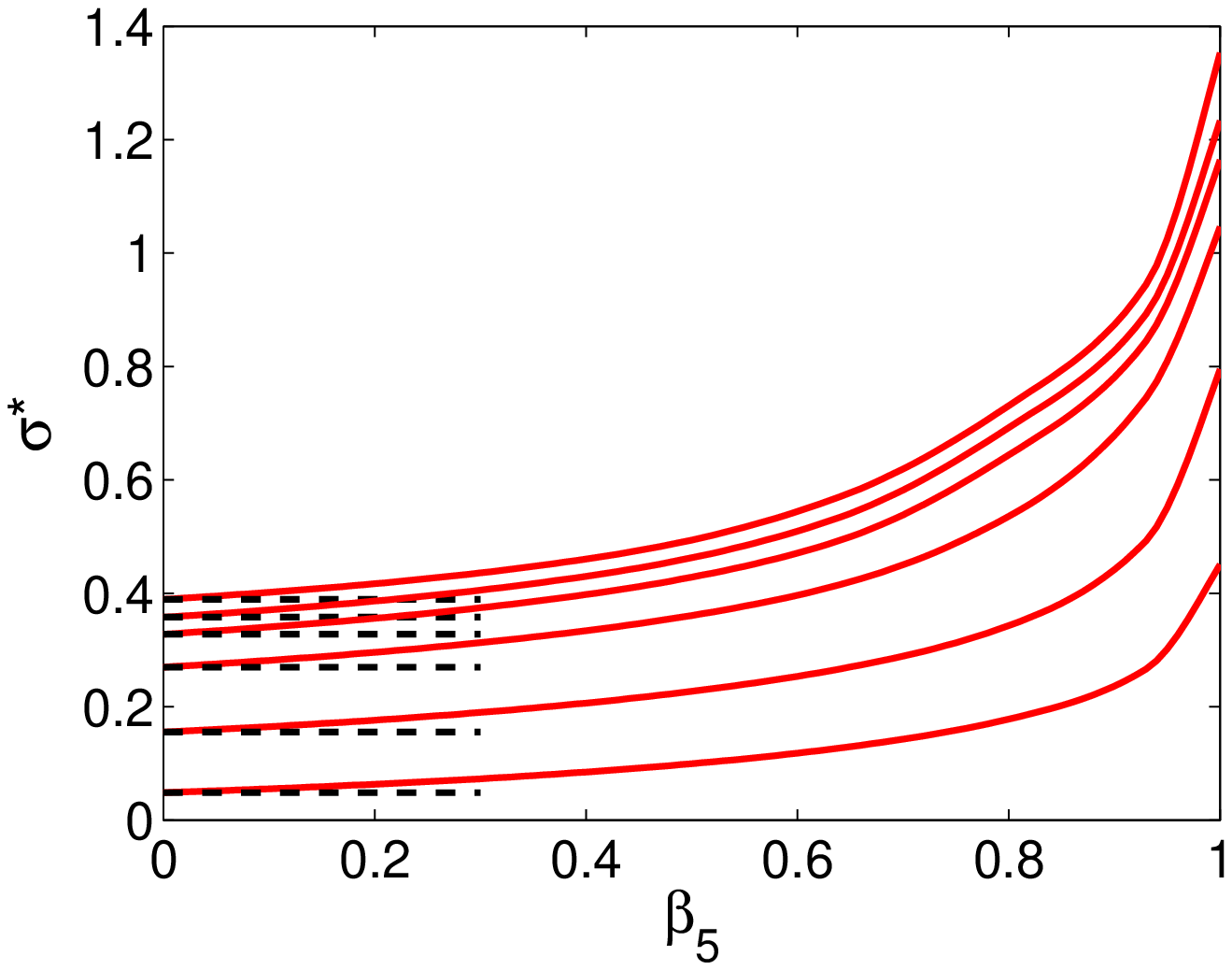}} 
    \end{tabular}
    \caption{Comparison of the analytical growth rate (black dashed lines) given by eq. (\ref{eq:sigWKB11}), obtained for $\varepsilon p \ll 1$, with the solution (red solid lines) of stability eq. (\ref{eq:lifshitz1})-(\ref{eq:lifshitz3}) for $\omega=3$ and $\varepsilon=0.1$, $0.5$, $0.7$, $1$, $1.5$, and $2$ (from the upper curve to the lowest one). (a) $n=2$. (b) $n=3$. (c) $n=4$. (d) $n=5$.}
         \label{fig:compWKB}                
  \end{center}
\end{figure}
In this section, we solve equations (\ref{eq:lifshitz1})-(\ref{eq:lifshitz3}) using a multiple scale analysis  \cite[e.g.][]{kevorkian1996multiple},  assuming that the product $\varepsilon p \ll 1$ remains small. Since the method is now rather classical \cite[e.g.][]{le2000three,herreman2009effects,cebronAA} we only give a short outline of the calculation. One first finds the trajectory as 
\be
\boldsymbol{X} (t) = \boldsymbol{X}^{(0)} (t)  + \varepsilon p \,  \boldsymbol{X}^{(1)} (t)  + \mathcal{O}(\varepsilon^2 p^2)  \, .
\ee
Here $ \boldsymbol{X}^{(0)} (t)$ is the circular trajectory induced by the solid-body rotation and $ \boldsymbol{X}^{(1)} (t)$ deviations induced by the multipolar deformation (see Appendix \ref{sec:localAnnexe} for the expressions of $\boldsymbol{X}^{(0)} (t)$ and $\boldsymbol{X}^{(1)} (t)$). With this, one can evaluate $\nabla \boldsymbol{U}$ on the perturbed trajectory, up to order $\mathcal{O}(\varepsilon p)$, allowing to solve for the wavenumber:
\ba
\boldsymbol{\mathcal{K}} (t) &=& \boldsymbol{\mathcal{K}}^{(0)} (t)  + \varepsilon p \,  \boldsymbol{\mathcal{K}}^{(1)} (t)  + \mathcal{O}(\varepsilon^2 p^2) \, .
\ea
At lowest order, we get 
\be
 \boldsymbol{\mathcal{K}}^{(0)} (t)  = \mathcal{K}_0 \Big [  \sin (\xi)  \cos ( t+ \phi)  \boldsymbol{e}_x + \sin (\xi)  \sin ( t+ \phi)  \boldsymbol{e}_y +  \cos (\xi) \boldsymbol{e}_z     \Big ].
\ee
Each wavenumber is specified by a magnitude $\mathcal{K}_0$, angle $\xi$ and initial phase $\phi$. This rotating wavenumber is actually stationary in the rotating frame and represents a plane wave there. The first order deviation $\boldsymbol{\mathcal{K}}^{(1)} (t)$ is important in the stability calculation and its expression is thus given in the Appendix \ref{sec:localAnnexe}.
At leading order, for $\varepsilon p = 0$, these equations can be reduced to an harmonic equation for the amplitude $a_z$ of the axial velocity plane wave perturbation,
\be
\frac{\dd^2 a_z^{(0)}}{\dd t^2}  + \Lambda^2 a_z^{(0)} = 0  \, , 
\ee 
showing that the amplitude oscillates with angular frequency $\Lambda = \pm 2 \cos \xi$. This can be identified as the usual inertial wave dispersion relation in an unbounded fluid domain. We modify the expansion to make a superposition of two waves
\be
a_z= \left[ c_1 \, \mathrm{e}^{ \mathrm{i} \Lambda t }  + c_2 \, \mathrm{e}^{-\mathrm{i}  \Lambda t}  + \varepsilon p \, a_z^{(1)} \right] \, \mathrm{e}^{\varepsilon p\,  \hat{\sigma} t}  \, 
\ee
and give them a common but small growth rate $\sigma = \varepsilon p\,  \hat{\sigma}$ ($\hat{\sigma}$ being our unknown). Injecting this in the $\mathcal{O}(\varepsilon p)$-balance, we get an equation of the form 
\be
\frac{\dd^2 a_z^{(1)}}{\dd t^2}  +  \Lambda^2 a_z^{(1)} = \sum_{j \in \mathbb{Z}} \left [  F_j \,  \mathrm{e}^{\mathrm{i} ( \Lambda + j \omega) t }  + G_j \,  \mathrm{e}^{\mathrm{i} (- \Lambda + j \omega) t }  \right ]  \, .
\ee

\begin{table}
  \begin{center}
\def~{\hphantom{0}}
  \begin{tabular}{lccc}
      Parameters  & Definition   \\[3pt]
       $\boldsymbol{a}$   & Amplitude of the velocity plane wave perturbation  \\
       $\boldsymbol{\mathcal{K}}$   & Local wavevector along a Lagrangian trajectory  \\
       $\xi$   & Initial angle of $\boldsymbol{\mathcal{K}}$ with the vorticity axis $\boldsymbol{e}_z$ \\
       $T$ & Turnover period along a pathline  \\
       $p$  & Multipolar deformation amplitude  \\
       $\varpi_i$  & Floquet exponents of the monodromy matrix $\boldsymbol{a}(T)$  \\
       $\Lambda = \pm 2 \cos \xi$  & Inertial wave angular frequency in an unbounded domain \\
       $j$ & Difference (in $\omega$ units) between the resonant inertial waves angular frequencies\\
       $\hat{\cdot}=\cdot/(\varepsilon p)$ & Operator (dividing the quantity by $\varepsilon p$)  \\
       $\sigma$ & Inviscid growth rate  \\
       $\sigma^*$ & Scaled maximum inviscid growth rate (defined by eq. \ref{eq:defsigstar})  \\
       $\sigma_v$ & Viscous growth rate \\
       $k,m,l$ & Inertial waves radial, azimuthal and axial wavenumbers in a bounded geometry \\
       $\lambda$ & Inertial wave angular frequency in a bounded domain \\
       $ \varsigma$ & Frequency detuning \\
       $\alpha_w$ & Viscous damping of the resonant inertial waves ($w=1, \, 2$)
  \end{tabular}
  \caption{Stability analysis parameters. }
  \label{tab:2}
  \end{center}
\end{table}

\begin{figure}                   
  \begin{center}
    \begin{tabular}{ccc}
      \setlength{\epsfysize}{5.0cm}
      \subfigure[]{\epsfbox{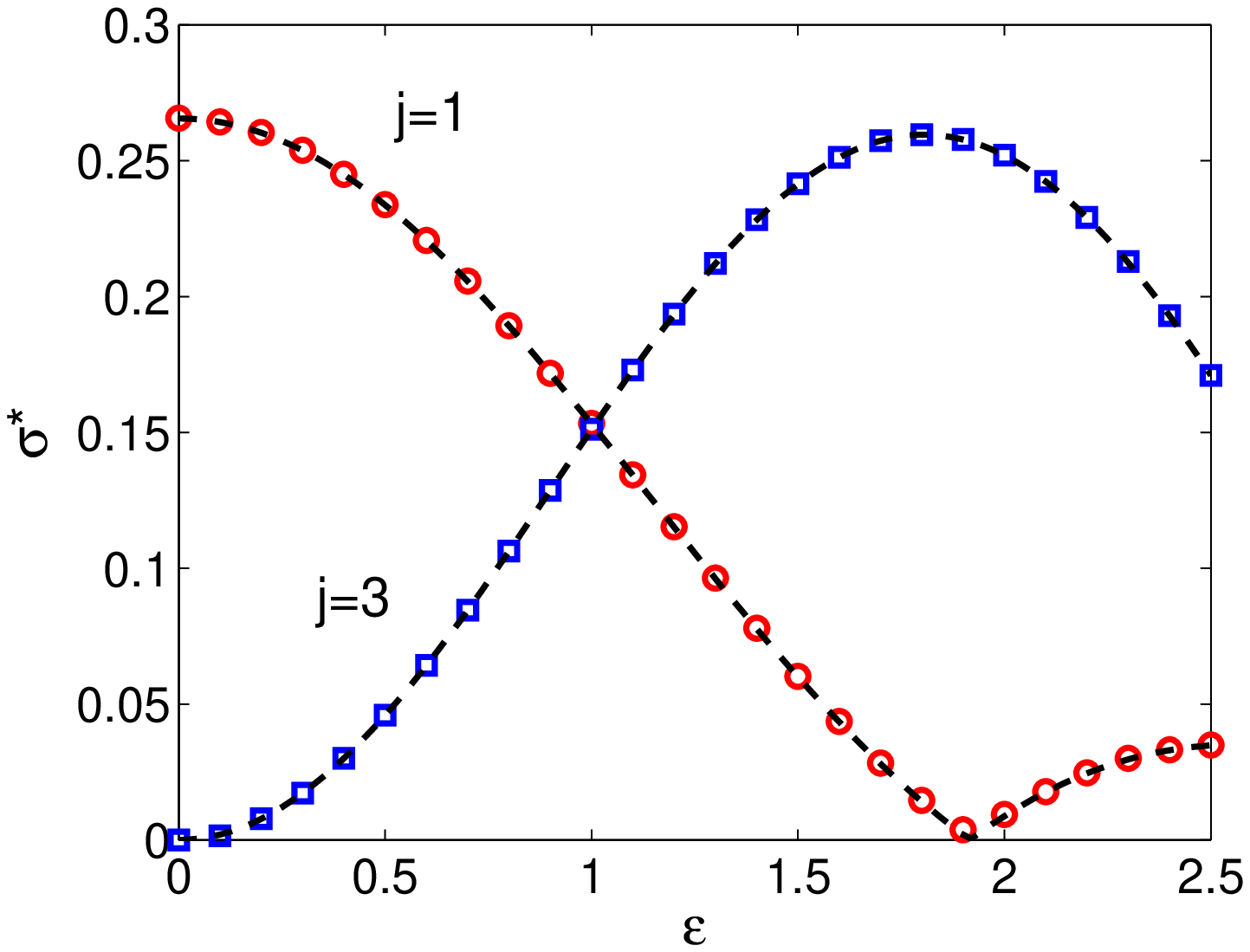}} &
      \setlength{\epsfysize}{5.0cm}
      \subfigure[]{\epsfbox{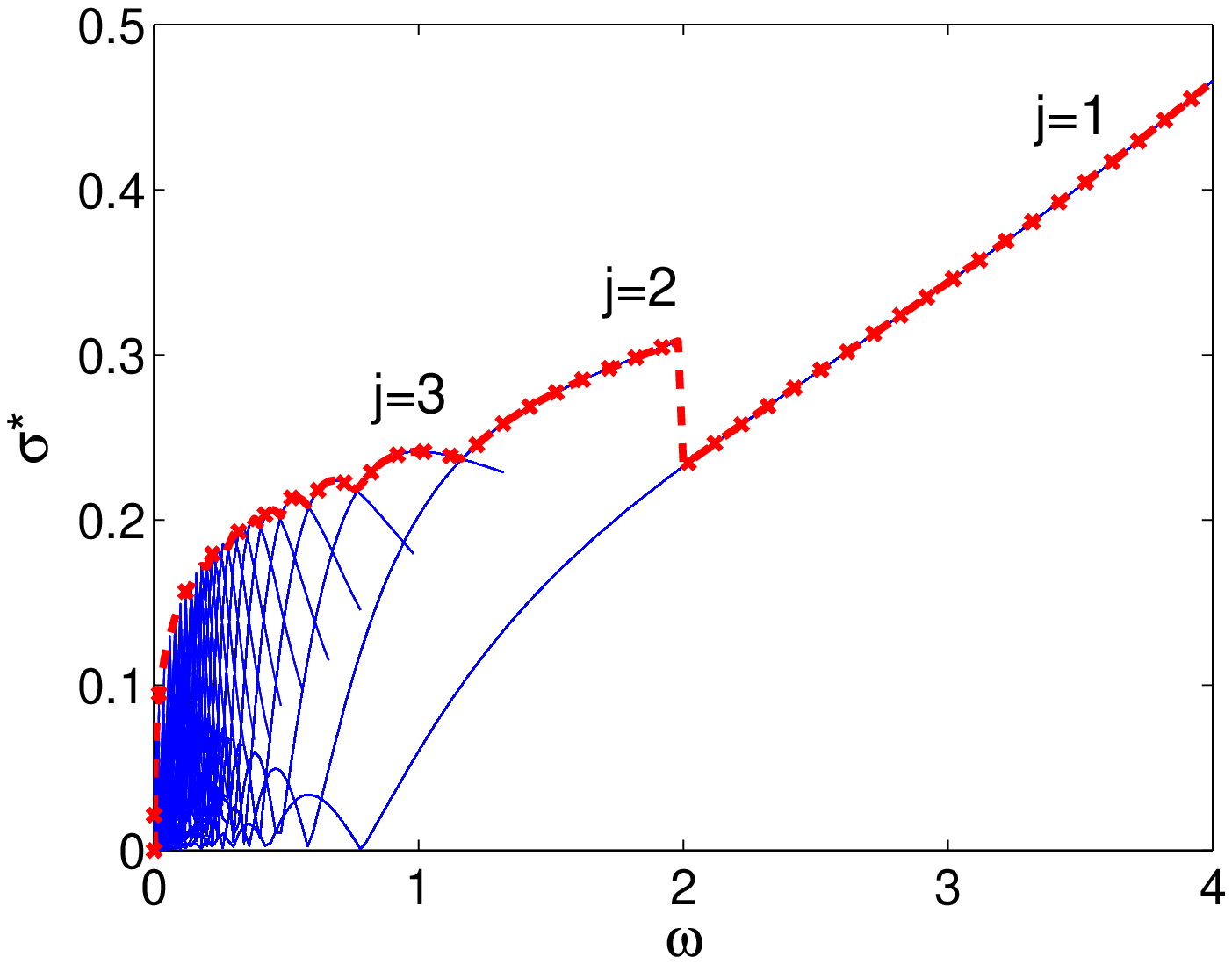}}
    \end{tabular}
    \caption{(a) Reduced growth rate $\sigma^*$ against the libration amplitude $\varepsilon$ for the two resonances $j=1$ and $j=3$ ($n=2$, $\omega=1$) in the limit of small deformation $p$. Symbols (circles for $j=1$, squares for $j=3$) are obtained by a direct numerical solving of the stability eq. (\ref{eq:lifshitz1})-(\ref{eq:lifshitz3}) with $p=0.015$, whereas the dashed lines correspond to the analytical formula (\ref{eq:sigWKB11}), obtained for $\varepsilon p \ll 1$. (b) $\sigma^*$ against the libration frequency $\omega$ ($n=3$, $\varepsilon=1$),  for all the resonances $j\in 1,\ldots, j_{max}$ (blue solid lines). The number of resonances $j_{max}$ changes with $\omega$. The envelope gives the maximum growth rate $\sigma^*$ for a given $\omega$ (red dashed line with cross symbols).}
    \label{fig:localfig}            
  \end{center}
\end{figure}

\noindent Secular terms on the right hand side, appear for all resonant frequencies 
\be
\Lambda_j = \pm j \omega / 2  \quad, \quad j \leq j_{max} \, , \label{eq:localres}
\ee
with $j \in \mathbb{N}^{*}$ and $0<4/\omega -j_{max} \leq 1$. Due to the inertial wave dispersion relation, each value of $j$ corresponds to a resonant angle $\xi_j = \arccos ( j \omega / 4)$. These resonant angles, found in the limit $\varepsilon p \ll 1$, are the points where the Floquet resonance tongues emanate. Note that $j_{max}=1$ for $\omega>2$, which means that only one Floquet tongue exists (see figure \ref{fig:tongues}). The bounds for $j_{max}$ also show that no instability is possible for $|\omega| > 4 +\mathcal{O}(\varepsilon p)$. This is called the forbidden zone of the LDMI \cite[e.g.][for the forbidden zone of the TDEI]{le2007coriolis}. Note that, at order $1$ in $\varepsilon p$, this band is extended to $|\omega| > 4+\varepsilon p +\mathcal{O}(\varepsilon^2 p^2)$ \cite[exactly as for the case considered by][]{le2000three}. The growth rate $\sigma$ is found after posing the solvability condition: multiply the right hand side with $\exp (\pm \mathrm{i}  \Lambda_j t)$ and integrate over a period $2 \pi / \Lambda_j$.  We then find a homogenous system of algebraic equations for $c_1$ and $c_2$, that defines the growth rate (valid for $\omega \neq 0$):  
\be
\sigma^* = \frac{\sigma}{\varepsilon p (n-1) |2C|^{n/2-1}}  = \frac{16+(j \omega)^2}{64}   \Big | \mathrm{J}_{j-1} ( n \Delta \varphi ) +\mathrm{J}_{j+1} ( n \Delta \varphi ) \Big | \,   .   \label{eq:sigWKB11}
\ee
The coefficient $C$ relates to the streamline (and thus initial position) under consideration. The maximum growth rate is obtained on the most deformed boundary streamline where $C=-1/2$. We recognize the Bessel function factor (equal to $4 f_j$) of (\ref{eq:fdef}), which shows that each frequency component in the basic flow can couple different pairs of resonant modes, a particularity of libration driven multipolar instabilities. Figure \ref{fig:compWKB} shows that the analytical asymptotic growth rate given by equation (\ref{eq:sigWKB11}) allows a quite accurate prediction of the growth rate for values of $\beta_n$ up to $0.3$. As expected, the results differ for large $\beta_n$ but in a rather small extent. In figure \ref{fig:compWKB}, the libration frequency is greater than $2$ ($\omega=3$), and only one resonance is thus possible ($j_{max}=1$). In most cases, several resonances are possible and figure \ref{fig:localfig}(a) compares the formula (\ref{eq:sigWKB11}) with the exact solution of stability equation in such a case ($\omega=1 \Rightarrow j_{max}=3$), showing that the $\varepsilon$-dependency is exactly captured. 
Note that, for a given $\omega$, the local growth rate can go to zero for particular values of $\varepsilon$. Again, this is a particular feature of the LMDI. When the libration angle $\Delta \varphi = \varepsilon / \omega$ is such that $\mathrm{J}_{j-1} ( n \Delta \varphi ) +\mathrm{J}_{j+1} ( n \Delta \varphi ) = 0$, the base-flow does not have $j \omega$ frequency content (see \ref{eq:fdef}) and instability is then impossible. 
Figure \ref{fig:localfig}(b) completes this description of multiple resonances by representing the results of equation (\ref{eq:sigWKB11}) in function of $\omega$. This clearly shows the decrease of the number of resonances when $\omega$ is increased.

We now consider two different interesting limit cases of the formula (\ref{eq:sigWKB11}): the case of large libration forcing $\varepsilon \gg \omega/n$, and then the case of small libration forcing $\varepsilon \ll \omega/n$. Note that the formula (\ref{eq:sigWKB11}) has been obtained in the limit of small forcing $\varepsilon p  \ll 1$ ($p$ or $\varepsilon$ can thus have finite values if the other one is very small). This imposes that the limit $\varepsilon \gg \omega/n$ is actually $\omega p /n \ll \varepsilon p \ll 1$ whereas the other one is given by $\varepsilon p \ll \min(1,\omega p /n )$. 

First, we focus on the scaling of $\sigma^*$ when $\omega p /n \ll \varepsilon p \ll 1$, which implies $n \Delta \varphi \gg 1$. This limit case is interesting because it allows us to test the possibility of a LDMI excited by a small-scale periodic pattern in the boundary roughness ($n \gg 1$). In this case, using $\mathrm{J}_{j-1}(x) + \mathrm{J}_{j-1} (x)=2j\, \mathrm{J}_{j}(x) / x  $, we obtain 
\be
\sigma^* = \frac{16+(j \omega)^2}{32}\, j\, \sqrt{\frac{2}{\pi (n \Delta \varphi)^3}}\,   \left| \cos (n \Delta \varphi- j \pi/2 - \pi/4) \right|\, + \mathcal{O} \left( \frac{1}{(n \Delta \varphi)^{5/2}} \right) \,  . \label{eq:sigGrand}
\ee
Then, the maximum growth rate is reached for $j=j_{max} \approx 4/\omega$ and reads thus
\be
\sigma^* \sim 4\,  \sqrt{\frac{2\, \omega}{\pi n^3\, \varepsilon^3}} \quad  \Rightarrow  \quad  \sigma \sim 4\, p\,  |2C|^{n/2-1}\, \sqrt{\frac{2\, \omega}{\pi n\, \varepsilon}}\, . \label{eq:sigGrand2}
\ee
The growth rate $\sigma$ decreases thus towards $0$ as $n^{-1/2}$ in the limit $n \gg \omega/\varepsilon$. Considering a boundary of arbitrary shape with many multipolar components ($n=2, 3, ...$), this shows that the instabilities excited by the lowest orders $n$ will be the more unstable.

The opposite limit case $ n \Delta \varphi \ll 1$ corresponds to the asymptotic regime of small libration forcings $\varepsilon \ll \omega / n$, which is the relevant limit in geophysics. In this case, only the first resonance $j=1$ can contribute (since $J_0 (0) = 1$ and  $\forall j \neq 0,\, \mathrm{J}_{j} (0) = 0$). We then have  a single unstable mode with rescaled growth rate
\be
  \sigma^* =  \frac{16 + \omega^2}{64} \,  ,     \label{eq:sigWKB12}
\ee
which is in agreement with the result previously obtained in the particular case $n=2$ \cite[][]{Cebron2012MHD,Cebron_Pof}. In the limit of $\omega \rightarrow 0$ (absence of libration), i.e. in the limit of the so-called synchronization state in an astrophysical context \cite[e.g.][]{cebronAA}, the growth rate tends toward $\sigma^*=1/4$. This limit can also be obtained from the results of \cite{le2000three} who considers the case of a constant non-zero differential rotation between the multipolar strain and the fluid. In the limit of infinitesimal multipolar deformation, this case leads to an instability only for $n \leq 4$, and \cite{le2000three} shows that the maximum inviscid growth rate is then given by (using the constant differential rotation as the timescale)
\begin{eqnarray}
\frac{\bar{\sigma}}{p}= \frac{[n+4(1+\Omega^G)]^2}{64\ (1+\Omega^G)^2}\ (n-1) \,  , \label{eq:steph}
\end{eqnarray}
where $\Omega^G=\Omega_{orb}/(\Omega-\Omega_{orb})$, with $\Omega_{orb}$ the rotation rate of the strain in the inertial frame. The expression (\ref{eq:steph}) is consistent with the expression (\ref{eq:sigWKB12}) in the limit of synchronized state:
$\Omega_{orb}/\Omega_{spin}=1-\varepsilon$ i.e. $ |1+\Omega^G| \sim
|\Omega^G|=1/\varepsilon $ and $\bar{\sigma} = \sigma / \varepsilon$ (to have the same timescale).

\subsection{Global stability analysis} \label{sec:global_stability}
 
 \subsubsection{Problem definition}
 
The linear stability problem for the perturbation flow $\boldsymbol{u}$ and modified pressure $\pi$ writes:
\be
 \pd_t \boldsymbol{u} + \boldsymbol{U} \cdot \nabla \boldsymbol{u}  + \boldsymbol{u} \cdot \nabla \boldsymbol{U}  + 2 \boldsymbol{e}_z \times \boldsymbol{u} = - \nabla \pi + E\, \Delta \boldsymbol{u}  \quad , \quad \nabla \cdot \boldsymbol{u} = 0 \, ,
\ee
in the rotating frame of reference (R) used for our global analysis. We will again omit the use of these superscripts. No-slip boundary conditions are imposed on the deformed cylinder's surface. We adopt a quadri-vector notation $\boldsymbol{Y} = [u_r , u_\theta, u_z, \pi]^T $, so that the problem may be rewritten in compact form 
\be
 \mathcal{L}   \boldsymbol{Y} = \varepsilon p\  \,  \left [ \mathrm{e}^{\mathrm{i} n \theta} \, f( t ) \,  \mathcal{N} + c.c.  \right ]    \boldsymbol{Y}  + E\, \mathcal{V} \boldsymbol{Y} \, .\label{eq:opdef}
\ee
The operators $ \mathcal{L} ,  \mathcal{N} ,  \mathcal{V} $ are defined in Appendix \ref{sec:annexOperator}. We can now investigate the stability of the flow using an asymptotic model, valid for small forcings $\varepsilon p\ll 1$ and small Ekman numbers $E \ll 1$. 

\subsubsection{Leading order solution: inertial waves in cylinders}

In absence of boundary deformation ($p=0$) and viscosity ($E=0$), the perturbation flow is solution of
\be
 \mathcal{L}  \boldsymbol{Y}^{(0)}   = 0 \quad , \quad u_{r}^{(0)} (1,\theta,z,t) = 0\, .
\ee
The solution to this problem is a general superposition of inertial waves in cylindrical geometry;
\be
\boldsymbol{Y}^{(0)}  (r,\theta,z,t)  = \sum_{m l \lambda}  A_{m l \lambda} \,  \boldsymbol{Q}_{m l \lambda}  (r) \,  \mathrm{e}^{\mathrm{i} m \theta} \,\mathrm{e}^{\mathrm{i} l z}  \mathrm{e}^{\mathrm{i} \lambda  t} \, .
\ee
Here $A_{m l \lambda}$ are arbitrary amplitudes. Each wave is specified by an angular frequency $\lambda \in [-2,2]$ an azimuthal wavenumber $m \in \mathbb{Z}$ and a axial wavenumber $l= 2 \pi n_z  / h$ with $n_z \in \mathbb{Z}$ the number of axial wavelengths. Note that both $\lambda$ and $\Lambda$ denote inertial wave frequencies, but the former is related to inertial waves in a radially bounded cylinder, whereas the latter to those in an unbounded medium. The radial profiles  $\boldsymbol{Q}_{m l \lambda}$ are solutions of  $\mathcal{L}_{m l \lambda } \boldsymbol{Q}_{m l \lambda} = 0 $, where $\mathcal{L}_{m l \lambda} $ corresponds to the operator $\mathcal{L}$ in which $\pd_t \rightarrow  \mathrm{i} \lambda$, $\pd_\theta \rightarrow \mathrm{i} m $, $\pd_z \rightarrow \mathrm{i} l $ have been replaced. We find
\be
 \boldsymbol{Q}_{m l \lambda} =
\left [ \begin{array}{c} 
\ \ \ \,  \left [  (2-\lambda)\, \mathrm{J}_{m+1} (k r) + (2+\lambda)\, \mathrm{J}_{m-1} (k r) \right ] / 2\\
 - \mathrm{i} \left [  (2-\lambda)\, \mathrm{J}_{m+1} (k r) - (2+\lambda)\, \mathrm{J}_{m-1} (k r) \right ] / 2 \\
- \mathrm{i} (k \lambda / l )\, \mathrm{J}_m ( kr)  \\
 \mathrm{i} (k \lambda^2/l^2)\, \mathrm{J}_m ( k r)
 \end{array} \right ]  \quad , \quad  k^2 = \frac{(4- \lambda^2)}{\lambda^2} \, l^2 \, .
\ee
%
The radial wavenumber $k$ and the frequencies are discretized by the boundary conditions on the surface $r=1$, that together with the previous definition of $k$, fixes the inertial wave dispersion relation:
\be\label{dispcyl}  
(2-\lambda)\, \mathrm{J}_{m+1} (k) + (2+\lambda)\, \mathrm{J}_{m-1} (k ) = 0  \quad , \quad \lambda =\pm  2 \sqrt{\frac{l^2}{k^2 +l^2}}\, .
\ee
For fixed $m,l$ and both possible signs of the frequency, this equation admits a countable infinite number of discrete radial wave numbers, that are easily identified numerically. We label these radial wave numbers of the frequency with a radial counter $n_r = 1,2,\ldots$. A wave is entirely determined by the set $(k,m,l,\lambda)$ or alternatively $(n_r,m,n_z,\lambda)$ in a cylinder with fixed height. The frequency $\lambda$ is such that $\lambda \in [-2,2]$, which is a general property of inertial waves.

\subsubsection{Inviscid growth rate $\sigma$ at resonance}

As in the local theory, we will now propose a asymptotic solution that is a superposition of two waves and a remainder:
\be 
\boldsymbol{Y} = \Big ( c_1\,  \boldsymbol{Q}_{1} (r)  \,   \mathrm{e}^{\mathrm{i} m_1 \theta} \mathrm{e}^{\mathrm{i} \lambda_1 t} + c_2\,   \boldsymbol{Q}_{2}  (r) \ \mathrm{e}^{\mathrm{i} m_2 \theta}   \mathrm{e}^{\mathrm{i} \lambda_2 t}  + \varepsilon p\,  \boldsymbol{Y}^{(1)}    \Big) \mathrm{e}^{\mathrm{i} l z} \mathrm{e}^{\varepsilon  p \hat{\sigma} t}  + \mathcal{O}(\varepsilon p^2) \, . \label{eq:ansatzglobal}
\ee
Here $\boldsymbol{Q}_{j}$ is shorthand for $\boldsymbol{Q}_{m_j l_{j} \lambda_j} $. We search for an expression for the growth rate $\sigma = \varepsilon p \, \hat{\sigma} $. Injecting this ansatz in the previous system of equations (\ref{eq:opdef}) we have:
\ba
 \mathcal{L} \boldsymbol{Y}^{(1)}   =  \left [- \hat{\sigma} \mathcal{J} +  \left ( \mathrm{e}^{\mathrm{i} n \theta} \, f( t ) \,  \mathcal{N} + c.c.  \right )  \right ]   \quad \quad \quad \quad \quad \quad \quad   \\
\times  \left [  c_1\,  \boldsymbol{Q}_{1} (r)  \,   \mathrm{e}^{\mathrm{i} m_1 \theta } \mathrm{e}^{\mathrm{i} \lambda_1 t} + c_2\,   \boldsymbol{Q}_{2}  (r) \ \mathrm{e}^{\mathrm{i} m_2 \theta }   \mathrm{e}^{\mathrm{i} \lambda_2 t}  \right ] \, . \nonumber
\ea
The right hand side is secularly forcing the left hand side, whenever two waves satisfy resonance conditions 
\be \label{resonance}
m_1 + n = m_2 \quad , \quad l_{1}  = l_{2}  = l   \quad , \quad \lambda_1 + j \, \omega = \lambda_2\, ,
\ee 
for $j \in \mathbb{Z}$. This is the global equivalent of (\ref{eq:localres}). Considering the inertial wave dispersion relation, this is a quite restrictive constraint that certainly not all pairs of inertial waves will be able to satisfy. Exact resonances cannot be found when $|j \omega| > 4$ because $\lambda_1, \lambda_2 \in [-2,2]$. We shortnote each resonance by a quintuplet $(m_1,m_2,n_r,n_z,j)$. Here $n_r$ is the radial wavenumber label, $n_z$ the number of vertical wavelengths. As in the study of Eloy, we systematically find the largest growth rates for central couplings that pair waves with the same radial label $n_{r,1} = n_{r,2}$, synonymous for $k_{1} \simeq k_{2}$. The field $\boldsymbol{Y}^{(1)}$ is necessarily composed of 
\be 
\boldsymbol{Y}^{(1)} = \boldsymbol{Z}_1 (r,z) \, \mathrm{e}^{\mathrm{i} m_1 \theta}  \mathrm{e}^{\mathrm{i} \lambda_1 t} + \boldsymbol{Z}_2 (r,z) \, \mathrm{e}^{\mathrm{i} m_2 \theta}  \mathrm{e}^{\mathrm{i} \lambda_2 t}  +   \boldsymbol{Z}_{NR} (r,\theta,z,t) \, .
\ee
The last term $\boldsymbol{Z}_{NR} $ absorbs non-resonant contributions. Injected in the previous equation, we get the secularly forced system:
\ba
 \mathcal{L}_{1}  \boldsymbol{Z}_1  & = &  - \hat{\sigma} \, c_1\,  \mathcal{J}    \boldsymbol{Q}_1 +   f_j  \,c_2\,  \mathcal{N}^\dagger  \boldsymbol{Q}_2  \label{eq:Z1} \,  ,   \\
 \mathcal{L}_{2}  \boldsymbol{Z}_2  & = &  - \hat{\sigma} \,  c_2\,   \mathcal{J}   \boldsymbol{Q}_2 +   f_{j}  \, c_1\, \mathcal{N}  \boldsymbol{Q}_1  . \label{eq:Z2}  \, .
 \ea
Here $\mathcal{L}_i = \mathcal{L}_{m_i l_{i} \lambda_i} $. We used $(f^\dagger)_{-j}  =  f_j $ in (\ref{eq:Z1}) (see eq. \ref{eq:fdef}).  A solvability condition fixes the growth rates but in order to write it, we need a well adapted scalar product. We choose here
\be
\langle  \boldsymbol{Q}_1 , \boldsymbol{Q}_2 \rangle =   \int_0^1 \left ( \sum_{\mu=1}^4  Q_{1,\mu}^\dagger Q_{2,\mu}^{\, } \right ) \,   r\, \dd r \, .
\ee 
This scalar product is advantageous for our calculations, because direct and adjoint inertial modes are then identical \cite[][]{eloy2003elliptic}. Indeed, if we consider that adjoint modes satisfy the same boundary conditions than direct modes (a zero normal velocity), an integration by parts gives (noting the adjoint with a superscript $A$):
\be
\langle  \boldsymbol{Q}_{m l \lambda}^A , \mathcal{L}_{m l \lambda} \boldsymbol{Q}_{m l \lambda} \rangle =  - \langle \mathcal{L}_{m l \lambda} \boldsymbol{Q}_{m l \lambda}^A ,  \boldsymbol{Q}_{m l \lambda} \rangle \, .
\ee 
The adjoint operator is thus the opposite of the direct operator ($ \mathcal{L}_{m l \lambda}^A= - \mathcal{L}_{m l \lambda}$), and direct and adjoint modes are identical with this scalar product. 

The solvability condition is expressed by projecting (\ref{eq:Z1}), (\ref{eq:Z2}) onto $\boldsymbol{Q}_1$ and $\boldsymbol{Q}_2$. On the left hand-side, we use a partial integration and the definition of the adjoint modes, e.g. for the first equation:
\ba
\langle  \boldsymbol{Q}_{1} ,  \mathcal{L}_{1}  \boldsymbol{Z}_1 \rangle  &=& -\langle   \underbrace{ \mathcal{L}_{1}  \boldsymbol{Q}_{1} }_{0},    \boldsymbol{Z}_1  \rangle +   Q_{1,4}^{\dagger} (1) \,Z_{1,1}(1) \, . \label{IP}
\ea
This process introduces boundary terms that appeal for first order corrections $Z_{1,1}$, $Z_{2,1}$ of the radial velocities of the modes. It is through these terms that $\mathcal{O}(p)$ couplings related to boundary deformation (see Appendix \ref{sec:annex_globalwall} for more details) arrive in the global stability analysis. The solvability condition thus results in a homogenous system of 2 algebraic equations: 
\ba
c_1 \left (  - \sigma   \, \mathcal{J}_{11}    \right )   + c_2 \left ( \varepsilon p \, f_{j} \,  \mathcal{N}_{12}  -  p  \, g_{j} \,  \mathcal{B}_{12} \right ) & = & 0 \, , \label{eqalg1} \\
 c_2  \left (  - \sigma   \, \mathcal{J}_{22}  \right ) + c_1 \left ( \varepsilon p \, f_{j} \,  \mathcal{N}_{21}  -  p  \, g_{j} \,  \mathcal{B}_{21} \right )    & = & 0 \,  ,  \label{eqalg2}
\ea
with matrix elements
\ba
&&\mathcal{J}_{11} = \langle  \boldsymbol{Q}_1 ,  \mathcal{J}   \boldsymbol{Q}_1 \rangle  \quad , \quad \mathcal{N}_{12} = \langle  \boldsymbol{Q}_1 ,  \mathcal{N}^\dagger   \boldsymbol{Q}_2 \rangle \,  \nonumber \\
&& \mathcal{J}_{22} = \langle  \boldsymbol{Q}_2 ,  \mathcal{J}   \boldsymbol{Q}_2 \rangle  \quad , \quad \mathcal{N}_{21} = \langle  \boldsymbol{Q}_2 ,  \mathcal{N}  \boldsymbol{Q}_1 \rangle \, ,
\ea
and boundary terms
\ba
\mathcal{B}_{12} &=& Q_{1,4}^\dagger (1)  \left ( -\frac{1}{n} \pd_r Q_{2,1} (1) - \mathrm{i} \,Q_{2,2} (1) \right ) \, , \nonumber \\
\mathcal{B}_{21} &=& Q_{2,4}^\dagger (1)  \left ( -\frac{1}{n} \pd_r Q_{1,1} (1) + \mathrm{i} \,Q_{1,2} (1) \right ) \, . \label{eq:BT}
\ea
We used $(g^\dagger)_{-j} = g_j$ in (\ref{eqalg1}).  Elimination of $c_1$ and $c_2$ results in an equation for the inviscid growth rate at resonance:
\ba
 \sigma^2  =   \frac{ \left ( \varepsilon  \, f_{j} \,  \mathcal{N}_{12}  -    \, g_{j} \,  \mathcal{B}_{12} \right )  \left ( \varepsilon  \, f_{j} \,  \mathcal{N}_{21}  -    \, g_{j} \,  \mathcal{B}_{21} \right )  }{\mathcal{J}_{11} \mathcal{J}_{22}} p^2 \,  .
\ea
We have instability only when the real part of the right hand side is positive. Note that the relation $c_1/c_2$ is fixed by one of the equations of (\ref{eqalg1}).

\subsubsection{Frequency detuning \& viscous damping: growth rate $\sigma_v$} \label{sec:detuning}

Pairs of waves that perfectly satisfy resonance conditions (\ref{resonance}) can only exist in cylinders with well chosen height $h$, or for well defined frequencies $j \omega$. These conditions may however be relaxed in order to admit imperfect resonance. In the present article, we build these detuning effects in through a frequency detuning. Imperfect resonance then supposes that
\be
\lambda_1 + j \omega = \bar{\lambda} + \varsigma \quad ,\quad 
\lambda_2 = \bar{\lambda} - \varsigma ,  \label{eq:detun_omega}
\ee
with $\varsigma = \varepsilon p\, \hat{\varsigma} \ll 1$ a small frequency detuning and $\bar{\lambda}$ a modified resonant frequency. This detuning is introduced in the model by modifying the asymptotic ansatz into 
\be 
\boldsymbol{Y} = \Big ( c_1\,  \boldsymbol{Q}_{1} (r)  \,   \mathrm{e}^{\mathrm{i} m_1 \theta} \mathrm{e}^{\mathrm{i}( \bar{\lambda} - j \omega ) t} + c_2\,   \boldsymbol{Q}_{2}  (r) \ \mathrm{e}^{\mathrm{i} m_2 \theta}   \mathrm{e}^{\mathrm{i} \bar{\lambda} t}  + \varepsilon p\,  \boldsymbol{Y}^{(1)}    \Big) \mathrm{e}^{\mathrm{i} l z} \mathrm{e}^{\varepsilon  p \hat{\sigma} t}  + \mathcal{O}(\varepsilon p^2) .  
\ee
Starting form this ansatz, the solvability condition leads to the set of equations (\ref{eqalg1})-(\ref{eqalg2}) in which  $- \sigma \rightarrow - \sigma + \mathrm{i} \varsigma $ in (\ref{eqalg1}) and  $- \sigma \rightarrow - \sigma - \mathrm{i} \varsigma $ in (\ref{eqalg2}) are modified. 

Viscous corrections induced by boundary layers, which scales as $\sqrt{E}$, formally enter the model through boundary terms as (\ref{eq:BT}) as boundary layer pumping modifies the radial velocity at order $\sqrt{E}$. Volume damping can be formally introduced through the operator $\mathcal{V}$. Still since viscous effects are not modified by the libration at lowest order, it is sufficient to use preexisting formula  \cite[][]{kerswell1995viscous}. Waves with positive frequencies $\lambda_j > 0$ have a damping 
\be
\alpha_w =\sqrt{E} \,   \frac{(1+\mathrm{i})}{4 \sqrt{2}}\frac{(\,4-\lambda_w^2 \,) (m_w^2+l_{w}^2) \sqrt{\lambda_w }      }{ \, \left (m_w^2+l_{w}^2-m_w \lambda_w / 2 \right )  }
   + E\, (k_{w}^2 +l_{w}^2)   \, ,
\ee
for $w=1,2$. If $\lambda_w <0$ we need to use the complex conjugate formula. Viscosity is introduced in the previous set of equations  (\ref{eqalg1})-(\ref{eqalg2}) by modifying $- \sigma \rightarrow - \sigma - \alpha_w $ in both equations. Combining both effects of viscous damping and frequency detuning into a growth rate  noted $\sigma_v$, we have 
\be
\tilde{\sigma_v} = - \frac{\alpha_1+\alpha_2}{2} + \frac{1}{2} \sqrt{(\alpha_1 - \alpha_2)^2   - 2 \mathrm{i} \varsigma (\alpha_1-\alpha_2) + 4 \, (  \sigma^2 - \varsigma^2)  }.
\ee
We have instability when the real part $\sigma_v=\mathrm{Re}(\tilde{\sigma_v})$ is positive.

\subsection{Parameter survey}

In the practical implementation of the global stability, we fix the cylinder height $h$ and vary the libration frequency $\omega$. Then, as described in section \ref{sec:detuning} (see eq. \ref{eq:detun_omega}), imperfect resonances are taken into account via a frequency detuning \cite[as in][]{gledzer1992instability,kerswell1993elliptical,lacaze2004elliptical,herreman2010magnetic}. Note that, as \cite{eloy2003elliptic} or \cite{Lagrange11}, we can also calculate imperfect resonances by fixing $\lambda_1+j \omega = \lambda_2$ and relaxing the resonant constraint on the axial wavenumber ($l_1 - l_2 = \mathcal{O}(p)$). But this method is less efficient for large coupling frequencies $j \omega \simeq 4$.

To calculate $\sigma_v$, we thus find all the waves  
\be
\mbox{mode 1} \: \  (n_r,m_1,n_z, \lambda_1) \quad, \quad  \mbox{mode 2} \: \  (n_r,m_2,n_z, \lambda_2)
\ee
with numbers
\be
n_r = 1, \ldots , n_{r,max}  \quad , \quad \left \{ \begin{array}{rcl} m_1 &=&  -n +1 , \ldots , m_{1,max} \\
m_2 & = & 1 , \ldots , m_{1,max} + n \end{array}\right .  \quad , \quad n_z = 1 , \ldots , n_{z,max} \, , \nonumber 
\ee
and both positive and negative frequencies, that solve the dispersion relation (\ref{dispcyl}). For each pair of waves, we then know exactly the resonant frequencies 
\be
\omega = (\lambda_2 - \lambda_1 ) / j  \quad , \quad |j| \in 1,\ldots,j_{max} \, ,
\ee
so that the resonance conditions (\ref{resonance}) are exactly satisfied. Since $|\lambda_2 - \lambda_1| \leq 4$ maximally, modes with $|j| \neq 1$ exist within bands $\omega \in [0,4/ |j|]$. Note also that only central couplings with $n_{r,1}= n_{r,2} =n_r$ are considered here; we tested that they always have significantly larger growth rates \cite[as in][]{eloy2003elliptic}. We take into account (through $j$) that a given pair of modes may be destabilized by different frequencies. For each of this wave-pairs, we also calculate all the necessary matrix elements and damping coefficients. All this information is stored for a post-processing phase in which we can vary $\varepsilon, \Delta \varphi, E , p $ and consider the effect of detuning $\varsigma$ (practical information on the implementation of the global instability analysis can be found in a series of commented Matlab scripts that are available online as supplementary material). Here we discuss some particular features of the global stability theory. Further, we will perform a systematical comparison with numerical results. 

\subsubsection{Inviscid growth rate $\sigma$ vs. $\omega$}

The non viscous stability of a given pair of modes entirely depends on the sign of both frequencies $\lambda_1$ and $\lambda_2$. We observe that when the frequencies of the waves have the same sign $\mbox{Sgn} (\lambda_1 \lambda_2 ) = +1$, then the inviscid growth rate $\sigma$ is purely imaginary, so that these pairs of modes can never be destabilized (with $\mbox{Sgn}$ the sign function). If on the contrary, frequencies have an opposite sign  $\mbox{Sgn} (\lambda_1 \lambda_2 ) = -1$, we always have a real $\sigma$ and so inviscid instability. With the convention that $\omega > 0$, we have also noticed that modes with $\lambda_1 < 0$ and $\lambda_2> 0$ and thus $j > 0$  are always much more unstable than the opposite case. We therefore concentrate on this type of modes.  We finally observe that the couplings with the lowest radial labels, $n_r=1$ but the highest azimuthal wavenumbers $m_1$ and the largest number of axial wavelengths $n_z$ are generally the most unstable in the inviscid limit. This explains why we used $n_{r,max} = 2 $, $m_{1,max} = 50 $, $n_{z,max} = 50 $ for this inviscid study. 

\begin{figure}                   
  \begin{center}
    \begin{tabular}{ccc}
      \setlength{\epsfysize}{5.5cm}
      \subfigure[]{\epsfbox{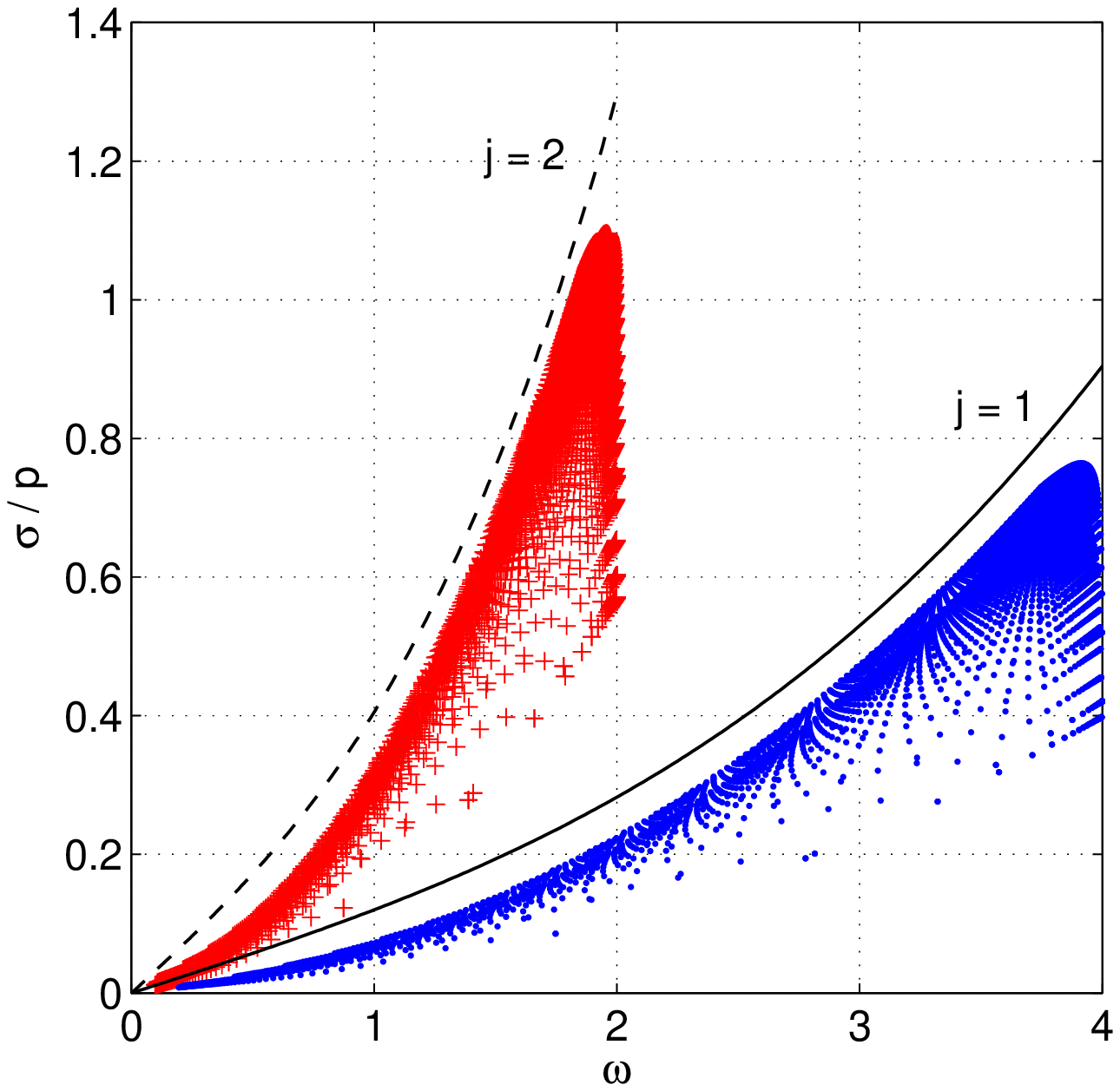}} 
       &
      \setlength{\epsfysize}{5.5cm}
      \subfigure[]{\epsfbox{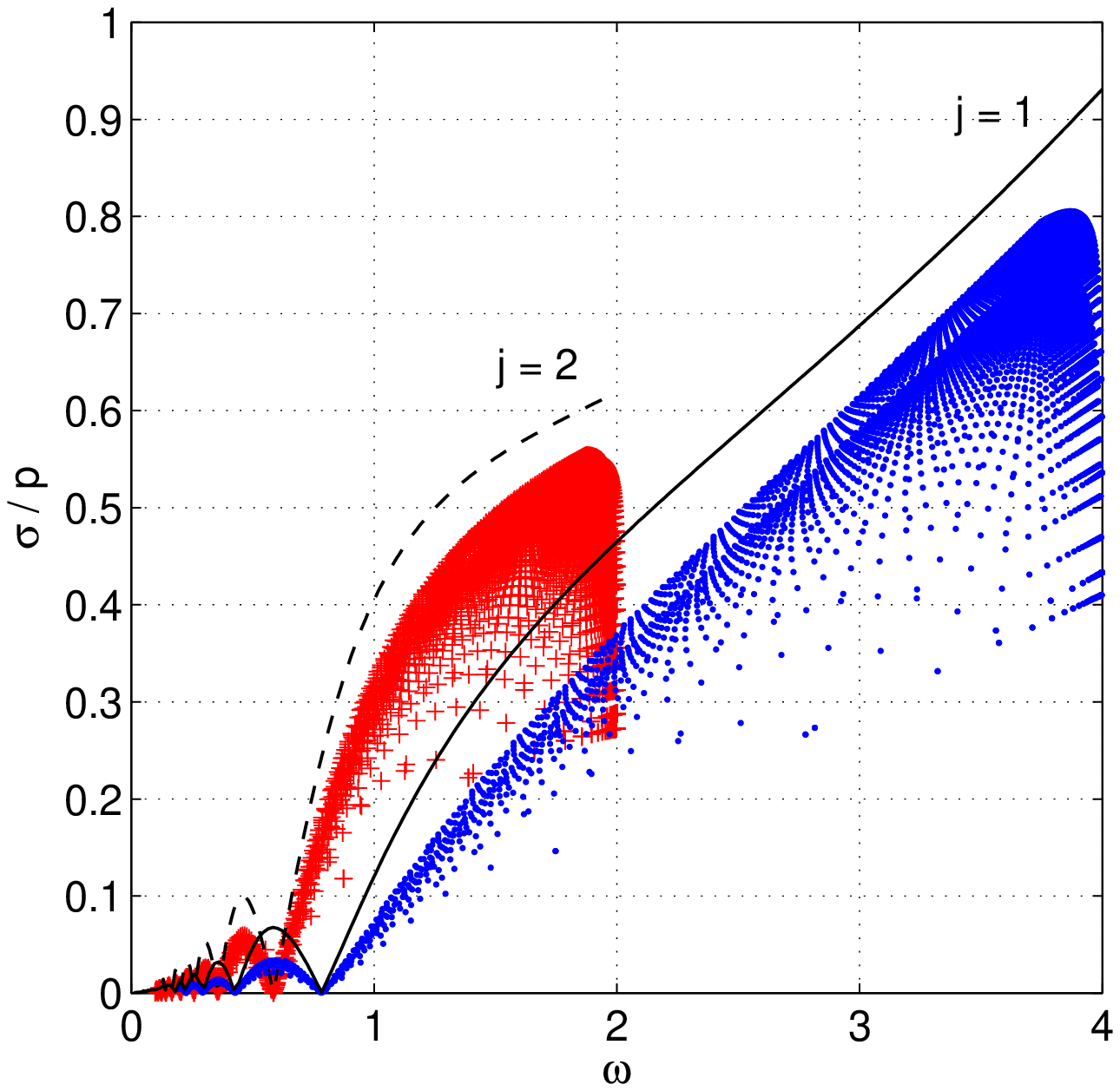}} 
    \end{tabular}
    \caption{Rescaled inviscid growth rates $\sigma / p $ at resonance as a function of $\omega$, for different values of $j=1, 2$  as marked in the figure and comparison with maximum local instability growth rate (eq. \ref{eq:sigWKB11} with $C=-1/2$) (full and dashed lines). We consider triangular deformation $n=3$ in a cylinder with height  $h=2$. In (a) we fix the libration angle  $\Delta \varphi =1$, in (b) we fix the libration amplitude $\varepsilon =1 $. The survey involves all modes up to $n_{r,max} = 2 $, $m_{1,max} = 50 $, $n_{z,max} = 50 $. }
         \label{fig:nonviscous}                
  \end{center}
\end{figure}

In figure \ref{fig:nonviscous}, we show inviscid growth rates at resonance for a triangular ($n=3$) deformation, at fixed libration angle $\Delta \varphi = 1$ (panel (a)) and for fixed libration amplitude $\varepsilon = 1$ (panel (b)). Each point represents a different unstable couple of modes at resonance. We only show 2 different $j =1,2$ to not overload the figures. It is particularly important to notice that couplings with $j=2$ may become more unstable than couplings with $j=1$ in the interval $\omega \in [0,2] $ a consequence of the large frequency content of the libration driven basic flow, at large $\varepsilon$ or large libration angles $\Delta \varphi$. In the numerical simulations, $\varepsilon$ is always large, so it is important to take this effect into account.  

The local instability analysis estimate (full and dashed lines) provide excellent upper bounds over the entire $\omega$-span and for both $j=1,2$.  Note also in panel (b), how the local growth rate of a given resonance may drop to zero for particular frequencies at fixed $j$ as a consequence of the Bessel function correction (see eq. \ref{eq:sigWKB11}).
This feature is well reproduced in the global growth rates.   

\subsubsection{Corrected growth rate $\sigma_v$ vs. $\omega$ }

\begin{figure}                   
  \begin{center}
    \begin{tabular}{ccc}
      \setlength{\epsfysize}{5.5cm}
      \subfigure[]{\epsfbox{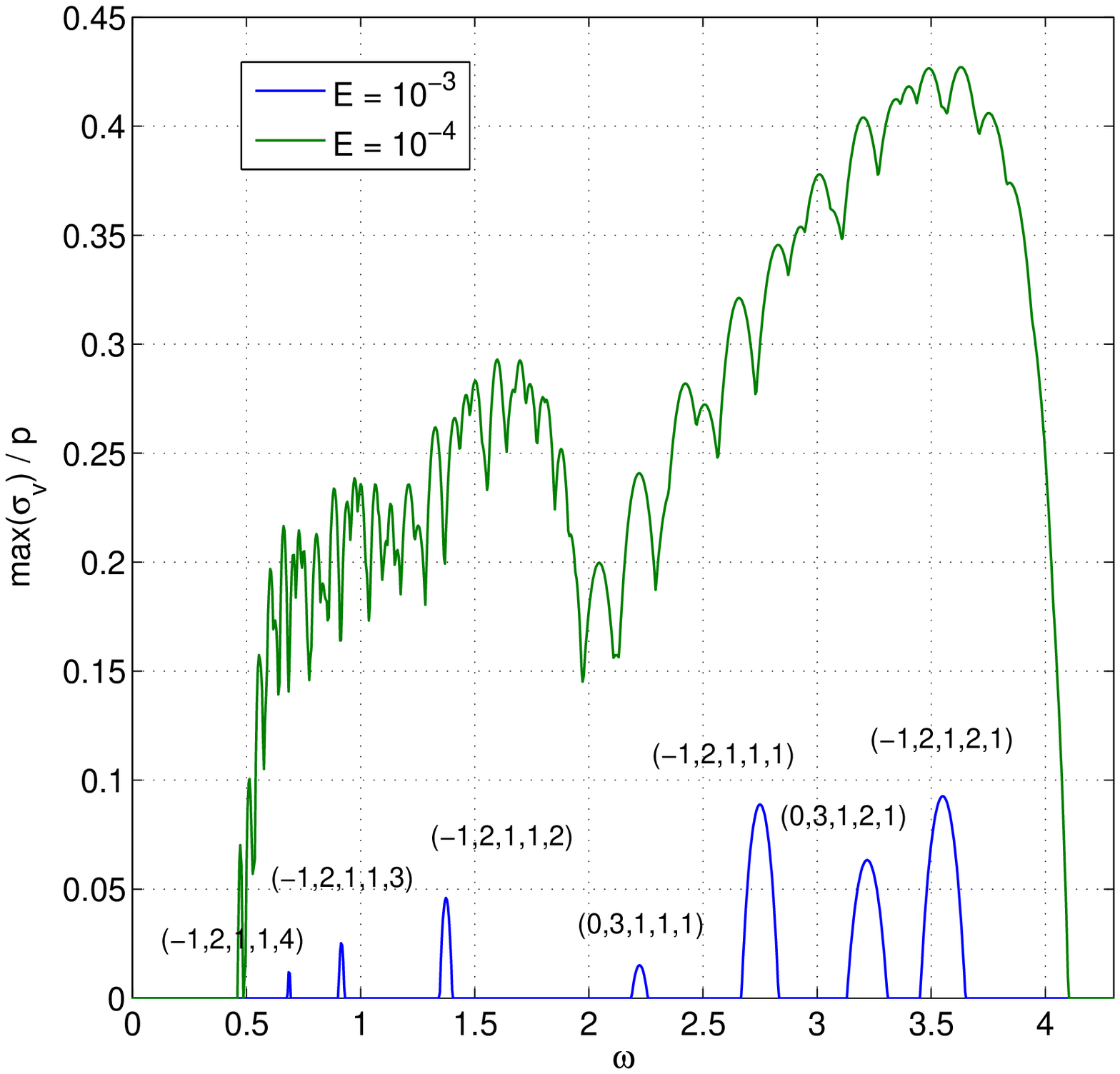}} 
       &
      \setlength{\epsfysize}{5.5cm}
      \subfigure[]{\epsfbox{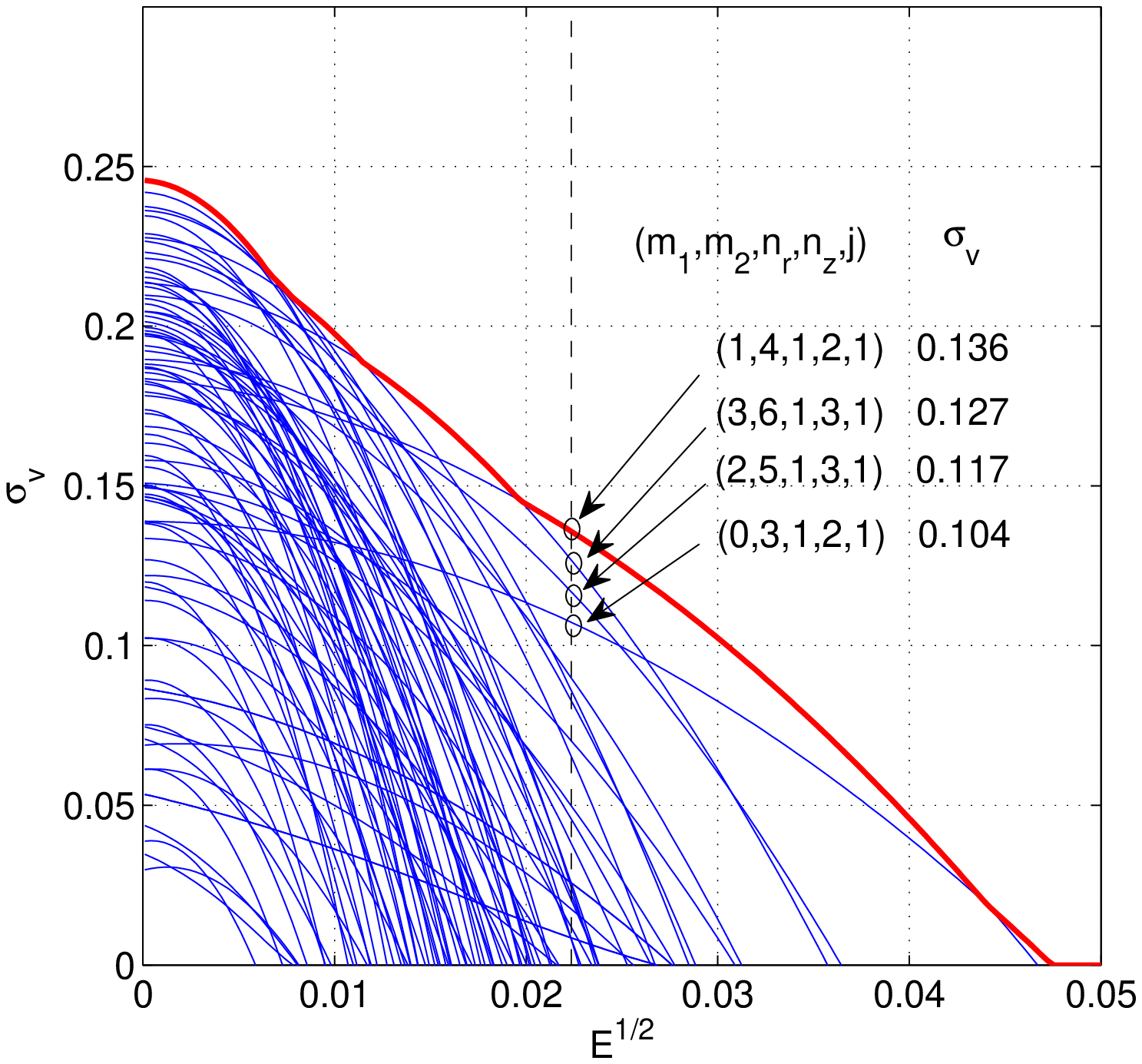}} 
    \end{tabular}
    \caption{LDMI growth rates in triangularly deformed cylinder ($n=3$) for $h=2$ and $\varepsilon=1$. (a) Maximal viscous growth rates $\max(\sigma_v) / p $ as a function of $\omega$ for $E=10^{-3} - 10^{-4}$ and $p=0.2$. For $E=10^{-3}$, we identify unstable couplings  $(m_1,m_2,n_r,n_z,j)$.  (b) Growth rate $\sigma_v$ as a function of $E^{1/2}$  for $h=2$ and $p=0.45$. We identify the 4 most unstable couplings and give the numerical values for $\sigma_v$ at $E=5 \cdot 10^{-4}$ (parameters studied numerically in section \ref{res}). The thick red line is the envelope of the resonance curves.}
         \label{fig:sigmamax}                
  \end{center}
\end{figure}

The inviscid growth rates at resonance do not give a very realistic picture as modes with high wavenumbers can be strongly damped by viscosity. In figure \ref{fig:sigmamax}(a), we show the maximal viscous growth rate $\max(\sigma_v)$ as a function of $\omega$ for $\varepsilon = 1$,  $p=0.2$, $h=2$ and different $E$ numbers (data from survey $n_{r,max} = m_{1,max} = n_{z,max} = 10, j_{max}=4 $). At low $E$, we see that the maximal growth rate follows a bumpy curve that has a shape that is quite close to the envelope of figure \ref{fig:nonviscous}(b). Above $\omega>2$ only $j=1$ couplings survive. The instability domain is slightly extended in the forbidden zone $\omega>4$. Close to threshold, here for $E=10^{-3}$, we can clearly identify different resonances. In the figure, we added $(m_1,m_2,n_r,n_z,j)$ to characterize the coupling. We see that $n_r=1$ for all modes, so the radial structures are large. Most resonant peaks combine $(m_1,m_2)=(-1,2)$, modes with 1 or 2 axial wavelengths ($n_z$) and for $j=1,2,3,4$. We also see two peaks involving modes $(m_1,m_2)=(0,3)$ with 1 or 2 axial wavelengths.

\subsubsection{Growth rate $\sigma_v$ vs.  $E$}

In figure \ref{fig:sigmamax}(b), each line follows the growth rates of different couplings with respect to $\sqrt{E}$, for fixed $\omega=3$, $\varepsilon=1$, $p=0.45$ and for $h=2$. The parabolic form of each of these curves is indicative of dominant volume damping for the range of $E$ considered here. But, note that in the limit of very small $E$, we naturally expect the surfacic viscous to be dominant (for $E \ll1$, we have $E^{1/2} \gg E $). Note that the situation is very different in a spherical or even a spheroidal container, where the volume damping exactly vanishes for any inertial mode \cite[][]{zhang2004inertial} and thus where the viscous damping can only be due to surface effects \cite[e.g.][]{lacaze2004elliptical}.

We further identify the 4 most unstable couplings at $E=5 \cdot 10^{-4}$, a configuration that will be studied numerically and mark $(m_1,m_2,n_r,n_z,j)$ in the figure, together with the numerical value of the growth rate $\sigma_v$. Modes with higher azimuthal wavenumbers dominate $(m_1,m_2)=(1,4),(3,6),(2,5),(0,3)$ dominate and we count $n_z=2,3$ axial wavelengths.

\section{Numerical simulations of libration-driven multipolar flow} \label{res}   
In this section, we present simulations of libration driven multipolar flows. First, we provide some details on the computational method (section \ref{sec:comp}). Then, we validate in section \ref{baseflow} that the basic flow (\ref{eq:basicflow3}) is indeed established for two different very simple and experimentally realizable forcings. In the last section, we demonstrate the existence of the libration driven multipolar instability, and characterize its properties such as the growth rate, saturation amplitude and viscous dissipation rate.     

\subsection{Numerical method} \label{sec:comp}
\begin{figure}                   
  \begin{center}

            \includegraphics[width=8cm]{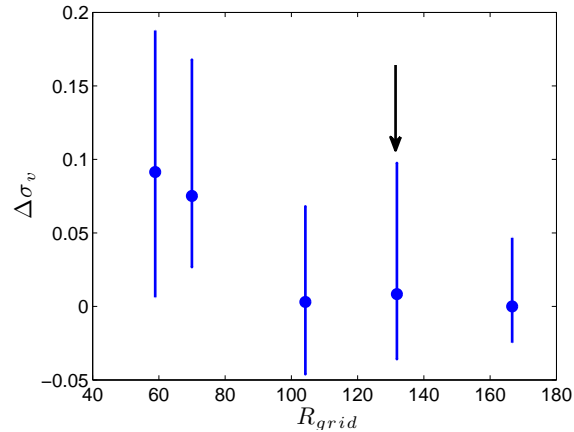}
  \end{center}
  \caption{Convergence of the growth rate of a 3D multipolar instability with increasing grid resolution (for $E=5\cdot10^{-4}$, $n=3$, $p=0.45$, $\varepsilon=1$, $\omega=3$). For each simulation at a given mesh resolution $R_{grid}$, we perform several measures of $\sigma$, which gives several measures of the relative difference $\Delta \sigma_v$. We report in this figure the mean value of $\Delta \sigma_v$ (dots), as well as the obtained extremum values (errorbars). The arrow indicates the minimal resolution used for the systematic study of the instability in subsection \ref{instab}.}
  \label{fig:convergence}
\end{figure}

To perform our numerical simulations, we use a parallel unstructured finite-volume code \cite[]{Vantieghem_phd}. It is based on a collocated arrangement of the variables, and a second-order centered-finite-difference-like discretization stencil for the spatial differential operators. The time advancement algorithm is based on a canonical fractional-step method \cite[]{Kim1985}.  More specifically, the procedure to obtain the velocity and reduced pressure ${\boldsymbol u}^{N+1},\Pi^{N+1}$ at time-step $t^{N+1}=t^{N}+\Delta t$, given the respective variables at time step $N$ is as follows:
\begin{enumerate} 
\item We first solve the intermediate velocity ${\bs u}^{\star}$ from the equation:  
\begin{eqnarray}
\frac{{\boldsymbol u}^{\star} - {\boldsymbol u}^{N}}{\Delta t} & = &  - {\boldsymbol u}^{N+1/2}_{AB} \cdot \nabla {\boldsymbol u}^{N+1/2}_{CN} - \nabla \Pi^{N} \\
&& - 2 \gamma^{N+1/2} {\boldsymbol e}_z \times {\boldsymbol u}^{N+1/2} + E \nabla^2 {\boldsymbol u}^{N+1/2}_{CN} - \frac{\mathrm{d}\left( \gamma^{N+1/2}\right)}{\mathrm{d}t}\,{\boldsymbol e}_z \times {\boldsymbol x} \nonumber, 
\end{eqnarray}
with no-slip boundary condition ${\boldsymbol u}^{\star}_{bnd} = {\boldsymbol 0}$. In this expression, ${\boldsymbol u}^{N+1/2}_{AB}$ and ${\boldsymbol u}^{N+1/2}_{CN}$ denote the velocity at time-step $N+1/2$ obtained using a second-order Adams-Bashforth, respectively Crank-Nicholson approach, i.e.:
\begin{eqnarray}
{\boldsymbol u}^{N+1/2}_{AB} & = &\frac{3}{2}{\boldsymbol u}^{N} - \frac{1}{2} {\boldsymbol u}^{N-1}, \\ 
{\boldsymbol u}^{N+1/2}_{CN}& = & \frac{1}{2}({\boldsymbol u}^{N} + {\boldsymbol u}^{\star}). 
\end{eqnarray}
The mixed Adams-Bashforth/Crank-Nicholson formulation for the advective term has the advantage of being kinetic-energy conserving and time-stable for any $\Delta t$ \cite[][]{ham2004}, and it does not require the solution of a non-linear system for the unknown ${\boldsymbol u}^{\star}$.
\item The new velocity ${\boldsymbol u}^{N+1}$ is then related to ${\boldsymbol u}^{\star}$ by:
\begin{equation}
 {\boldsymbol u}^{N+1}={\boldsymbol u}^{\star} - \Delta t \left( \Delta \Pi^{N+1} \right),
 \end{equation}
with $\Delta \Pi^{N+1} = \Pi^{N+1} - \Pi^{N}$. Imposing the incompressibility constraint on ${\boldsymbol u}^{N+1}$ leads to a Poisson equation for  $\Delta \Pi^{N+1}$,
\begin{equation}
\nabla^2 \left( \Delta \Pi^{N+1} \right) = (\Delta t)^{-1} \nabla \cdot {\boldsymbol u}^{\star},
\end{equation}
with boundary condition ${\boldsymbol e}_n \cdot \nabla \left( \Delta \Pi^{N+1} \right) = 0$. This Poisson equation is solved with the algebraic multigrid method BoomerAMG \cite[]{Henson00}.
\end{enumerate}
To discretize the equations in space, we use a grid that is shaped such that its boundary coincides with a streamline of the flow (\ref{eq:basicflow3}). To this end, we transform a circular mesh into one bounded by a streamline (see figure \ref{fig:mesh}). This requires an explicit parametrization of the streamlines $r=F(\theta)$, which is derived in Appendix \ref{sec:app2D}. As shown in figure \ref{fig:mesh}(a), we start from a grid whose nodes occur on curves of constant $\theta$ or constant $r$, except for a smaller inner core of radius $r < 0.15$. The outer part of the grid (i.e. for $0.15 \le r \le 1$) consists of highly regular regions of quadrilateral elements, separated by transition layers of triangular elements; these transition layers allow to decrease the number of grid points in azimuthal direction as $r \rightarrow 0$,  so to avoid any clustering of the grid points near the origin. The grid consists of quadrilateral elements and is unstructured for $r<0.15$. Furthermore, the grid points are more closely spaced in wall-normal direction in the vicinity of $r=1$ in order to account for the presence of thin viscous boundary layers of thickness $\delta = \sqrt{2E/\omega}$ \cite[]{wang1970cylindrical}; more precisely, the grid spacing in the near-wall region is such that there are at least 5 grid nodes within a distance $\delta$ from the wall. Then, we gradually deform the outer quadrilateral elements, starting at $r_0=0.5$ towards $r=1$. The radial coordinate of the grid nodes is transformed from $r$ into $r'$ according to the following formula:
\begin{eqnarray}
r' =  r &  \mbox {for} &  r \le r_0 \label{grid_deformation1}, \\
r' = \left[1 + \left(F(\theta)-1\right)\frac{r-r_{0}}{1-r_{0}}\right]r  & \mbox{for} & r_0<r \le 1. \label{grid_deformation2}
\end{eqnarray} 
This results in a smooth transformation, in which the grid elements are not too distorted, as shown in figure \ref{fig:mesh}(b) for $p=0.2$. Moreover, we also wish to avoid grid distortion at interfaces between zones of triangular and quadrilateral elements because these interfaces are more sensitive to numerical stability and accuracy problems. Therefore, we choose the value $r_0=0.5$ such that the grid deformation only affects the elements in the outer shell of quadrilateral elements. To determine the required grid resolution for the simulations of the LDMI three-dimensional flows, we have first performed a grid convergence study. More specifically, for a given set of typical parameters ($E=5\cdot10^{-4}$, $n=3$, $p=0.45$, $\varepsilon=1$, $\omega=3$), we have investigated the dependency of the growth rate $\sigma_v$ on the number of control volumes $N_{CV}$. We characterize the spatial resolution using the resolution number $R_{grid}$, defined by $R_{grid} = N_{CV}^{1/3}$. To estimate the growth rate $\sigma_v$, we use the procedure outlined in subsection \ref{instab}.  Our results are summarized in figure \ref{fig:convergence}, where we show the relative difference $\Delta \sigma_v$ between $\sigma_v$ for a given value of $R_{grid}$ and $\sigma_v$ for the largest value of $R_{grid}$ we have considered ($R_{grid}=167$):
\begin{eqnarray}
\Delta \sigma_v=\frac{ \sigma_v(R_{grid})-\sigma_v(R_{grid}=167)}{\sigma_v(R_{grid}=167)}.
\end{eqnarray} 
Figure \ref{fig:convergence} shows that the relative difference of $\sigma_v$ is smaller than $0.5\%$ for $R_{grid} \gtrsim 105$, and much smaller than uncertainties (errorbars) associated to the measure of $\sigma_v$. For the systematic study of the multipolar instability discussed below, we systematically work at $R_{grid} \approx 131.8$ (indicated by an arrow in figure \ref{fig:convergence}) for $E \ge 5 \cdot 10^{-4}$, which corresponds to approximately 2.2 million control volumes. For lower Ekman number, we employ grids with up to 2.9 million control volumes (i.e. $R_{grid} \approx 142.7$) to ensure numerical convergence. Finally, one can notice that no instability is observed for $R_{grid}  \le 34$, i.e. when the grid is too coarse.

The typical time step is of the order $\Delta t = 5 \cdot 10^{-3}$, and the integration time $t_{max} \approx 750$. The time step was systematically chosen such that the CFL number remained smaller than 0.9 during the entire computation. The simulations were carried out using 64 CPUs on the Cray-XE6 machine `Monte Rosa' of the Swiss Supercomputing Center (CSCS).
\begin{figure}                   
  \begin{center}
     \begin{tabular}{ccc}
          \setlength{\epsfysize}{5.0cm}
            \subfigure[]{\epsfbox{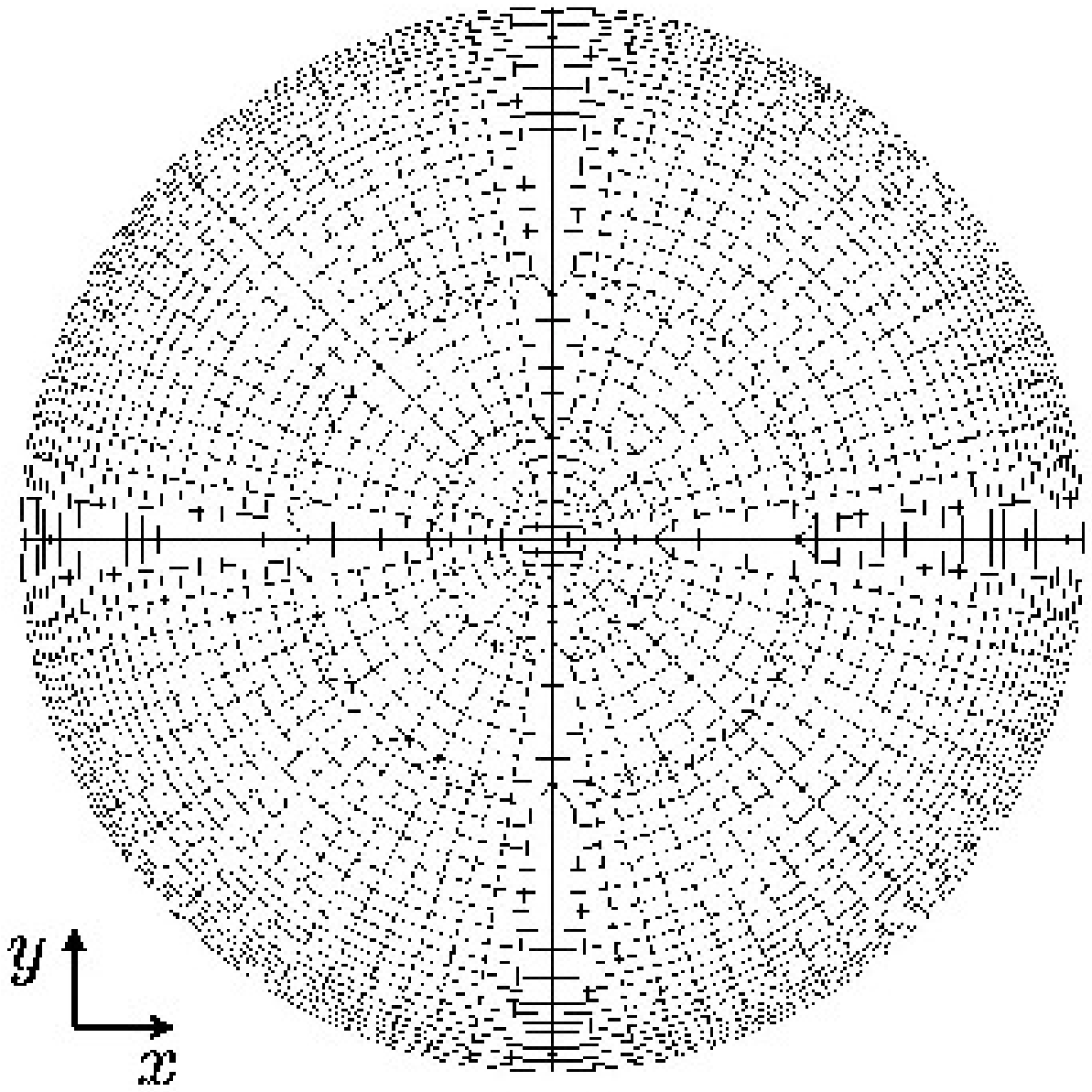}} &
            \setlength{\epsfysize}{5.0cm}
            \subfigure[]{\epsfbox{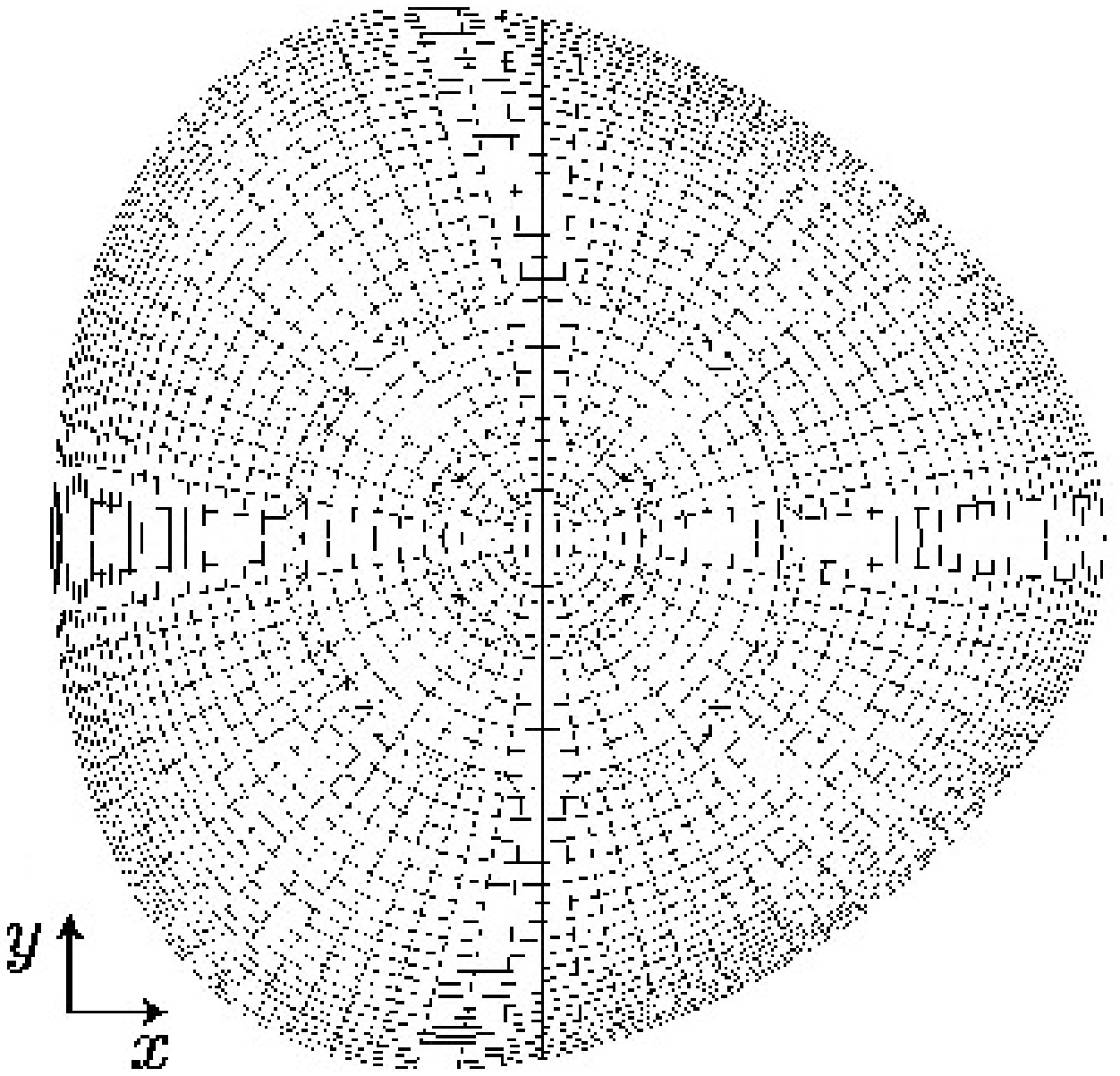}} \\
       \end{tabular}
  \end{center}
  \caption{Illustration of the deformed numerical grid in the $xy$-plane for $p=0.2$. For clarity, the closer spacing in the boundary layer region has been left away. (a) Initial circular grid before tripolar deformation. (b) Final grid after tripolar deformation (\ref{grid_deformation1}-\ref{grid_deformation2}). }
  \label{fig:mesh}
\end{figure}
\subsection{Numerical validation of the forced two-dimensional basic flow}\label{baseflow}
\begin{figure}                   
  \begin{center}
   \includegraphics[width=\textwidth]{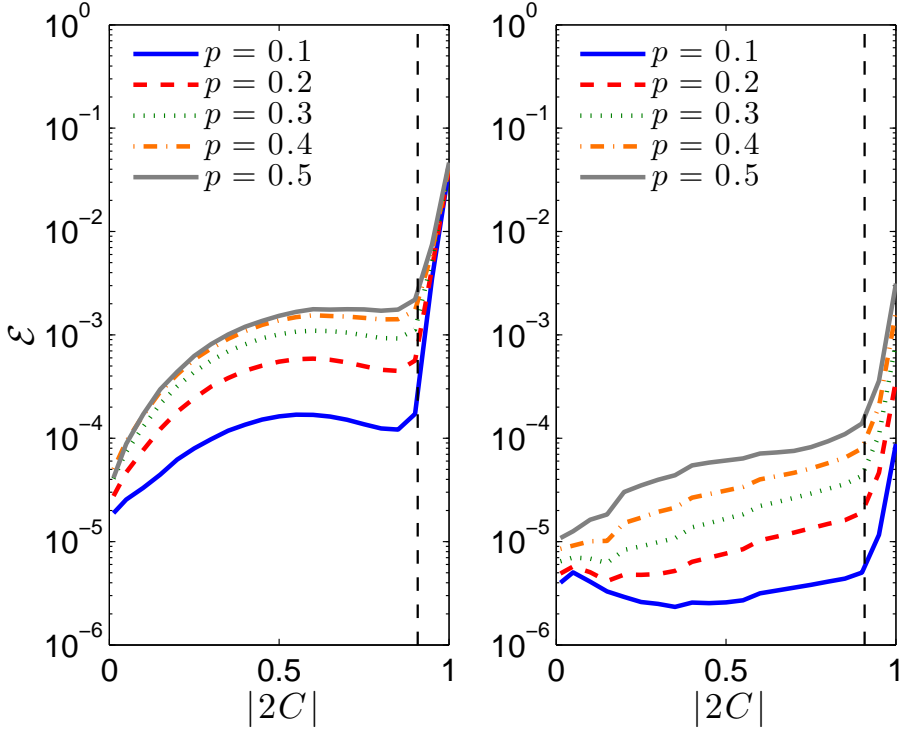}
    \caption{Relative $L2$ error norm $\mathcal{E}$ between the established flow and the exact basic flow (\ref{eq:basicflow3}), for the case of a rigid librating container (left) and a rotating container deformed by librating rollers (right). Parameters: $n=3$, $E=5\cdot{10}^{-4}$, $\varepsilon = 1.0$ and $\omega=2.5$. The dashed vertical line indicates a distance of $4.6 \delta$, with $\delta=\sqrt{2E\omega^{-1}}$ the thickness of the viscous layer, for which the theoretical boundary layer correction should be smaller than $1\, \% \approx \exp(-4.6)$.}
    \label{fig:validation_base_flow}            
  \end{center}
\end{figure}

Here, we investigate how to easily establish the basic flow (\ref{eq:basicflow3}). An obvious choice would be to solve the Navier-Stokes equations in an inertial frame of reference, in a domain bounded by a streamline, and to impose a boundary velocity ${\boldsymbol u}_{bnd} = \nabla \times \left(\Psi {\boldsymbol e}_z\right)$. However, this would require a numerical technique that can take into account a moving boundary, such as the Arbitrary Lagrangian-Eulerian method. Moreover, this numerical approach does not have a simple experimental counterpart. Therefore, we have considered two alternatives that are expected to generate the basic flow (\ref{eq:basicflow3}) in the bulk. A first possible realization is the one of a librating rigid container whose boundary takes the form of a streamline of the basic flow. An ingenuous, but somehow more complex alternative to obtain streamline deformation was devised by \cite{eloy2003elliptic} for steady rotation along deformed streamlines. They performed experiments in a cylindrical container, deformed by the compression of $2$ or $3$ rollers. To extend this approach towards libration mechanical forcing, one can librate the rollers, while rotating the container at constant speed. In both cases, the dynamics of the system is the most easily expressed in the librating frame of reference attached to the deformation, because the boundary is stationary in this frame. As such, the flow is governed by the equation (\ref{syst:eqNS_sl})-(\ref{syst:eqNS_s2l}). However, in the former case the boundary condition is ${\boldsymbol u}_{bnd} = {\bs 0}$, whereas in the latter case it is  ${\boldsymbol u}_{bnd} = -\varepsilon \cos (\omega t){\boldsymbol e}_z \times {\boldsymbol e}_n$, where ${\boldsymbol e}_n$ denotes the outward unit normal vector of the boundary.  
To assess if the basic flow is correctly established, we consider the following error estimate:
\begin{equation}
\mathcal{E}(C) = \frac{\int_{0}^{T} \int_{C>-r^2/2} ||{\boldsymbol u} - {\boldsymbol U}||^2 \,\mathrm{d}{\boldsymbol r} \,\mathrm{d}t}{\int_{0}^{T} \ \int_{C > -r^2/2} ||{\boldsymbol U}||^2 \,\mathrm{d}{\boldsymbol r} \,\mathrm{d}t},
\end{equation}
where the integration is performed within a domain enclosed by a contour $C = cst$. This expression can be interpreted as follows: it is the relative $L2$ error norm of the deviation between the numerically established flow and the exact basic flow (\ref{eq:basicflow3}), within a domain bounded by a streamline of (\ref{eq:basicflow3}), time-averaged over an interval $T$ of ten libration periods. We have evaluated $\mathcal{E}(C)$ for 20 equidistant values of $C$ in the interval $[0.025,1]$. In figure \ref{fig:validation_base_flow}, we show $\mathcal{E}$ for the two considered cases (librating rigid container and librating rollers on a deformable container rotating at constant speed) and for several values of $p$.
\par
\begin{figure}                   
  \begin{center}
     \begin{tabular}{ccc}
          \setlength{\epsfysize}{5.0cm}
            \subfigure[]{\epsfbox{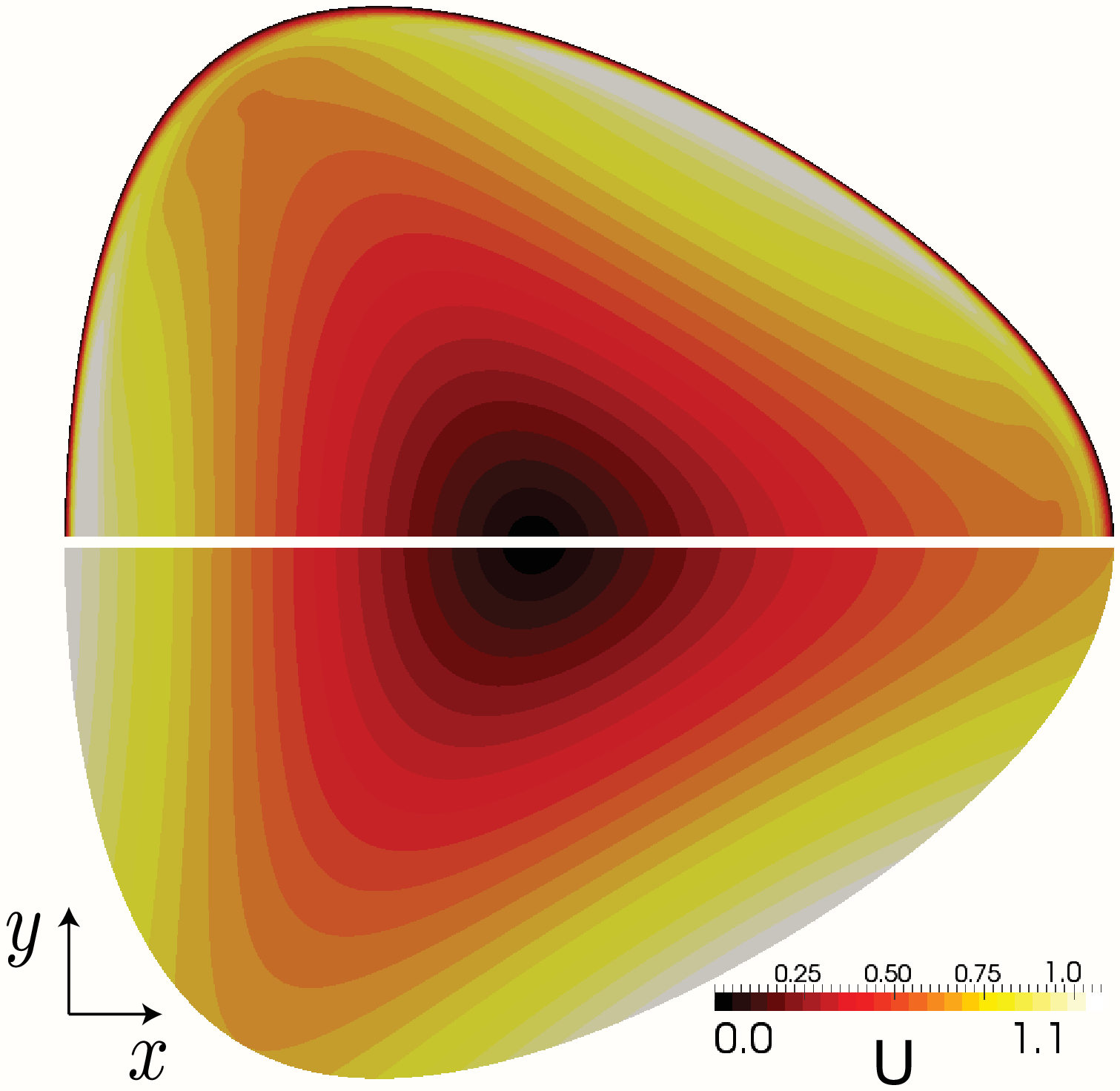}} &
            \setlength{\epsfysize}{5.0cm}
            \subfigure[]{\epsfbox{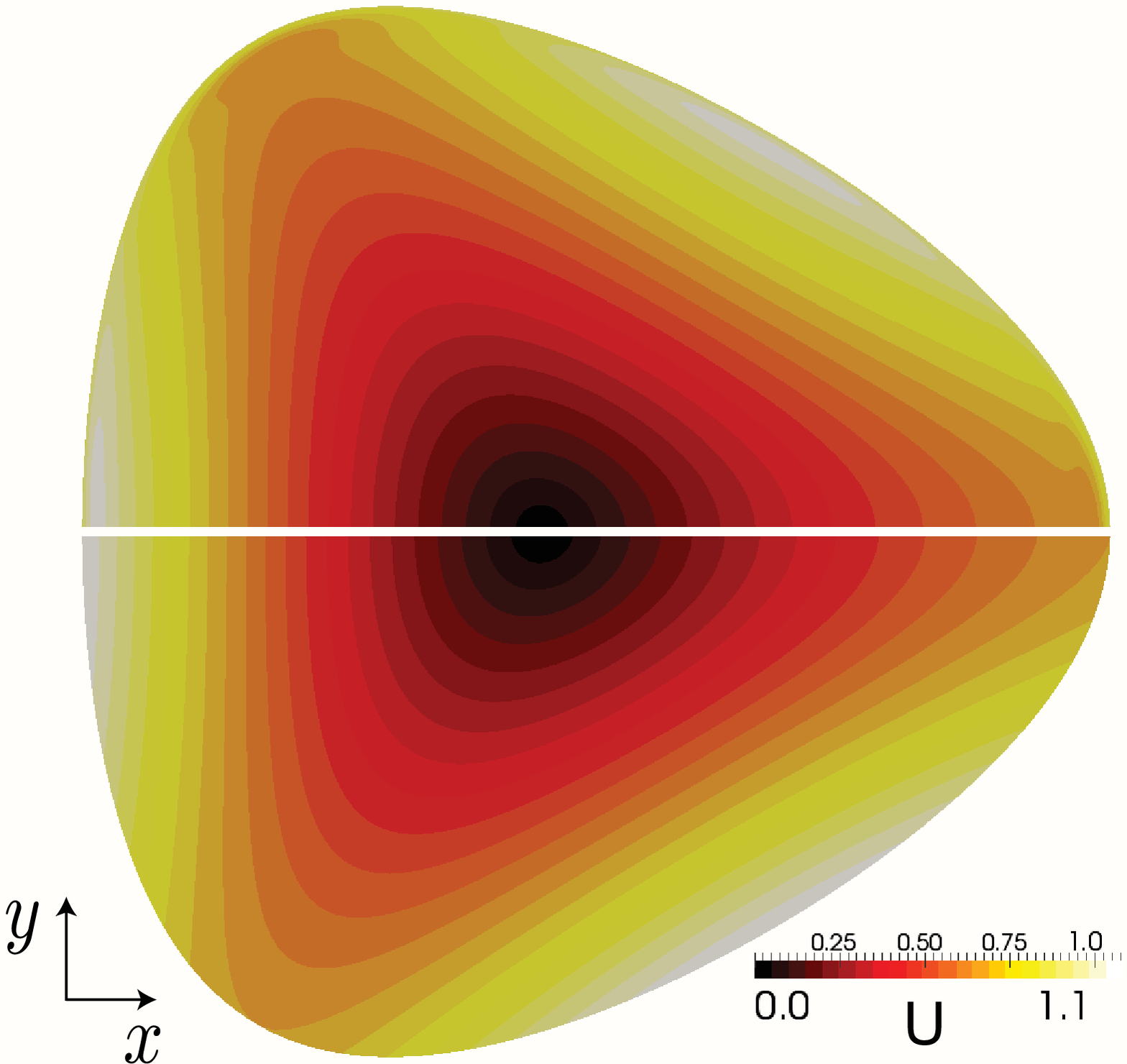}} \\
       \end{tabular}
       \caption{Comparison for $n=3$, $p=0.3$, $E=5 \cdot 10^{-4}$, $\omega = 2.5$ and $\varepsilon=1$ between the exact basic flow (\ref{eq:basicflow3}) (bottom half) and the numerically established 2D basic flow (top half) for the case of a rigid tripolar ($n=3$) container (a), and a deformable container, deformed by rollers (b).}
      \label{fig:validation_snapshot}            
   \end{center}
\end{figure}

We see that $\mathcal{E}(C)$ remains small for both forcing mechanisms and for all investigated values of $p$, except in a viscous boundary layer that emerges to accommodate the difference between the bulk flow (close to the exact basic flow) and the boundary velocity ${\boldsymbol u}_{bnd}$. The thickness of these viscous layers is \cite[]{wang1970cylindrical}: 
\begin{equation}
\delta = \sqrt{2E \omega^{-1}}. 
\end{equation}
We assume that the velocity matching in this layer (between the bulk and the wall) is of the form of $1 - \exp(r^{\ast} \delta^{-1})$, where $r^{\ast} = r - F(\theta)$ denotes the distance from the wall. The boundary layer correction should thus remain smaller than 1 $\%$ outside an annular-like region where $r^{\ast} \delta^{-1} <  -\log(0.01) \approx 4.6$. This distance is indicated by a dashed axial line in figure \ref{fig:validation_base_flow}.  The excellent agreement within the bulk is also confirmed in figure \ref{fig:validation_snapshot}, which compares the exact basic flow and the numerical solutions at time $t=M \pi /\omega$ with $M$ an integer, i.e. when the basic flow has maximum strength. Moreover, $M$ is such that we are beyond the spin-up regime, i.e. such that $t > 5 E^{-1/2}$. Comparing both realizations, we find the discrepancy between the established and basic flow within the boundary layer is considerably larger in the case of a rigid container. This can be explained as follows. For the case of a deformable container, the boundary velocity of the system with rollers is much closer to the exact basic flow (\ref{eq:basicflow3}). We thus expect viscous effects to be much less important. Hence, the discrepancy $\mathcal{E}$ to be much smaller near the walls, than when a rigid container is used to establish the desired basic flow.      

\par
Nevertheless, as the bulk flow in both realizations is very close to (\ref{eq:basicflow3}), all the theoretical results obtained in section \ref{sec:stability} are valid in both cases. In the following, we will only consider the case of a rigid container because of its experimental convenience.  

\subsection{Onset and development of the LDMI} \label{instab}
In this section, we discuss three-dimensional non-linear simulations of the libration-driven tripolar instability. The simulation domain is a cylinder that is periodic in $z$-direction with a tripolar cross-section (in the $x-y$ plane) like the ones discussed in the previous section. Furthermore, we keep the aspect ratio $h$ between the height of the cylinder and the mean radius of the cross-section fixed at a value of $2$. The choice of periodic boundary conditions is motivated by the fact that we avoid the presence of thin Ekman layers at the top and bottom of the cylinder, which have two important drawbacks: a) They are at the origin of three-dimensional flows, even before the multipolar instability emerges.  b) The proper numerical resolution of these layers would lead to a considerable increase in CPU time. 
\par
We will first present some general features of this instability. Then, we will validate the theoretical results obtained in section \ref{sec:stability} through a systematic study of the dependency of the viscous growth rate $\sigma_v$ on the flow parameters $E,\omega,p$ and $\varepsilon$. Finally, as numerical simulations allow us to go beyond the linear theory, we will investigate two characteristics of the non-linear regime that are of interest, namely the amplitude and viscous dissipation of the instability at saturation. 
\begin{figure}                   
  \begin{center}
    \begin{tabular}{ccc}
      \setlength{\epsfysize}{5.0cm}
      \subfigure[]{\epsfbox{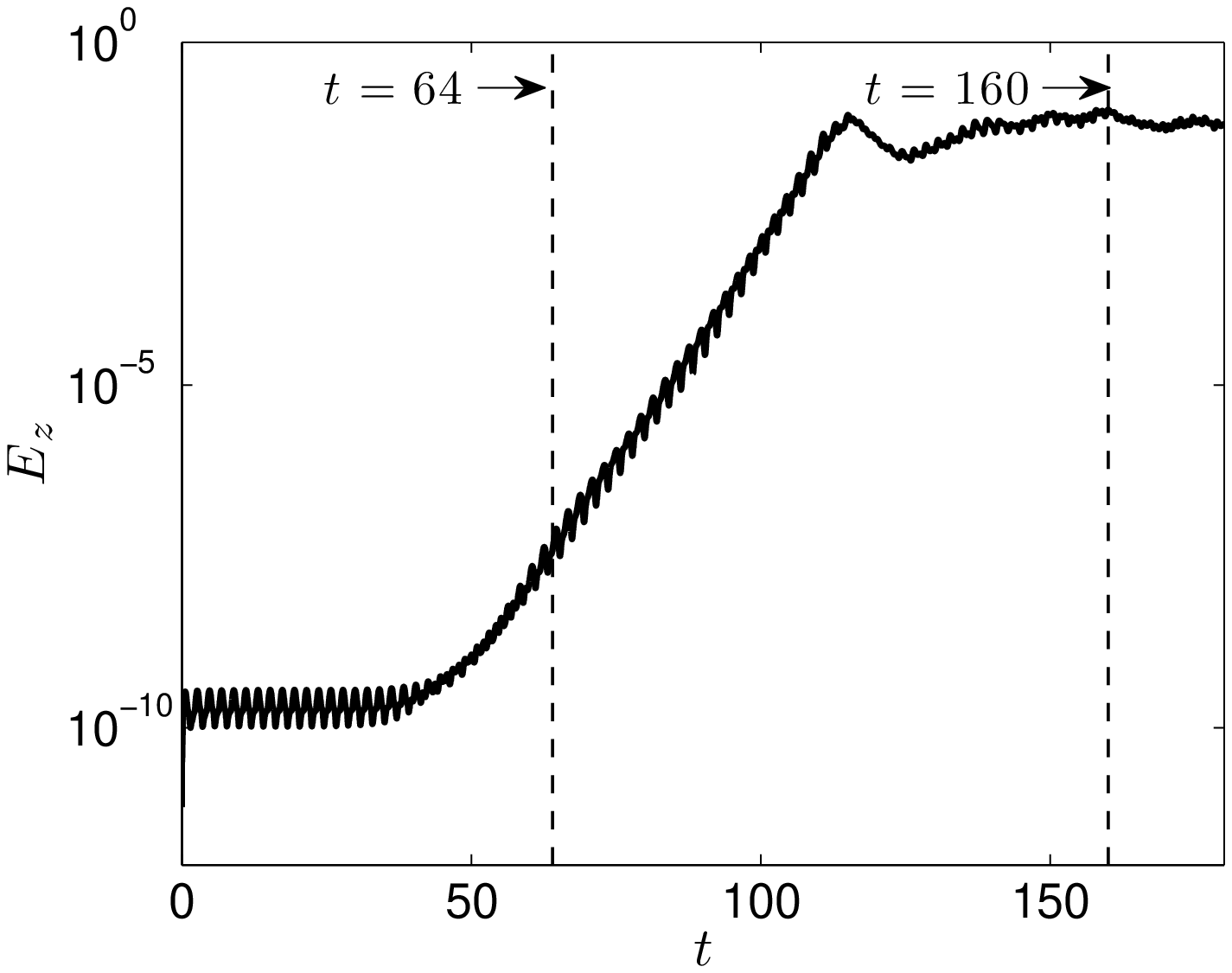}} &
      \setlength{\epsfysize}{5.0cm}
      \subfigure[]{\epsfbox{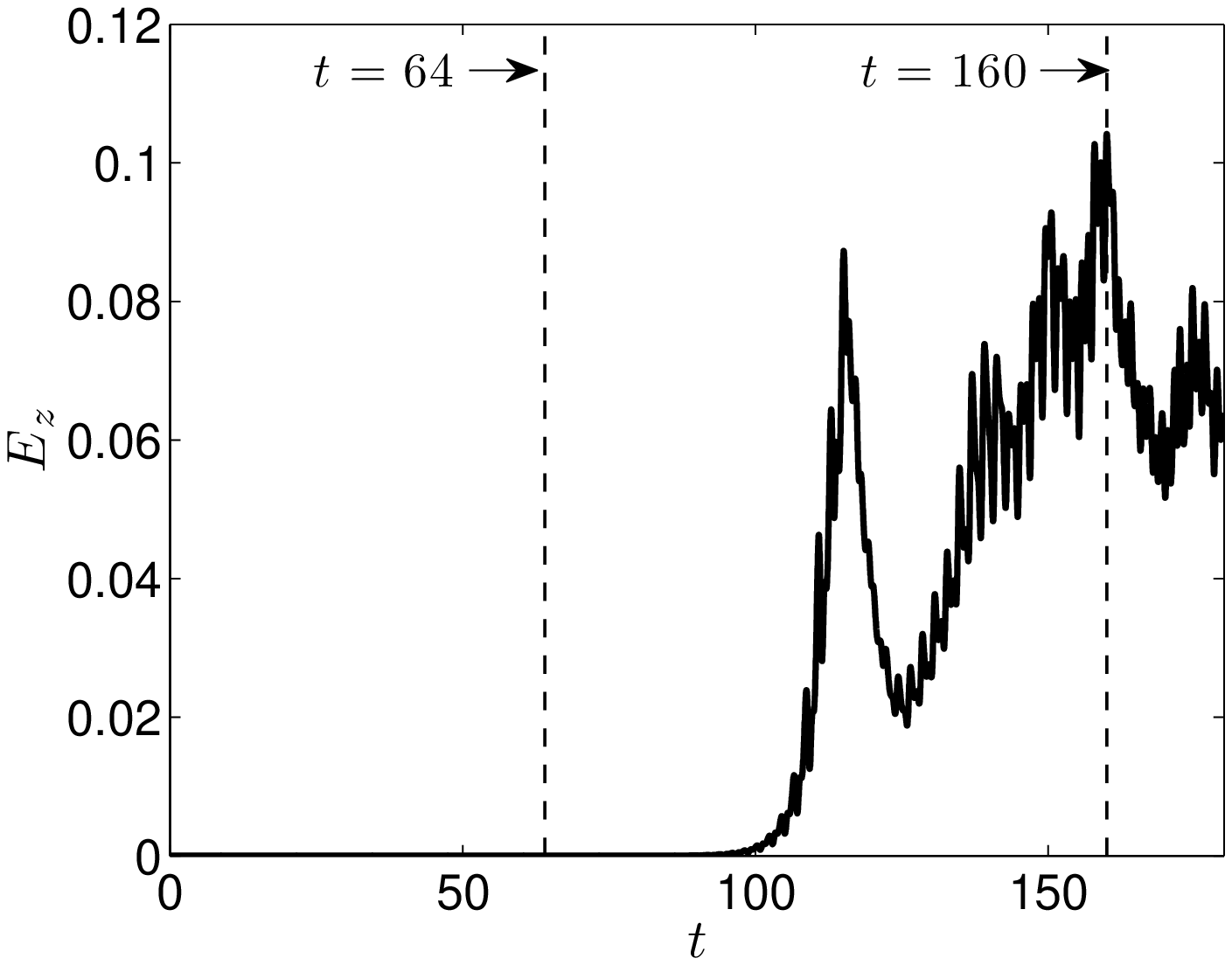}} \\
       \setlength{\epsfysize}{5.0cm}
      \subfigure[]{\epsfbox{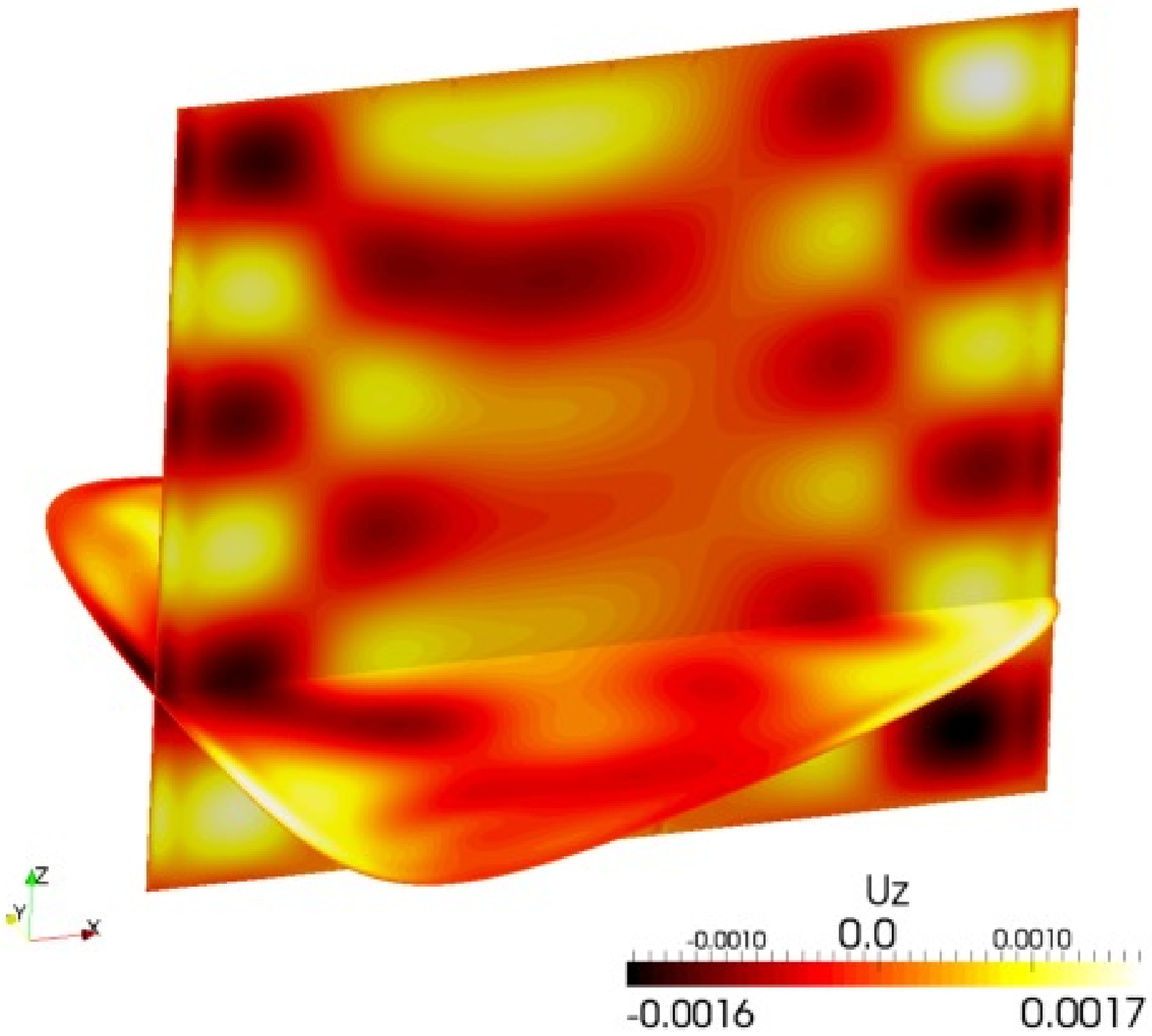}} &
      \setlength{\epsfysize}{5.0cm}
      \subfigure[]{\epsfbox{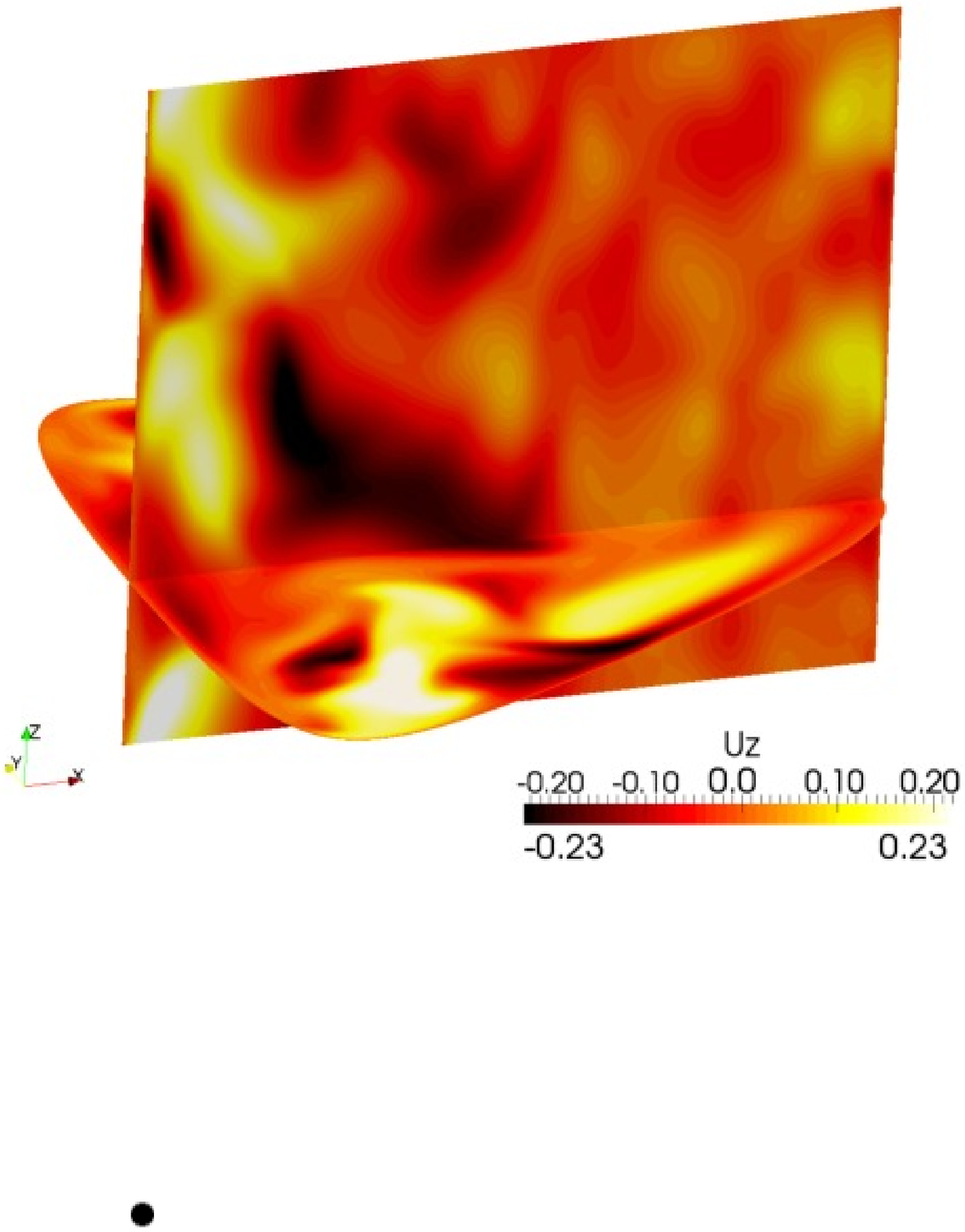}} \\
    \end{tabular}
    \caption{Time series of the axial kinetic energy $E_z$ in a 3D numerical simulation of the LDMI for $n=3$, $\omega=3$, $\varepsilon=1$, $p=0.45$, $E=5\cdot 10^{-4}$, both on a logarithmic scale (a), illustrating the exponential growth, and on a linear scale (b), giving evidence of the subsequent intermittent behaviour. Snapshots of $u_z$ during the exponential growth stage at $t=64$ (c), and the intermittency stage at $t=160$ (d).}
    \label{fig:time_series_LDTI}            
  \end{center}
\end{figure}

\subsubsection{General characteristics of the LDMI flows}
The basic flow being 2D, and the stability analysis showing instability for 3D perturbations, we can expect that the axial kinetic energy $E_z$,
\begin{equation}
E_z = \frac{1}{2}\int_{V} u_z^2 \,\mathrm{d}V,
\label{eq:axial_energy}
\end{equation}
is a good proxy for the development of the instability ($V$ being the volume of the container). Figures \ref{fig:time_series_LDTI}(a) and (b) show typical time series of $E_z$, which exhibit three distinct stages. Until $t\approx 40$, $E_z$ is negligibly small, and hence, the basic flow is virtually 2D. From $t \approx 40$, the axial kinetic energy undergoes exponential growth over many decades. During this stage, $u_z$ has a wavy structure, as highlighted by a snapshot of $u_z$ at $t=80$ (see figure \ref{fig:time_series_LDTI}c). Eventually $E_z$ saturates at a value of approximately $0.08$ around $t \approx 110$. In the last stage, $E_z$ exhibits chaotic intermittent behaviour, which is related to the appearance of small-scale turbulence; this is illustrated in figure \ref{fig:time_series_LDTI}(d) by the snapshot of $u_z$ at $t=160$. The turbulence is space-filling, and is thus not related to the presence of a boundary layer instability. This contrasts with previous studies of libration-driven flows in axisymmetric containers  \cite[]{noir2009experimental,noir2010experimental,calkins2010axisymmetric,sauret2012fluid}, where the observed turbulence was triggered by a Taylor-G\"ortler instability and remained limited to the near-wall region.

\begin{figure}                   
  \begin{center}
    \begin{tabular}{cc}
      \setlength{\epsfysize}{5.0cm}
      \subfigure[]{\epsfbox{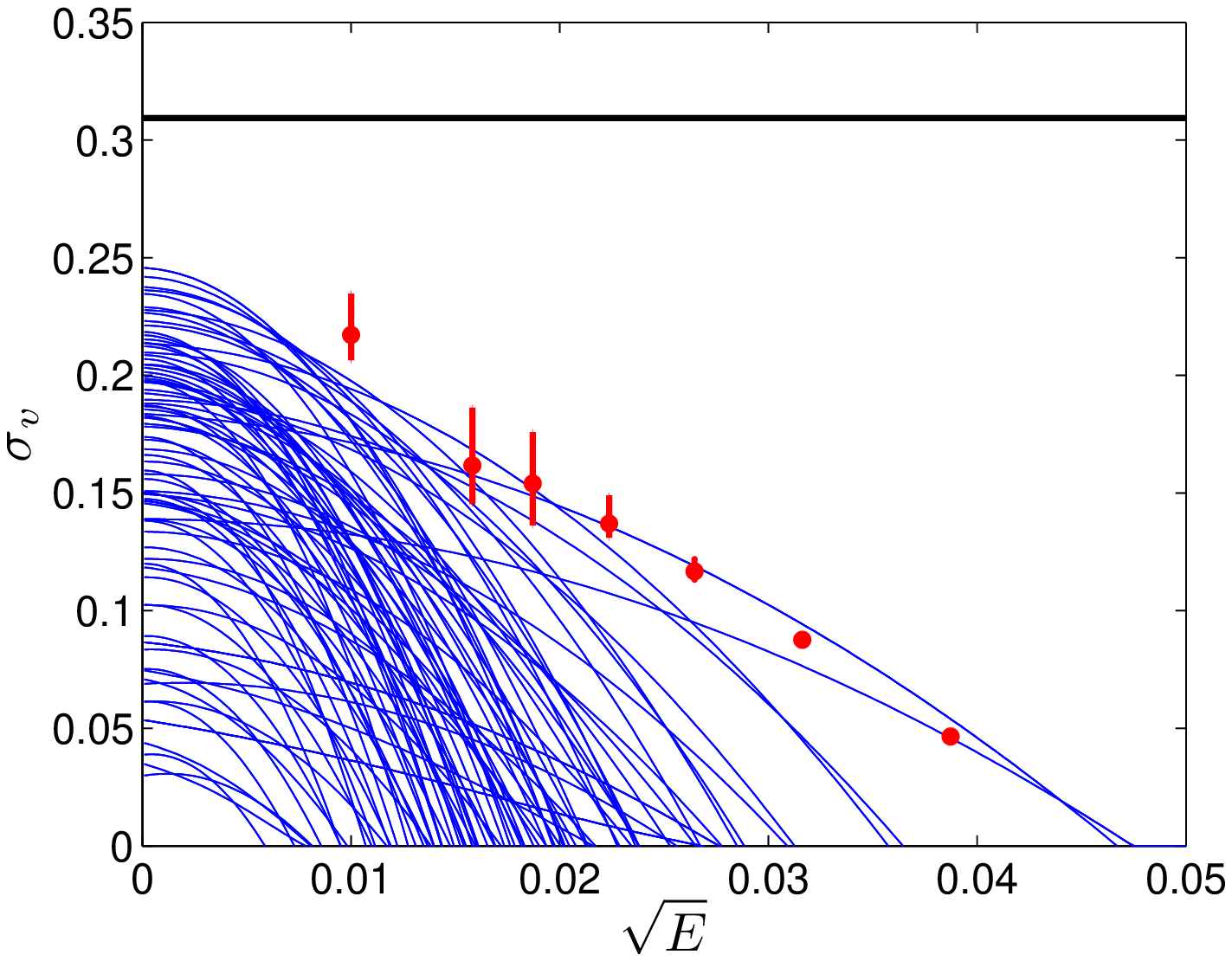}} &
      \setlength{\epsfysize}{5.0cm}
       \subfigure[]{\epsfbox{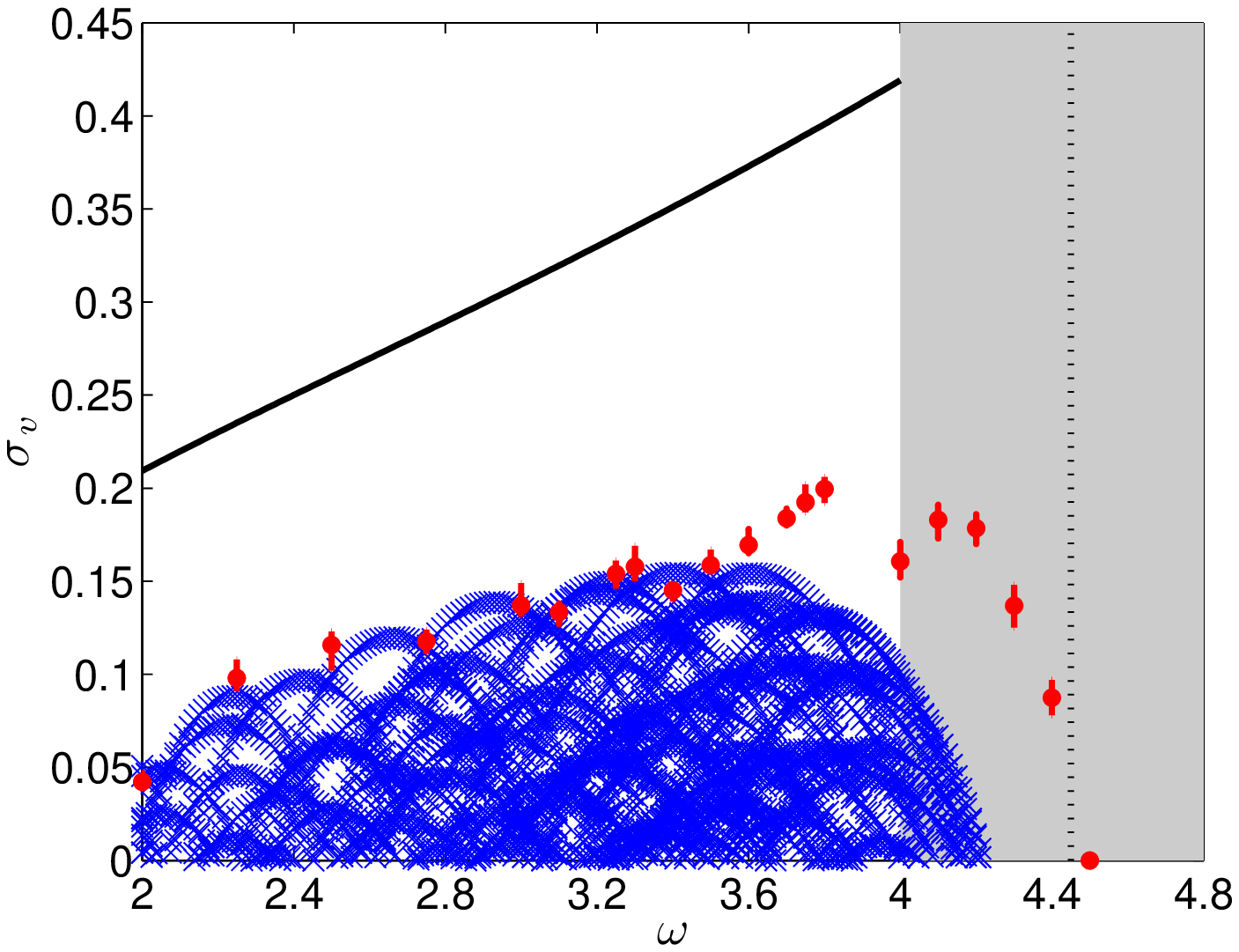}} \\
      \setlength{\epsfysize}{5.0cm}
      \subfigure[]{\epsfbox{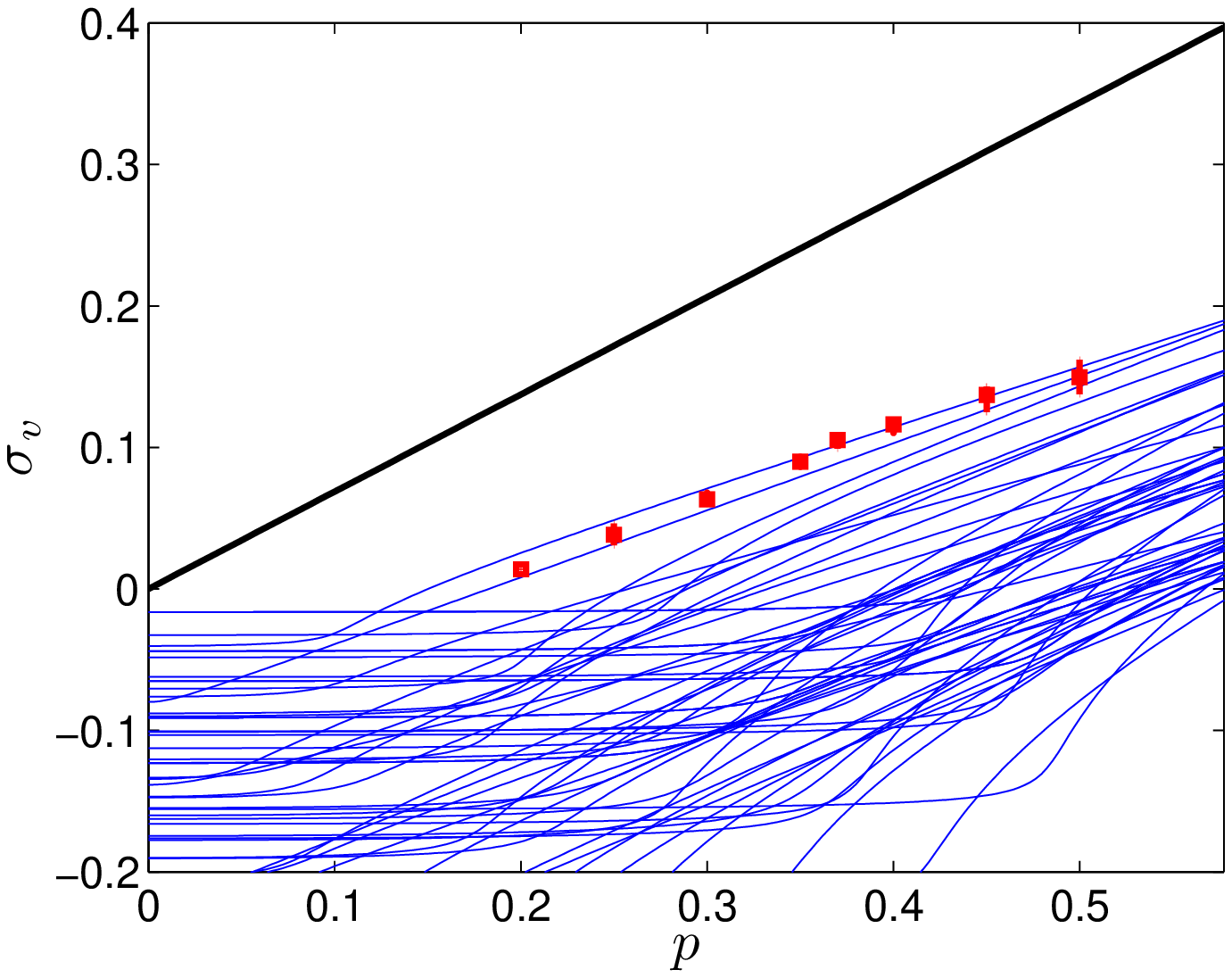}}  &
            \setlength{\epsfysize}{5.0cm}
      \subfigure[]{\epsfbox{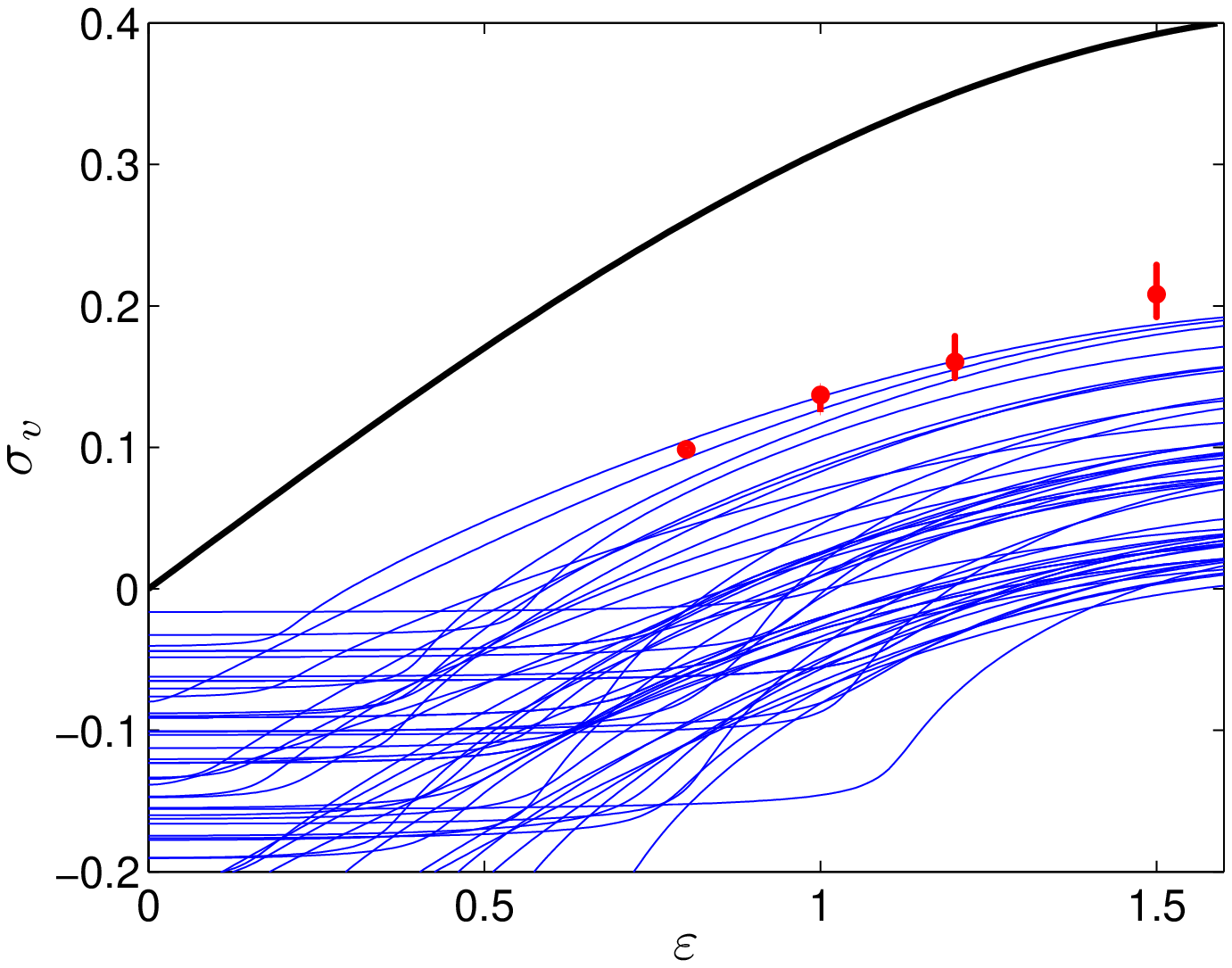}} 
    \end{tabular}
    \caption{Growth rate $\sigma_v$ in function of the flow parameters $E$, $p$, $\varepsilon$ and $\omega$ for $n=3$. Results of the asymptotic WKB analysis (thick black line) given by (\ref{eq:sigWKB11}) for $C=-1/2$ (boundary pathline), global analysis (thin blue lines and blue crosses) and numerical simulations (red circles). (a) In function of $E$ for $\varepsilon=1$, $p=0.45$, and $\omega=3$, (b) in function $\omega$ for $\varepsilon=1$, $p=0.45$ and $E= 5\cdot 10^{-4}$, (c) in function of $p$ for $\omega=3$, $\varepsilon=1$ and $E=5 \cdot 10^{-4}$, and (d) in function of $\varepsilon$ for $\omega=3$, $p=0.45$ and $E=5 \cdot 10^{-4}$.}
    \label{fig:sigma}            
  \end{center}
\end{figure}
\subsubsection{Thresholds and viscous growth rates of the LDMI}
 We now investigate systematically how $\varepsilon$, $p$ and $\omega$ and $E$ affect the growth rate $\sigma_v$ of the instability. Libration frequencies $\omega < 2$ are left out of consideration to avoid any direct forcing of inertial modes. Note however, that the LDMI is nevertheless expected for $\omega < 2$, as shown by figure \ref{fig:nonviscous}. In order to extract growth rates from time series of the axial kinetic energy $E_z$ such as the ones plotted in figure  \ref{fig:time_series_LDTI}(a), (b), we proceed as follows. First, we use a moving average procedure to filter out the frequency component at $2 \omega$ from $E_z$. Subsequently, we fit a function of the form $A\exp(2\sigma_v t) + B$ to the filtered signal within a certain time window $[t_1,t_2]$. The growth rates $\sigma_v$ obtained in this way are slightly dependent on the choice of $t_1$ and $t_2$. For the robustness of the results, we have repeated the procedure described above for several choices of $t_1$ and $t_2$. In the following figures, the growth rates displayed correspond to the mean of the measured values, whereas error bars indicate the maximum and the minimum value.

\par
The thick solid lines in figure \ref{fig:sigma} show the (inviscid) asymptotic WKB formula (\ref{eq:sigWKB11}) for $C=-1/2$ (boundary pathline), whereas each of the crosses or thin lines represent a (viscous) resonance between a pair of inertial modes. The red circles finally, correspond to the numerically obtained growth rates, and are in good agreement with the values of $\sigma_v$ of the most unstable resonances. The slight numerical discrepancy between the simulations and global analysis may be attributed to the following two factors: (ii) the global theory is, strictly speaking, only valid in the limit $\varepsilon p \ll 1$, and (ii) the seed perturbation on which the instability grows consists of pure numerical noise, which implies that we do not control whether inertial modes are equally represented within this seed perturbation. As such, the most unstable resonance does not necessarily dominate at the onset of instability and during its initial exponential growth. Finally, we see that the asymptotic WKB analysis provides a correct upper bound for the growth rates, but the results are not close to this bound. This is naturally due to the fact that the WKB theory is an inviscid theory, whereas the range of Ekman numbers under consideration is not asymptotically small. Note also that we represent in figure \ref{fig:sigma} the maximum local inviscid WKB growth rate, which is reached on the boundary pathline ($C=-1/2$), i.e. in a zone dominated by viscosity (viscous boundary layer) in the simulations. The growth rate provided by the local stability analysis can thus only be an upper bound. Nevertheless, we observe that the WKB captures reasonably well the trend of the dependency of $\sigma_v$ on $\omega$, $\varepsilon$ and $p$.
 
\par
Figure \ref{fig:sigma}(b) shows us that the instability tends to disappear for $\omega>4$. Indeed, the LDMI is the result of parametric resonances of inertial waves that do not exist for $\omega > 4$ at zeroth order in $\varepsilon$ and $p$: this is the forbidden zone (see section \ref{sec:asympt}). Note that the finite values of $\varepsilon p$ impose us to consider the first order in $\varepsilon p$, which gives a forbidden zone for $\omega > 4+\varepsilon p$ \cite[as in][]{le2000three}. The global analysis and the numerical simulations give evidence of the existence of resonant frequencies around which the growth rate peaks as e.g. at $\omega=3$, as already observed by \cite{Cebron_Pof}. Near $\omega=3.75$, there is some disagreement between the global theory and the simulations. The increased growth rates in this frequency range are due to the proximity of the forbidden zone, which leads to numerous resonances involving higher-order inertial modes, as already seen in \cite{le2010tidal} for instance. Hence, these become increasingly difficult to capture in the global analysis.  
\par
We can furthermore decompose the flow in its Fourier components (along z-direction):
\begin{equation}
{\boldsymbol u}({\bs r},t) = \sum_{n_z=-\infty}^{\infty} {\boldsymbol U}_{n_z}(x,y,t)\,\mathrm{e}^{\mathrm{i}n_z z},
\end{equation}
and consider the energies $E^{n_z}(t)$ associated with these modes:
\begin{equation}
E^{n_z}(t) = \frac{1}{2}\iint  {\bs U}_{n_z}^2(x,y,t) + {\bs U}_{-n_z}^2(x,y,t) \,\mathrm{d}x \mathrm{d}y.
\end{equation}
\begin{figure}                   
  \begin{center}
   \includegraphics[width=\textwidth]{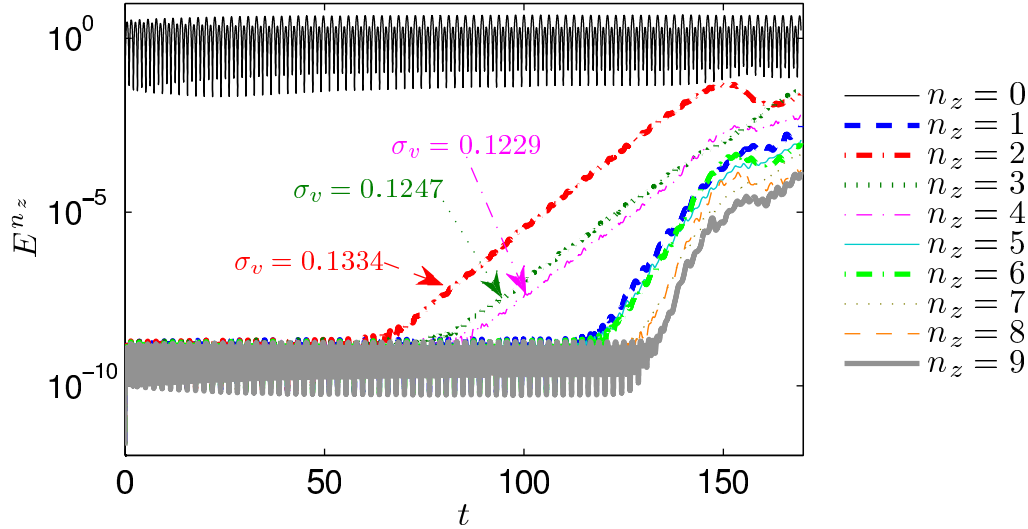}
    \caption{Time series of the axial Fourier components $n_z=0-9$ of the total kinetic energy for  $n=3$, $\omega=3$, $p=0.45$, $\varepsilon=1$ and $E=5 \cdot 10^{-4}$. Numerically obtained growth rates for the components $n_z=2,3,4$ are shown in the figure (and can be compared with the values given in figure \ref{fig:sigmamax}(b), i.e. $\sigma_v=0.136$ for $n_z=2$ and $\sigma_v=0.127$ for $n_z=3$). }
    \label{fig:time_series_Fourier}            
  \end{center}
\end{figure}
Figure \ref{fig:time_series_Fourier} shows the time evolution of the different components $E^{n_z}$ for $\omega=3$, $p=0.45$, $\varepsilon=1$ and $E=5 \cdot 10^{-4}$. We observe resonances that are associated with axial wave numbers $n_z=2,3,4$. Note that multiple resonances may coexist for each single value of $n_z$, which leads to a simultaneous growth of all the resonances. The growth rates $\sigma_v$ corresponding to $n_z=2,3,4$ are displayed as well. For $n_z=2,3$, these are in excellent agreement with the theoretically predicted growth rates given in figure \ref{fig:sigmamax}(b). However, we also find a resonance for $n_z=4$. Finally, we see that, for $t \gtrsim 120$, the flow contains a broad range of axial wavenumbers. This is a clear signature of the emergence of non-linear effects and the generation of turbulence observed in figure \ref{fig:time_series_LDTI}(d).  

\subsubsection{Amplitude of the flow driven at saturation} \label{sec:amplitude}
We have shown previously (see e.g. figure \ref{fig:time_series_LDTI}d) that the LDMI may generate vigorous flows that contain a broad range of length scales. An important measure of this regime is the amplitude $\mathcal{A}$ of the flow, defined by: 
\begin{equation}
\mathcal{A}(t) = {\sqrt{V^{-1}_{bulk}\iiint_{V_{bulk}} ({\boldsymbol u} - {\boldsymbol U})^2 \,\mathrm{dV}}} \label{def:amplitude}.
\end{equation}
This definition is based on the following considerations: the amplitude of the instability is related to the difference between the total driven (unstable) flow ${\boldsymbol u}$ and the exact (laminar) basic flow ${\boldsymbol U}$. However, as we have shown in section \ref{baseflow}, important differences between ${\boldsymbol u}$ and ${\boldsymbol U}$ exist before the instability sets in due to the boundary viscous layers. To discard the effect of these boundary layers, we limit the integration domain to a volume $V_{bulk}$ that only contains points for which $C(r,\theta)>-(1-5\delta)^2/2$. 
\par
\begin{figure}                   
  \begin{center}
     \begin{tabular}{ccc}
          \setlength{\epsfysize}{5.0cm}
            \subfigure[]{\epsfbox{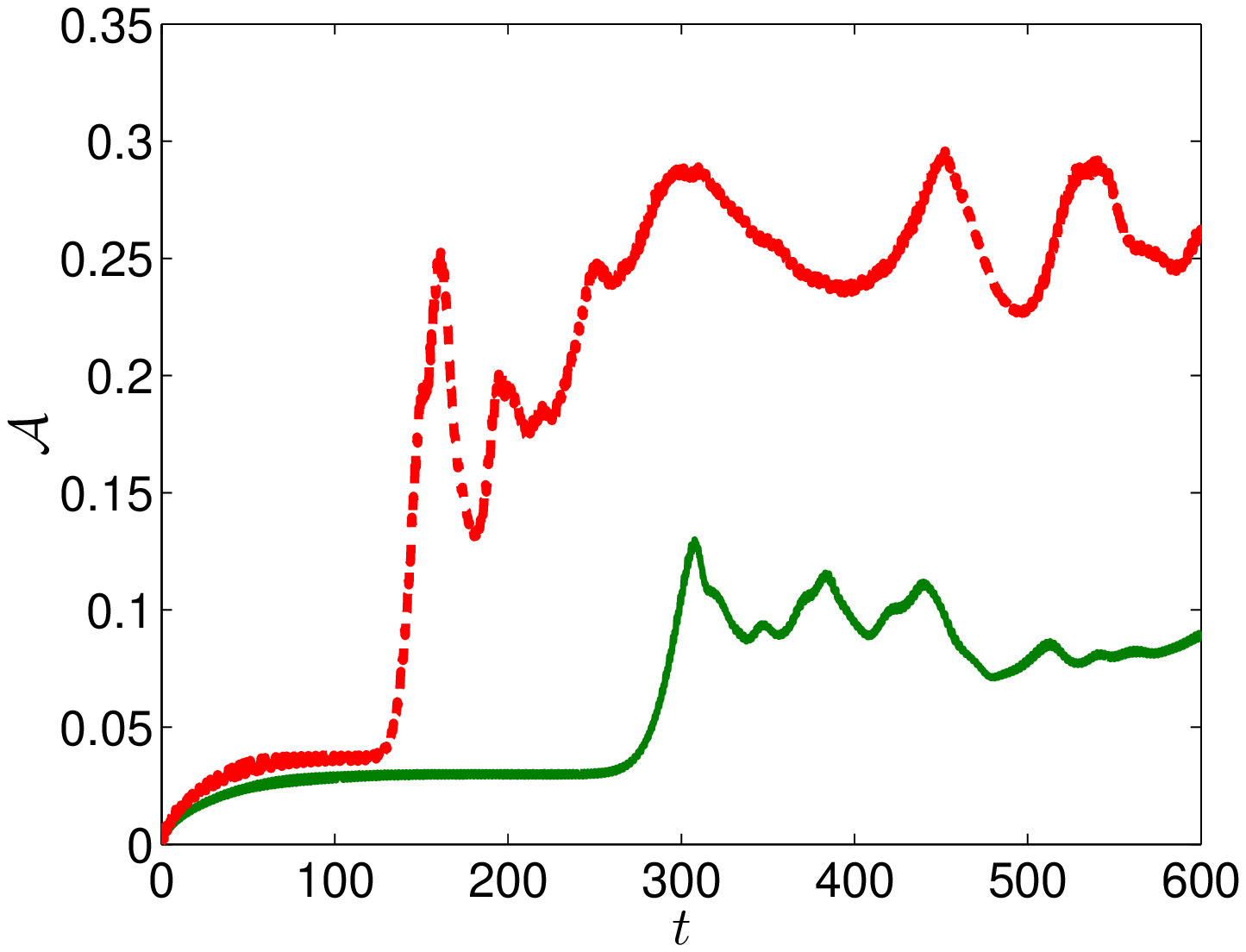}} &
            \setlength{\epsfysize}{5.0cm}
            \subfigure[]{\epsfbox{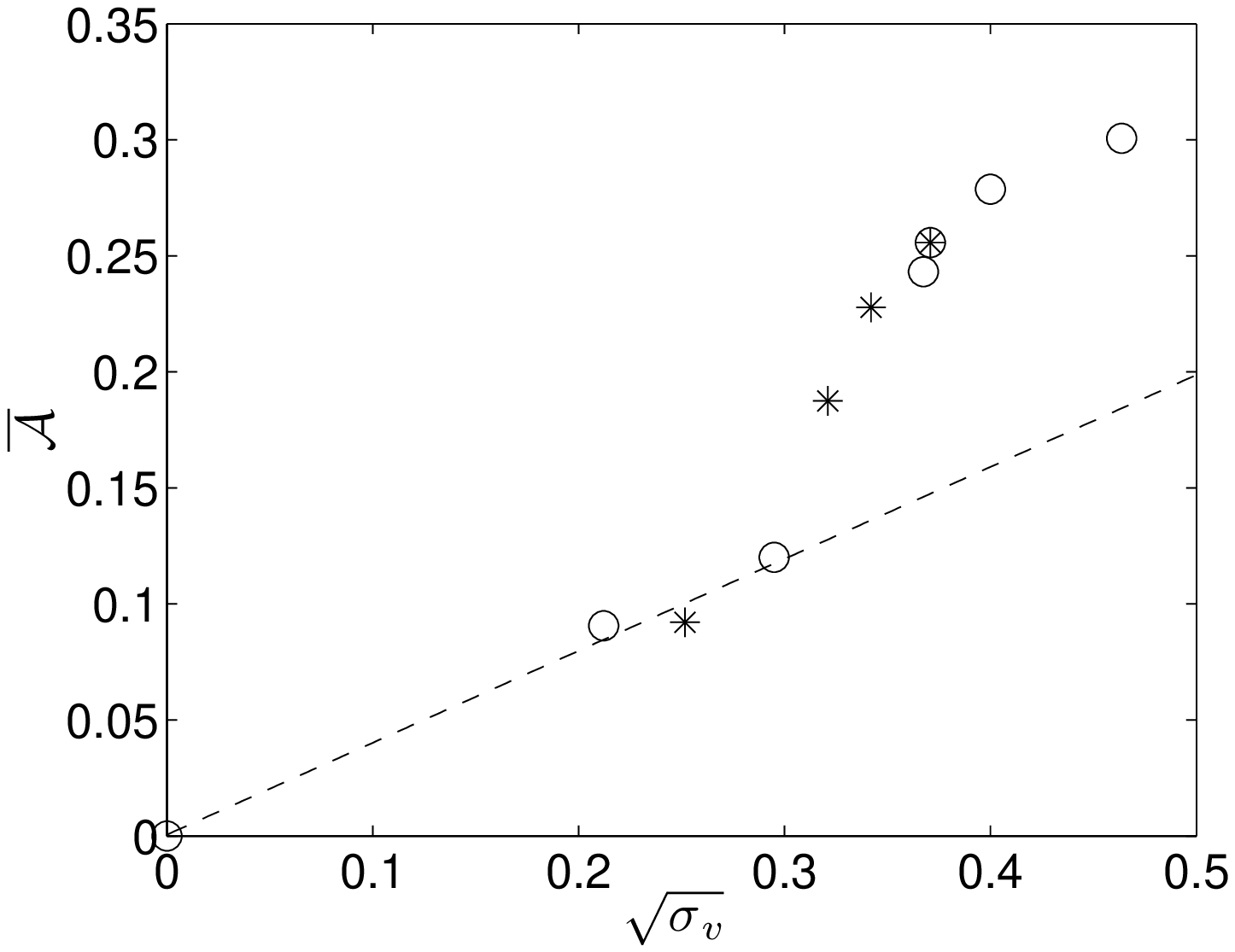}}
       \end{tabular}
       \caption{Amplitude of the LDMI for $n=3$, $\omega=3$, $\varepsilon =1$. (a) Time series of $\mathcal{A}$ (see definition \ref{def:amplitude}) for $p=0.3$, $E=5 \cdot 10^{-4}$ and $p=0.45$, $E=2.5 \cdot 10^{-4}$ (dashed line). (b) Time-averaged amplitudes $\overline{\mathcal{A}}$ against growth rates $\sigma_v$ for $p=0.45$ and $E=2.5 \cdot 10^{-3}$. $1.5 \cdot 10^{-3}$, $1 \cdot 10^{-3}$, $7 \cdot 10^{-4}$, $5 \cdot 10^{-4}$, $2.5 \cdot 10^{-4}$, $1 \cdot 10^{-4}$ ($\circ$) and $E = 5 \cdot 10^{-4}$ and $p=0.3$, $0.37$,$0.4$, $0.45$ $(\ast)$. Linear fit through the four leftmost data points with a slope of 0.40 (dashed line).}
      \label{fig:amplitude}            
   \end{center}
\end{figure}

In figure \ref{fig:amplitude}(a), we show time series of $\mathcal{A}(t)$ for the following parameter sets: $n=3$, $\omega=3$, $p=0.30$, $\varepsilon=1$, $E=5 \cdot 10^{-4}$ and $n=3$, $\omega=3$, $p=0.45$, $\varepsilon=1$, $E=2.5 \cdot 10^{-4}$. In both cases, we can identify three distinct stages. Prior to the presence of the LDMI, ${\mathcal A}(t)$ is almost constant and remains smaller than 0.05. Then, ${\mathcal A}(t)$ increases exponentially, and evolves in a complex way. Eventually, ${\mathcal A}(t)$ reaches a saturated state, in which it fluctuates around some time-averaged value.
\par
To study the effect of the flow parameters more systematically, we consider temporal averages $\overline{{\mathcal A}}$ of ${\mathcal A}(t)$, where the averaging interval typically consists of 150-200 time units. For the values $E = 10^{-4} - 2.5 \cdot 10^{-3}$ considered, this corresponds at least to 1.5 spin-up times. In figure \ref{fig:amplitude}(b), we display $\overline{{\mathcal A}}$ against  $\sigma_v^{1/2}$ for a large number of parameter combinations. We observe that 
\begin{equation}
{\overline{\mathcal A}} \approx 0.4\, \sigma_v^{1/2} \,
\end{equation}
for $ \sigma_v^{1/2} \lesssim 0.3$. This finding is consistent with previous studies of the non-linear evolution of the elliptical instability \cite[e.g.][]{mason1999nonlinear,lacaze2004elliptical,Cebron2010PEPI}, and, in a more general sense, the theory of supercritical pitchfork instabilities. In these previous studies, it was possible to define a single control parameter $\kappa$ that governs the onset of instability. It has been observed that, close to threshold, the amplitude scales as $(\kappa - \kappa_c)^{1/2}$, where $\kappa_c$ is the critical value for the onset of instability. In our present study, we may thus interpret $\sigma_v$ as an equivalent to $\kappa - \kappa_c$. This seems indeed justified as both measures are proxies for the distance from threshold.
\par
For larger values of $\sigma_{v}$, this simple scaling law does not hold anymore. This is in agreement with \cite{kerswell2002elliptical}, who argues that the primary instability only saturates and is stable for a small range of parameters near the threshold. Finally, we also observe that the ${\overline {\mathcal{A}}}$ tends to saturate to a maximum value of approximately 0.3 for $\sigma^{1/2}_v \gtrsim 0.45$. 

\subsubsection{Viscous dissipation of the instability}
\begin{figure}                   
  \begin{center}
     \begin{tabular}{ccc}
          \setlength{\epsfysize}{5.0cm}
            \subfigure[]{\epsfbox{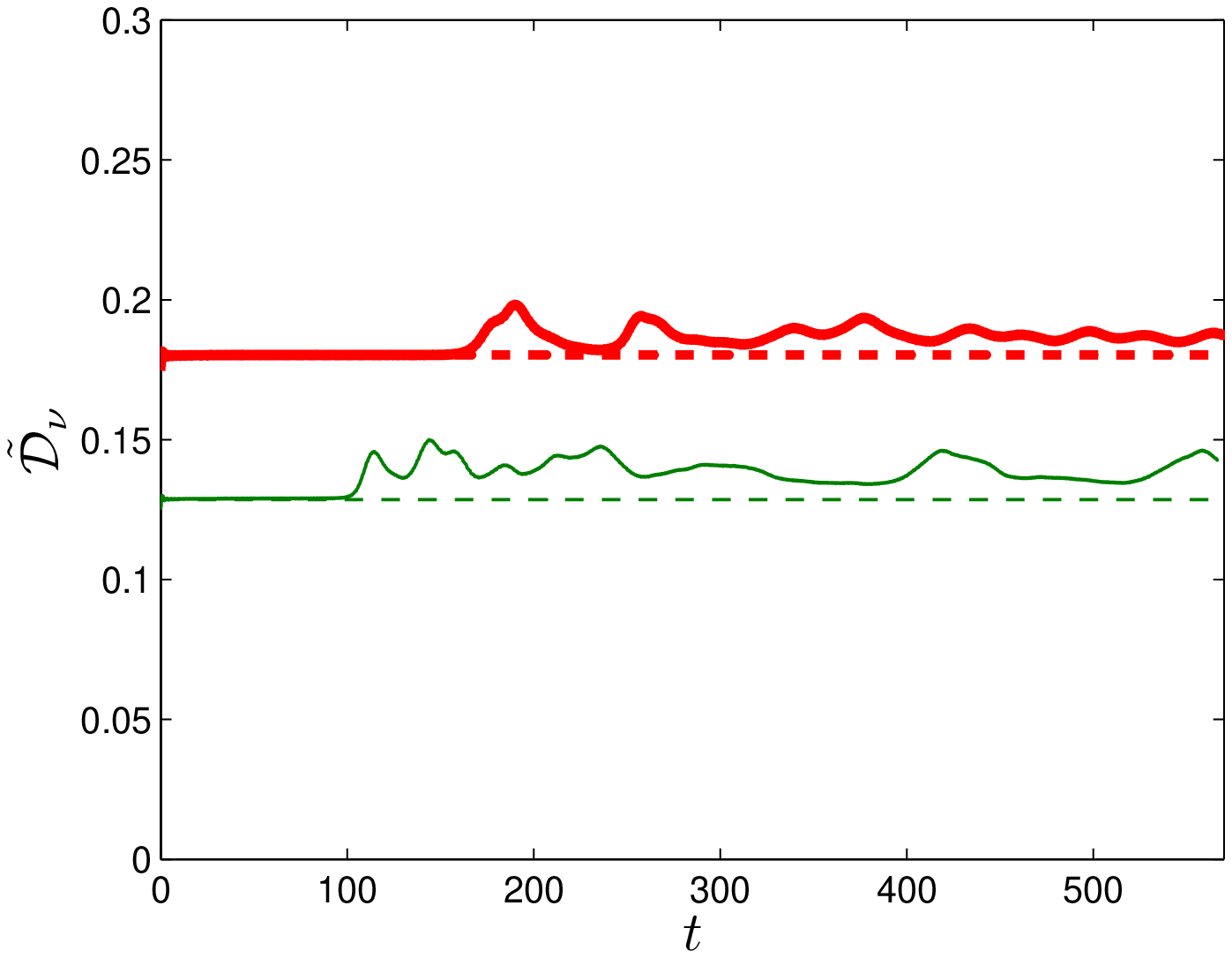}} &
            \setlength{\epsfysize}{5.0cm}
            \subfigure[]{\epsfbox{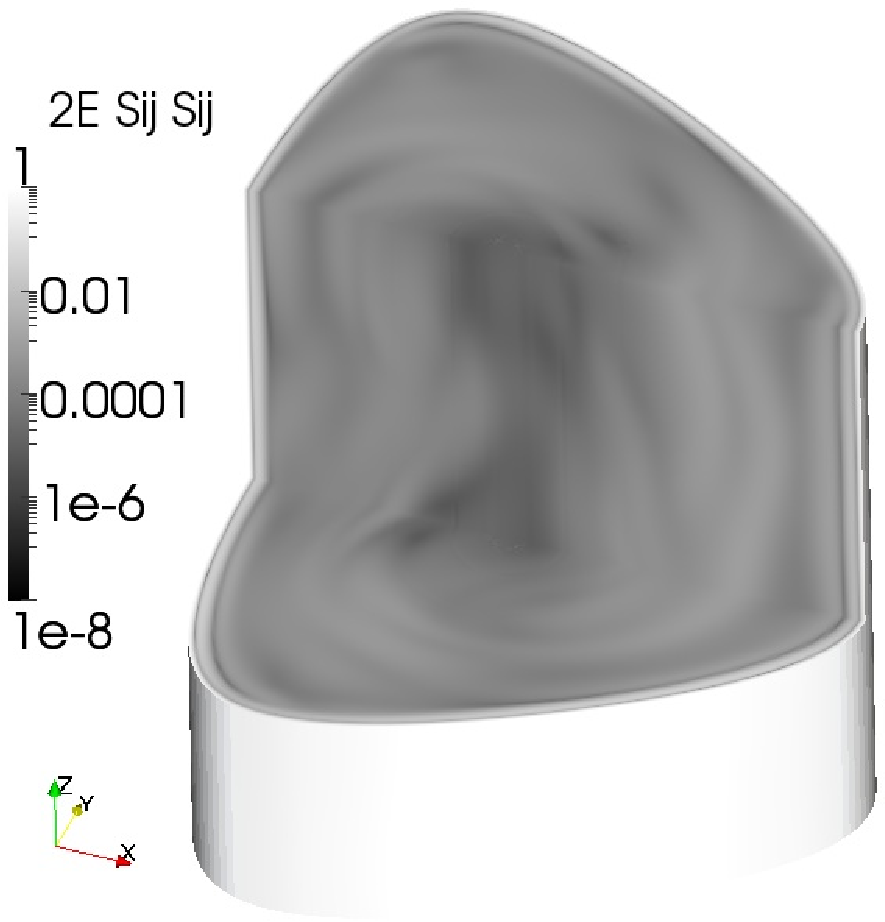}} 
       \end{tabular}
       \caption{(a) Moving average $\tilde{\mathcal{D}_{\nu}}$ of the viscous dissipation of the flow for $n=3$, $\omega=3$, $\varepsilon=1$, $p=0.4, E=5 \cdot 10^{-4}$ (thick) and $p=0.45, E=2.5\cdot 10^{-4}$ (thin). Dashed lines indicated the average dissipation $\mathcal{D}_{\nu}^{L}$ of the laminar base state. (b) Snapshot of the local viscous dissipation rate $2E S_{ij} S_{ij}$ for $n=3$, $\omega=3$, $\varepsilon=1$, $p=0.4$, $E=5 \cdot 10^{-4}$ at $t=424$.  }
      \label{fig:dissipation_time_snapshot}            
   \end{center}
\end{figure}

The viscous dissipation rate  $\mathcal{D}_{\nu}$ is defined by
\begin{eqnarray}
\mathcal{D}_{\nu} & = & 2 E\,  \iiint S_{ij} S_{ij} \,\mathrm{d}V ,
\label{def_dissipation}
\end{eqnarray}
where $S_{ij}=(\nabla{\boldsymbol u} + \nabla{\boldsymbol u}^\textrm{T})/2$ is the strain-rate tensor. This quantity however is strongly oscillating, and therefore, we show in figure \ref{fig:dissipation_time_snapshot}(a), for two set of parameters, a moving-average of $\mathcal{D}_{\nu}$ with an averaging window of two libration periods, and denote it $\tilde{\mathcal{D}_{\nu}}$. Clearly, even before the onset of instability, $\tilde{\mathcal{D}_{\nu}}$ takes significant values and is constant. The dissipation in this stage is mainly due to the presence of viscous boundary layers. In Appendix \ref{sec:app_dissipation}, we have modeled this dissipation with a simple theoretical model based on the boundary layer theory of \cite{wang1970cylindrical}. This model shows reasonable agreement with simulation results of the laminar base state. We denote this dissipation of the laminar flow $\mathcal{D}_{\nu}^{L}$ and indicate its average value by a dashed line in figure \ref{fig:dissipation_time_snapshot}(a). After the onset of instability, the dissipation slightly increases. In figure \ref{fig:dissipation_time_snapshot}(b), we show a snapshot of the local viscous dissipation rate $2E S_{ij}S_{ij}$ at $t=424$ for the case $n=3$, $\omega=3$, $\varepsilon=1$, $p=0.4$ and $E=5 \cdot 10^{-4}$. As can be seen, the dissipation rate is up to three orders of magnitude larger in the boundary layer region. Since the volume fraction occupied by this region is of the order of $E^{1/2} \approx 0.022$, we expect that boundary layer contributions will also dominate the total viscous dissipation rate in the non-linear regime. 
\par
We may now define the dissipation only due to the instability $\mathcal{D}_{\nu}^{I}$ as: 
\begin{equation}
\mathcal{D}_{\nu}^{I} = \mathcal{D}_{\nu} -  \mathcal{D}_{\nu}^{L}.
\end{equation}

\begin{figure}                   
  \begin{center}
     \begin{tabular}{ccc}
          \setlength{\epsfysize}{5.0cm}
            \subfigure[]{\epsfbox{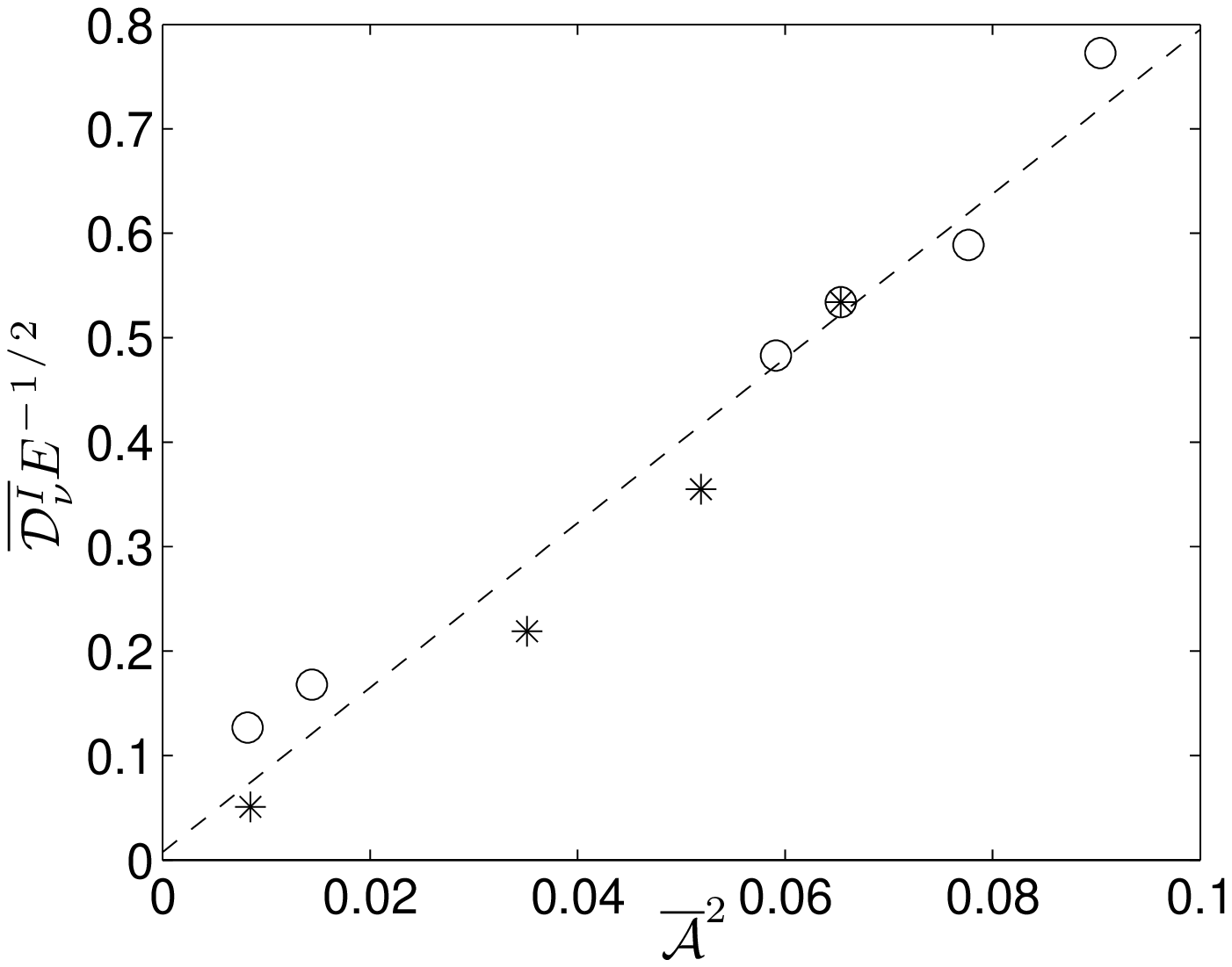}} &
            \setlength{\epsfysize}{5.0cm}
            \subfigure[]{\epsfbox{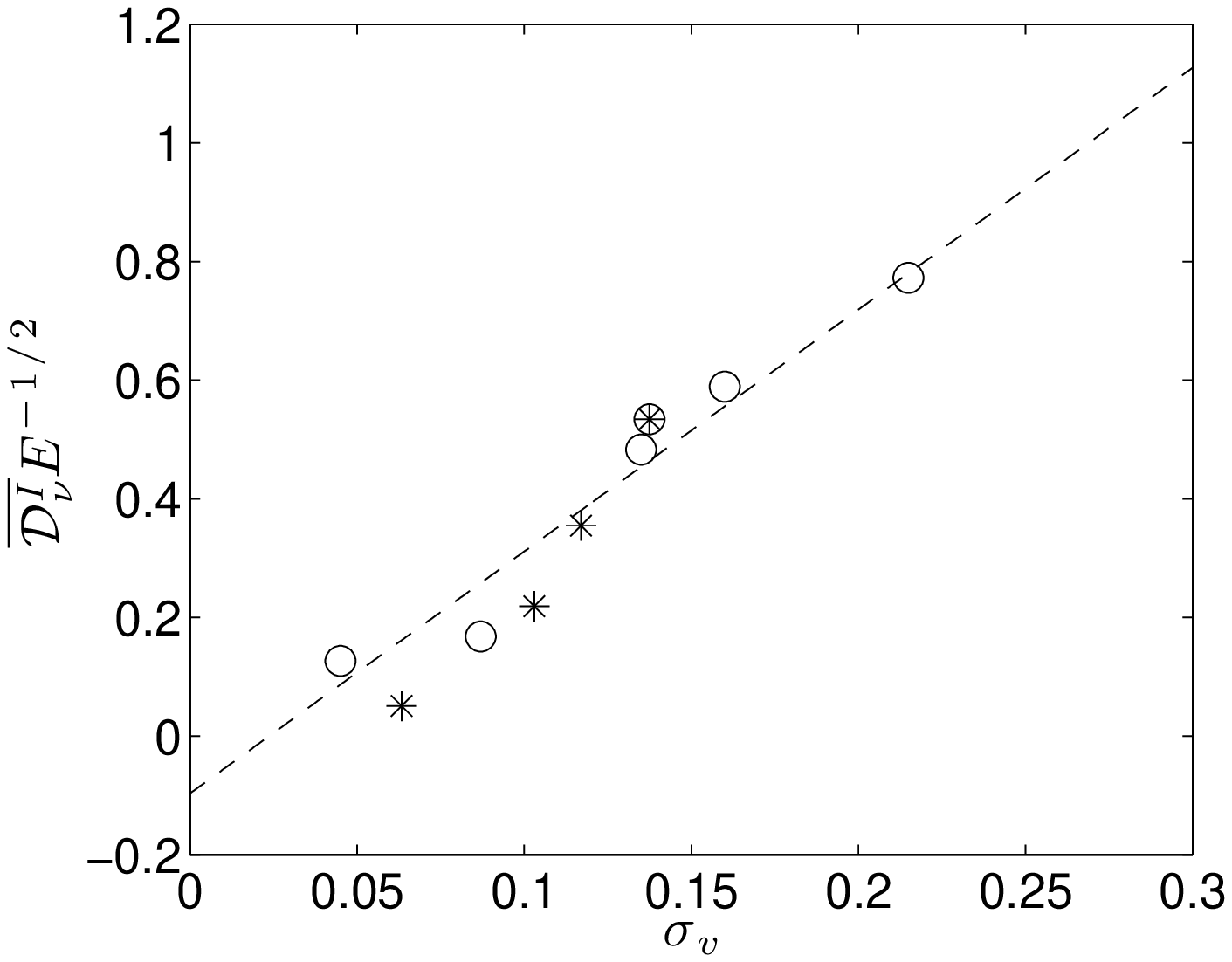}}
       \end{tabular}
       \caption{Evolution of the viscously rescaled dissipation $E^{-1/2} \overline{\mathcal{D}_{\nu}^{I}}$ for $n=3$, $\omega=3$, $\varepsilon=1$, $p=0.45$ and $E=1.5 \cdot 10^{-3}, 10^{-3}, 7 \cdot 10^{-4}, 5 \cdot 10^{-4}, 2.5 \cdot 10^{-4}, 10^{-4}$ $(\circ)$ and $p=0.3, 0.37,0.4, 0.45$, $E=5\cdot 10^{-4}$ $(\ast)$, dashed lines being linear fits of the data points. (a) In function of $\overline{\mathcal{A}}^2$, i.e. the square of the amplitude of the equilibrated flow driven by the LDMI. (b) In function of the growth rate $\sigma_v$.}
      \label{fig:dissipation_amplitude_sigma}            
   \end{center}
\end{figure}

\noindent As for the amplitude of the instability (see section \ref{sec:amplitude}), we now consider time-averages of $\mathcal{D}_{\nu}^{I}$ over long time intervals in the saturated non-linear regime, and investigate how this quantity scales with respect to other characteristics of the instability. In figure \ref{fig:dissipation_amplitude_sigma}, we find that $\overline{\mathcal{D}_{\nu}^{I}}$ scales as:
\begin{equation}
\overline{\mathcal{D}_{\nu}^{I}} \approx 7.8\, \overline{\mathcal{A}}^2  \sqrt{E}.
\end{equation}
This scaling law is in agreement with previous studies \cite[e.g.][]{williams2001lunar,lebarsNature}, and is consistent with (\ref{def_dissipation}). Indeed, as the viscous dissipation is quadratic in the velocity, we also expect it to scale quadratically in $\overline{\mathcal{A}}$. Since we have established previously that, close to the threshold, $\overline{\mathcal{A}}$ scales as $\sigma_v^{1/2}$, we expect $\overline{\mathcal{D}_{\nu}^{I}} \propto \sigma_v \sqrt{E} $. This is indeed the case, as illustrated in figure \ref{fig:dissipation_amplitude_sigma}, where we see that all data points approximately collapse on a straight line given by
\begin{equation}
\overline{\mathcal{D}_{\nu}^{I}} \approx 4.1 \,\sigma_v \sqrt{E} \, .
\end{equation} 
It is remarkable that the viscous growth rate, a result of the linear stability analysis, is still a relevant parameter to characterize the non-linear regime. It may indicate that the non-linear regimes we have explored are not very far from the instability threshold.

\subsection{LDMI, a generic instability (simulation in a spherical geometry)}

Because of its possible geophysical relevance and to show that the LDMI is a generic mechanism, we now investigate numerically whether the libration-driven tripolar instability can also take place in deformed spherical containers. We thus consider a spherical container, and move each point of its boundary at a cylindrical radius $r$ towards a point at the cylindrical radius $r'$ following
\begin{equation}
r' = \left[1+\frac{p}{n} \cos(n \theta) \right] r \, ,
\end{equation}
in each plane perpendicular to the rotation axis (using here $n=3$). This deformation corresponds to a multipolar shape in the limit $p \ll 1$ (see eq. \ref{eq:explicitstream}).

\begin{figure}                   
  \begin{center}
     \begin{tabular}{ccc}
        \setlength{\epsfysize}{5.0cm}
            \subfigure[]{\epsfbox{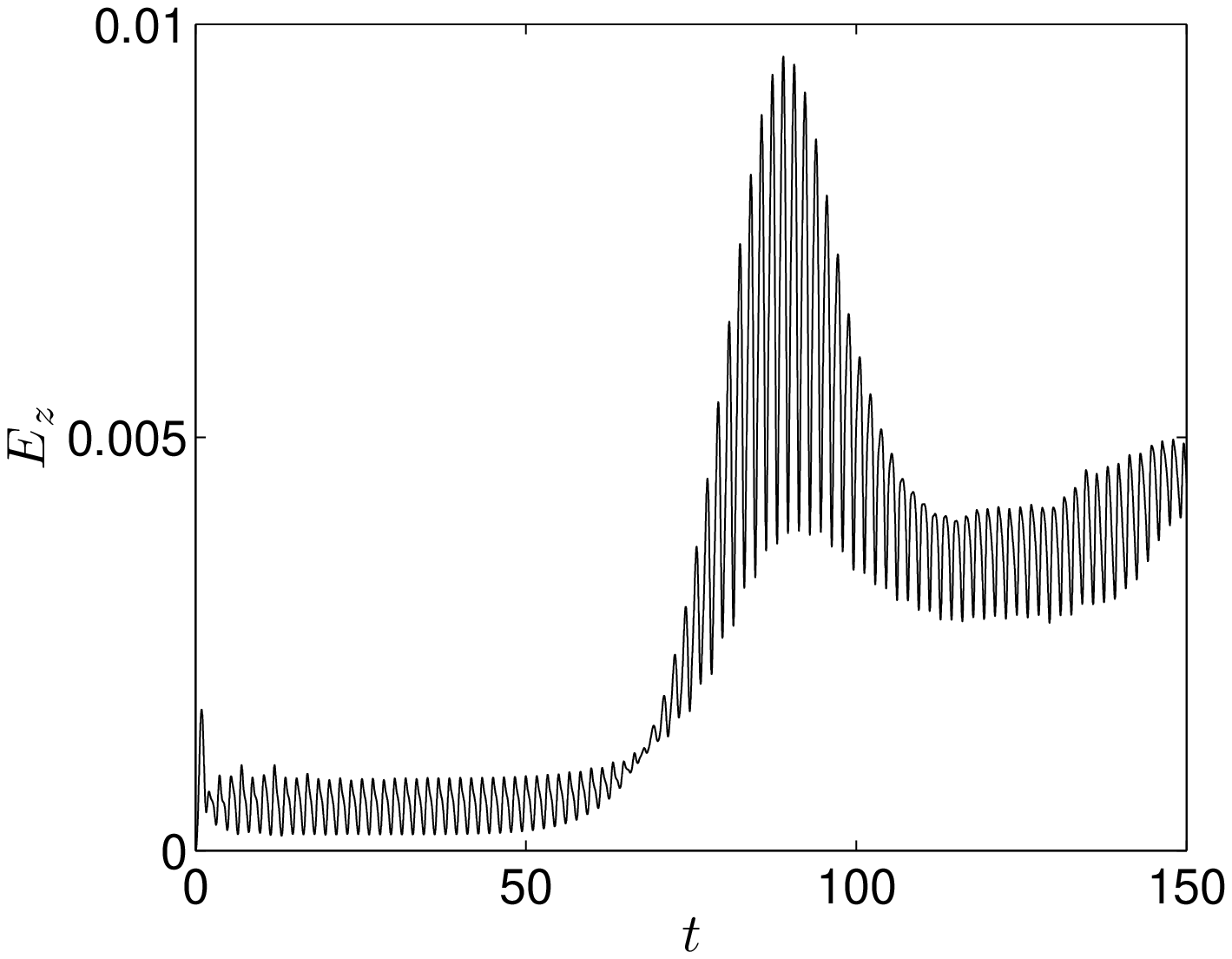}} &
            \setlength{\epsfysize}{5.0cm}
            \subfigure[]{\epsfbox{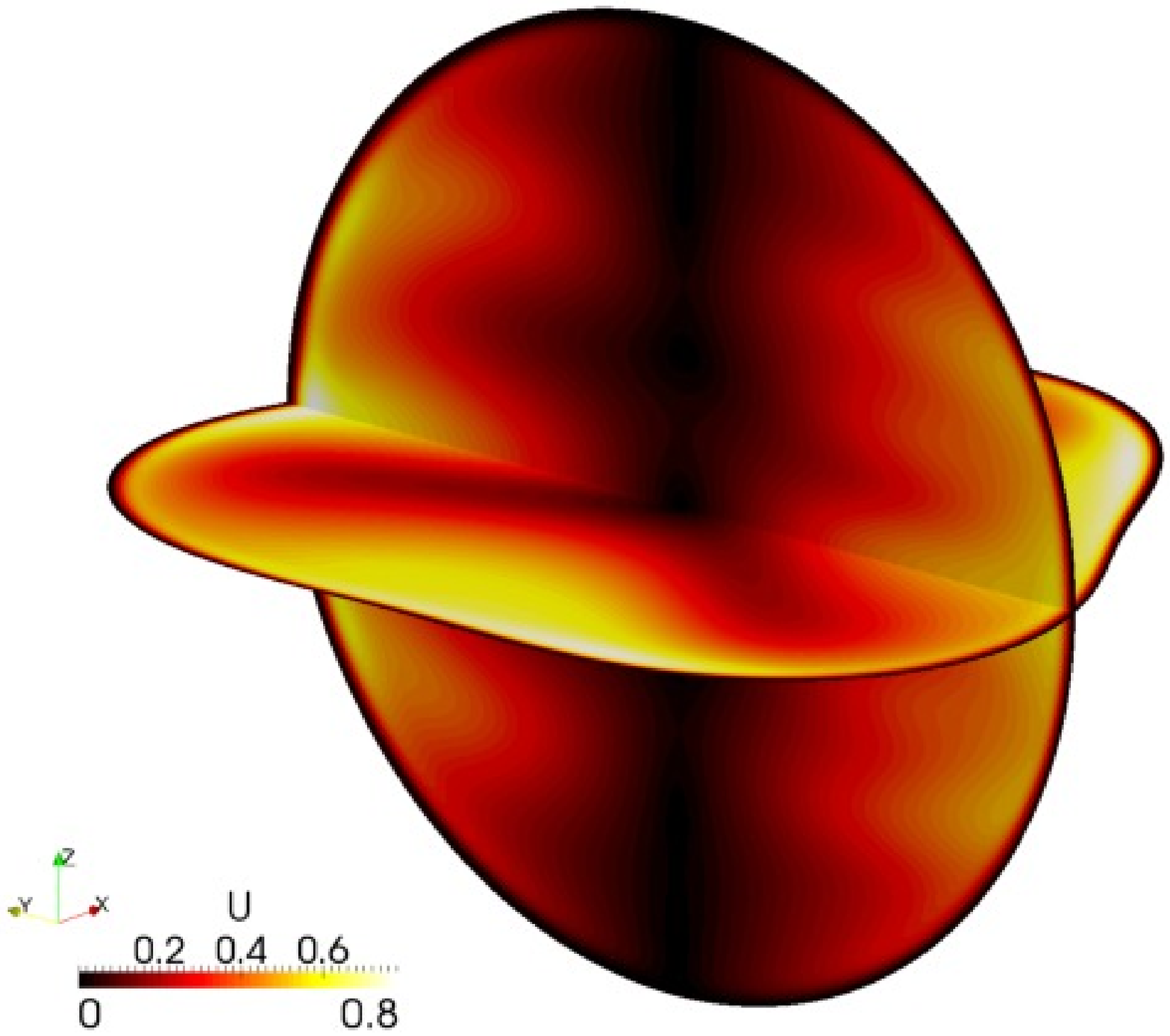}} \\
       \end{tabular}
       \caption{Libration-driven tripolar instability in a deformed spherical container for the parameters $p=0.25$, $E=5 \cdot 10^{-4}$, $\omega = 3.8$ and $\varepsilon=1$. Time series of axial kinetic energy (a) and 3D-snapshot of the velocity magnitude at $t=84$.}
      \label{fig:LDTI_sphere}            
   \end{center}
\end{figure}

Figure \ref{fig:LDTI_sphere} displays results of a simulation for parameters $n=3$, $p=0.25$, $\omega=3.8$, $\varepsilon=1$ and $E= 5 \cdot 10^{-4}$, which are values on the same order of magnitude as the ones used in previous sections on the cylindrical geometry. In figure \ref{fig:LDTI_sphere}(a), the time series of the axial kinetic energy (\ref{eq:axial_energy}) again exhibits three distinct stages. Prior to the onset of instability (for $t<60$), $E_z$ oscillates around a small but non-negligeable value of $0.001$. The corresponding velocity component $u_z$ is related to the Ekman pumping due to the viscous Ekman layers. Starting from $t \approx 60$, $E_z$ undergoes an exponential growth over a short time interval (until $t \approx 90$): a LDMI is thus excited. Further evidence for this is given in figure \ref{fig:LDTI_sphere}(b), where we observe that the velocity magnitude $||{\bs u}||$ is characterized by an oscillatory spatial pattern in the bulk of the fluid. Moreover, we find that the growth rate of the instability is $\sigma_v \approx 0.108$. We can compare this value to the corresponding values for cylindrical geometry, shown in figure \ref{fig:sigma}(b). For $p=0.45$ and all other parameters equal as in the present spherical case, we find that the growth rate in cylindrical geometry is $\sigma_v \approx 0.2$. Hence, for $p=0.25$, we can estimate a growth rate that is approximately $0.2 \cdot 0.25/0.45 \approx 0.111$, which is in good agreement with the measured value of $\sigma_v \approx 0.108$.
We can thus conclude that the LDMI, as a local instability, can be excited in any geometry with a non-zero multipolar component in its cross-section if the ratio $\varepsilon p/E$ is large enough.

\section{Conclusion and discussion} \label{sec:discussion}

Given the planetary relevance of libration driven flows, a number of studies has been devoted to librating axisymmetric containers in order to investigate the role of the viscous coupling  \cite[e.g.][]{busse2010mean,busse2010zonal,calkins2010axisymmetric,sauret2010experimental,noir2009experimental,noir2010experimental,Noir2012,sauret2012fluid}. These works show that in this case, libration does not lead to significant power dissipation or angular momentum transfer. As shown by \cite{Cebron_Pof}, these conclusions should be re-addressed in elliptical containers, since space-filling turbulence may be observed in numerical and laboratory experiments. In this work, we have shown that this space-filling turbulence is actually due to a particular case of a generic instability, the Libration Driven Multipolar Instability (LDMI), which can be excited in any librating non-axisymmetric container. For instance, in librating synchronized moons \cite[see e.g.][for details]{Noir2012}, the Ekman numbers of fluid layers are so small ($E=10^{-12}-10^{-10}$) that a LDMI can be expected, even if the libration amplitudes and the deformations are very small ($p=10^{-5}-10^{-3}$, $\varepsilon=10^{-5}-10^{-3}$, depending on the compressibility of the fluid and the rigidity of the solid layer). This may question the usual spherical geometry approximation used to study numerically planetary flows.

In the present study, we have first performed a short--wavelength Lagrangian local stability analysis of the basic flow. This has allowed us to compute the inviscid growth rates of the LDMI for arbitrary deformations and libration amplitudes. Then, in the limit of small deformations, we have obtained an analytical expression for the growth rate using a multiple-scale analysis, and we have successfully compared it to the exact stability results. This local stability analysis shows that the LDMI can be excited as soon as a flow perdiodic trajectory has a multipolar shape. 

To complete our understanding of the LDMI, we have then carried out a global stability analysis, which allows us to take confinements and viscous effects into account, and thus to predict accurate onsets of the LDMI. This analysis has shown that the LDMI can also be seen as the parametric resonance between two inertial waves of a rotating fluid and a librating multipolar strain (which is not an inertial wave or mode). Seldom compared in the literature, we have shown that the local and the global stability results are consistent and lead to similar growth rates in the inviscid limit.

Numerical simulations are then used to demonstrate the existence of the LDMI in librating systems. After confirming that the considered basic flow is indeed established in the bulk of librating multipolar containers, we have systematically compared the simulations with the theoretical stability results. The quantitative agreement bewteen the two is excellent, even for the details of simultaneous growths of several inertial waves parametric resonances. The simulations are then used to explore the non-linear regimes of the LDMI, which are difficult to describe theoretically. This allows to confirm that, in the equilibrated state, LDMI driven flows are of significant amplitude, which are almost of the same order of magnitude than the basic flow \cite[as previously observed for the elliptical instability; e.g.][]{Cebron2010PEPI}. Subsequently, the viscous dissipation of the libration-driven flows is carefully quantified and compared with previously established scaling laws. Finally, we confirm that the LDMI is a generic instability by showing one simulation of the excitation of the instability in a spherical container deformed with a multipolar shape.

To conclude, we would like to point out that the experimental setup needed to study the LDMI may be one of the simplest of those devoted to inertial instabilities. Indeed, we do not need deformable containers \cite[as][for the study of elliptical or triangular instabilities]{eloy2003elliptic}, or two motors \cite[as][for the study of the precessional instability]{Lagrange11}. To experimentally study the LDMI, only a rigid deformed container and a rotating table are needed. The range of parameters where the instability is excited are easy to reach: considering for instance a small tripolar cylinder with a radius $R=15\, \mathrm{cm}$, a height $H=30\, \mathrm{cm}$ and a deformation $p=0.45$, slowly rotating at $0.9\, \mathrm{rpm}$ and librating with a period of $22\, \mathrm{s}$, a LDMI is excited as soon as the libration angle is larger than $6^{\circ} $ (any larger rotation rate would be strongy destabilizing). Then, in spite of its simplicity, such a setup easily allows, via the LDMI, the generation of strong three-dimensional space-filling flows within a rigid container.

\begin{acknowledgements}
D. C\'ebron is supported by the ETH Z\"urich Postdoctoral fellowship Progam as well as by the Marie Curie Actions for People COFUND Program. S. Vantieghem is supported at ETH Z\"urich by ERC grant 247303 (MFECE). This work was supported by a grant from the Swiss National Supercomputing Centre (CSCS) under project ID s369. We also wish to thank three anonymous referees for their appreciated comments.
\end{acknowledgements}

\begin{appendix}

\section{Local and global stability analysis:  details}
\label{Appendix:global}


\subsection{Asymptotic local stability analysis for small forcings $\varepsilon p \ll 1$} \label{sec:localAnnexe}

Assuming that the product $\varepsilon p \ll 1$ remains small, we first calculate the trajectory in the inertial frame. At leading order, we obtain the circular trajectory $ \boldsymbol{X}^{(0)} (t)$ due to the solid-body rotation. For a given initial position, e.g. $(R,0)$, $ \boldsymbol{X}^{(0)} (t)$ can be written as
\begin{eqnarray}
\boldsymbol{X}^{(0)}= R \cos (t)\, \boldsymbol{e}_x +R \sin (t)\, \boldsymbol{e}_y\, .
\end{eqnarray}
This allows to obtain the deviations $ \boldsymbol{X}^{(1)} (t)$ induced by the multipolar deformation:
\begin{eqnarray}
 \boldsymbol{X}^{(1)} (t) &= & \frac{R^{n-1}}{2\,  \omega}  \left[ \cos((\omega-1)t) - \cos((\omega+1)t) \right]\, \boldsymbol{e}_x \nonumber \\
 {}\, & & \hspace*{6mm} +\,   \frac{R^{n-1}}{2\,  \omega}  \left[ \sin((\omega-1)t)- \sin((\omega+1)t) \right]\, \boldsymbol{e}_y \, . 
\end{eqnarray}
With this, one can evaluate $\nabla \boldsymbol{U}$ on the perturbed trajectory, up to order $\mathcal{O}(\varepsilon p)$, allowing to solve for the wavenumber $\boldsymbol{\mathcal{K}} (t) $. At lowest order, we obtain $\boldsymbol{\mathcal{K}}^{(0)} (t) $, given in section \ref{sec:asympt}, which allows to obtain the next order
\begin{eqnarray}
\boldsymbol{\mathcal{K}}^{(1)} (t) &= & \frac{(n-1) R^{n-2}}{2\, \omega} \left[ \cos((\omega+1)t-\phi)-\cos((\omega-1)t+\phi) \right]\, \boldsymbol{e}_x \nonumber \\ 
 {}\,   & & \hspace*{6mm}  +\, \frac{(n-1) R^{n-2}}{2\, \omega}  \left[ \sin((\omega+1)t-\phi)+\sin((\omega-1)t+\phi) \right]\, \boldsymbol{e}_y\,  .
\end{eqnarray}
Note that we recover the expressions of $\boldsymbol{X} (t)$ and $\boldsymbol{\mathcal{K}} (t)$ given in the appendix of \cite{herreman2009effects} by considering the particular case they study, i.e. $n=2$ and $\omega=1$.

\subsection{Global stability analysis}

\subsubsection{Definition of operators} \label{sec:annexOperator}

In the global stability analysis, we have used the operators
\ba
\mathcal{L} &=&  \left [ \begin{array}{cccc}
\pd_t & - 2 & 0 & \pd_r \\
2 & \pd_t & 0 & r^{-1} \pd_\theta \\
0 & 0 & \pd_t & \pd_z \\
\pd_r + r^{-1} & r^{-1} \pd_\theta & \pd_z  & 0  
\end{array} \right ]  \\ 
\ \nonumber \\
\mathcal{N} &= & \textrm{i}  \left [ \begin{array}{cccc}
D_1 - (n-1) r^{n-2} & -\textrm{i}\, (n-2)r^{n-2} & 0 & 0 \\
- \textrm{i}\,  n r^{n-2}  &D_1 + (n-1) r^{n-2}  & 0 & 0 \\
0 & 0 & D_1& 0 \\
0 &0 &0  & 0  
\end{array} \right ] \\
\ \nonumber \\
\mathcal{V} &= & \left [ \begin{array}{cccc}
D_2 -  r^{-2}  & - 2\,  r^{-2} \pd_\theta & 0 & 0 \\
2\,  r^{-2} \pd_\theta  &D_2 -  r^{-2}   & 0 & 0 \\
0 & 0 & D_2 & 0 \\
0 &0 &0  & 0  
\end{array} \right ] \\
\ \nonumber \\
\mathcal{J} &=& \left [ \begin{array}{cccc}
1& 0 & 0 & 0 \\
0 & 1 & 0 & 0 \\
0 & 0 & 1& 0 \\
0 &0 &0  & 0  
\end{array} \right ] \\
\ \nonumber \\
\ea
with
\be
D_1 = -r^{n-1} \pd_r - \textrm{i} r^{n-2} \pd_\theta 
\quad , \quad D_2 = r^{-2} \pd^2_{rr} +r^{-1} \pd_r  +r^{-2}  \pd^2_{\theta \theta} + \pd_{zz}^2 
\ee 

\subsubsection{Inviscid boundary condition correction induced by wall deformations.} \label{sec:annex_globalwall}

The deformation of the cylindrical boundary introduces flow corrections that are necessary to account for in the global stability calculation. To find these corrections, we express the kinematic boundary condtion on the moving boundary:
\be
\frac{\pd \zeta^R}{\pd t} + \boldsymbol{U}^R \cdot \nabla \zeta^R +  \boldsymbol{u}^R \cdot \nabla \zeta^R \Big |_{surface} = 0  \, .
\ee
Here $\zeta^R$ is defined by (\ref{eq:zetarotdef}) and the basic flow $\boldsymbol{U}^R$ is given by (\ref{eq:Urotdef}). The basic flow is such that this equation reduces to 
\be
 \boldsymbol{u}^R \cdot \nabla \zeta^R \Big |_{surface} = 0  \label{CLpert}
\ee
We will now find an asymptotic form of this condition for small $p$, expressed at the unperturbed boundary surface $r=1$. When $p\ll 1$, the lateral surface is written in cylindrical coordinates as the place where $r=r_s (\theta,t)$, with
\be
r_s  (\theta,t) = 1+ \frac{p}{n} \cos n (\theta + \Delta \varphi \sin \omega t )  + \mathcal{O}(p^2) \, ,
\ee
for $\theta \in [0,2 \pi]$. Taylor expanding (\ref{CLpert}) around $r=1$, we then get 
\be
u_r^R (1) = - p  \left  [  \frac{1}{n} \cos n (\theta + \Delta \varphi \sin \omega t ) \, \right ]  \pd_r u_r (1)  - p \Big [   \sin n (\theta + \Delta \varphi \sin \omega t )  \big ] u_\theta (1)     + \mathcal{O}(p^2) \,  \nonumber
\ee 
which using (\ref{eq:gdef}) becomes
\ba
u_r^R (1) &=&   p \left [  \mathrm{e}^{\textrm{i} n \theta}  g(t)  \left (-  \frac{1}{n} \, \pd_r u_r (1)  + \textrm{i} u_\theta (1)   \right )  \right .\nonumber \\
& & \hspace*{.5cm} +  \left . \mathrm{e}^{-\textrm{i} n \theta}  g^\dagger(t)  \left (-  \frac{1}{n} \, \pd_r u_r (1)  - \textrm{i} u_\theta (1)   \right )   \right ]   + \mathcal{O}(p^2) \, . \label{urcorr}
 \ea
These $\mathcal{O}(p)$ modifications of the radial velocity enter in the growth rate calculation through the boundary terms of the 2 equations that express the solvability condition (\ref{IP}). Considering the form of the asymptotic ansatz (\ref{eq:ansatzglobal}), we have
\be
Q_{1,4}^{\dagger} (1) \,Z_{1,1}(1) = p \,  A_2 \,  g^\dagger_{-j} \underbrace{\left [Q_{1,4}^\dagger (1)  \left ( -\frac{1}{n} \pd_r Q_{2,1} (1) - \textrm{i} \,Q_{2,2} (1) \right ) \right ] }_{\mathcal{B}_{12}} \, ,
\ee
\be
Q_{2,4}^{\dagger} (1) \,Z_{2,1}(1) = p \,  A_1 \,  g_{+j} \underbrace{\left [Q_{2,4}^\dagger (1)  \left ( -\frac{1}{n} \pd_r Q_{1,1} (1) + \textrm{i} \,Q_{1,2} (1) \right ) \right ] }_{\mathcal{B}_{21}} \, .
\ee
From a physical point of view, these boundary terms correspond to the power exchanged between two interacting modes (because of boundary deformations). 

%

\section{Explicit streamlines parametrisation for $n=3$} \label{sec:app2D}
For the case of the tripolar and quadrupolar basic flows, the streamlines defined by the stream function (\ref{eq:basicflow3}) can be expressed as an analytic functional relationship between $r$ and $\theta$, i.e. $r=F(\theta)$. We have used this formulation in our numerical approach to generate a grid whose boundary coincides with the shapes of the streamlines (see section \ref{sec:comp}) $\Psi \equiv C$. We haven chosen the value of $C$ such that the boundary contour tends to the unit circle as $p$ goes to zero, i.e. $C = -1/2$:
\begin{equation}
\label{eq:psi_tripolar}
\Psi = -\frac{r^2}{2} + p \frac{r^3}{3}\cos(3\theta) = -\frac{1}{2}.
\end{equation}\\
This implicitly defines $r=F(\theta)$, and can be recast as a cubic equation for $r$:
\begin{equation}
\tau r^3 - r^2 + 1 = 0,
\label{eq:cubic_streamline}
\end{equation}
where we have defined $\tau = 2\, p \cos(3 \theta)/3$. This equation can only have positive roots for all values of $\theta$ if $\Delta = 4 - 27\tau^2 > 0$. This implies that $p \le 1/\sqrt{3}$, which is equivalent to the condition $\beta_n \le 1$ for $n=3$ (see section \ref{sec:def_problem}). Upon the introduction of $\tilde r = r - 1/(3 \tau)$, we can transform (\ref{eq:cubic_streamline}) into: 
\begin{equation}
{\tilde r}^3 - \frac{1}{3\tau^2} {\tilde r} + \frac{1}{\tau}\left(1 - \frac{2}{27\tau^2}\right) = 0.
\end{equation}
Following the general theory for the solution of cubic equations, the solutions for $\tilde r$ can now be written as follows:
\begin{eqnarray}
{\tilde r}_k = \frac{2}{3|\tau|} \cos \left[\frac{1}{3} \arccos \left(\mathrm{Sgn}(\tau) \left(1 - \frac{27}{2}\tau^2 \right) \right) - \frac{2\pi k}{3} \right] & \mbox{for } k=1,2,3,
\end{eqnarray}
and hence:
\begin{eqnarray}
r_k = \frac{1}{3\tau} + \frac{2}{3|\tau|} \cos \left[\frac{1}{3} \arccos \left(\mathrm{Sgn}(\tau) \left(1 - \frac{27}{2}\tau^2 \right) \right) - \frac{2\pi k}{3} \right] & \mbox{for } k=1,2,3.
\end{eqnarray}
\par
The choice of $k$ is now determined by the requirement that $r_k \rightarrow 1$ in the limit of vanishing $\tau$ (i.e. for infinitesimally small streamline deformation). For $\tau>0$ (respectively $\tau<0$), we find that the only acceptable solution is the one corresponding to $k=1$ (respectively $k=3$). In both cases, the streamline can be parametrised explicitly, up to the leading order in $\tau$, as:

\begin{equation}
r = 1 + \frac{\tau}{2} + \mathcal{O}(p^2) = 1 + \frac{1}{3}p \cos \theta + \mathcal{O}(p^2)= F(\theta).
\label{streamline_approx}
\end{equation}
We now compute the surface area $\mathcal{S}$ of a small annular-like region of relative thickness $\delta \ll 1$. We may write:
\begin{eqnarray}
\mathcal{S} \approx  \int_{\theta = 0}^{2\pi} \int_{r=(1-\delta)F(\theta)}^{F(\theta)} r \,\mathrm{d}r \,\mathrm{d}\theta \approx  2\pi \delta \left(1 + \mathcal{O}(p^2) + \mathcal{O}(\delta) + + \mathcal{O}(\delta p^2) \right)
\end{eqnarray}

\section{Viscous dissipation rate of the basic flow \label{sec:app_dissipation}}
In this appendix, we derive a simple model to estimate the viscous dissipation of the laminar base flow based on the boundary layer theory of \cite{wang1970cylindrical}. Since, on average, the flow is steady, the mean viscous dissipation $\overline{{\mathcal D}_{\nu}^{L}}$ should be equal to the time-averaged power of the Poincar\'e force \cite[sometimes called Euler force, see e.g.][]{eckart1960}, i.e.
\begin{equation}
\overline{{\mathcal D}_{\nu}^{L}}  =   \overline{ \iint  {\boldsymbol u}\cdot \left({\boldsymbol r} \times \frac{\mathrm{d}\gamma}{\mathrm{d}t}{\boldsymbol e}_z \right) \,\mathrm{d}{\boldsymbol r} },
\end{equation}
where $\gamma$ is given by $1-\varepsilon \cos (\omega t)$ (see section 2.1).
Since $\omega \gg \sqrt{E}$, we may adapt the local tangential boundary layer correction provided by \cite{wang1970cylindrical} to account for the non-circular shape of the container. It can be expressed in the librating frame as follows:
\begin{equation}
{\boldsymbol u} = \varepsilon\left[ \exp\left(\frac{r-F(\theta)}{\delta}\right)\cos\left(\omega t - \frac{r - F(\theta)}{\delta}\right) \right]{\boldsymbol e}_{T} -  \varepsilon \cos (\omega t){\boldsymbol e}_{\theta}.
\end{equation}
Here, ${\boldsymbol e}_T$ is a unit vector tangential to the streamline, $F(\theta)$ is a parametrization of the boundary (see Appendix \ref{sec:app2D} for the particular case $n=3$), which can be approximated by $F(\theta)=1 + pn^{-1} \cos (n \theta) + \mathcal{O}(p^2)$ in the limit $p \ll 1$, and the last term in this expression comes from our librating frame. As such, we obtain:
\begin{equation}
\overline{\mathcal{D}_{\nu}^L}=  \frac{\pi}{\sqrt{2}}\, \varepsilon^2 \sqrt{E \omega}\, 
\left[1 - 2\delta+ \delta^2 \left(1+ e^{ -\delta^{-1}} (\sin \delta^{-1} - \cos \delta^{-1} )\right)+ \mathcal{O}(p^2)\right].
\label{dissipation_poincare}
\end{equation}
Given that $n$ is not present in eq. (\ref{dissipation_poincare}), the viscous dissipation rate of the basic flow is independent of $n$ in the limit of small deformation.
\begin{figure}                   
  \begin{center}
     \begin{tabular}{ccc}
          \setlength{\epsfysize}{5.3cm}
            \subfigure[]{\epsfbox{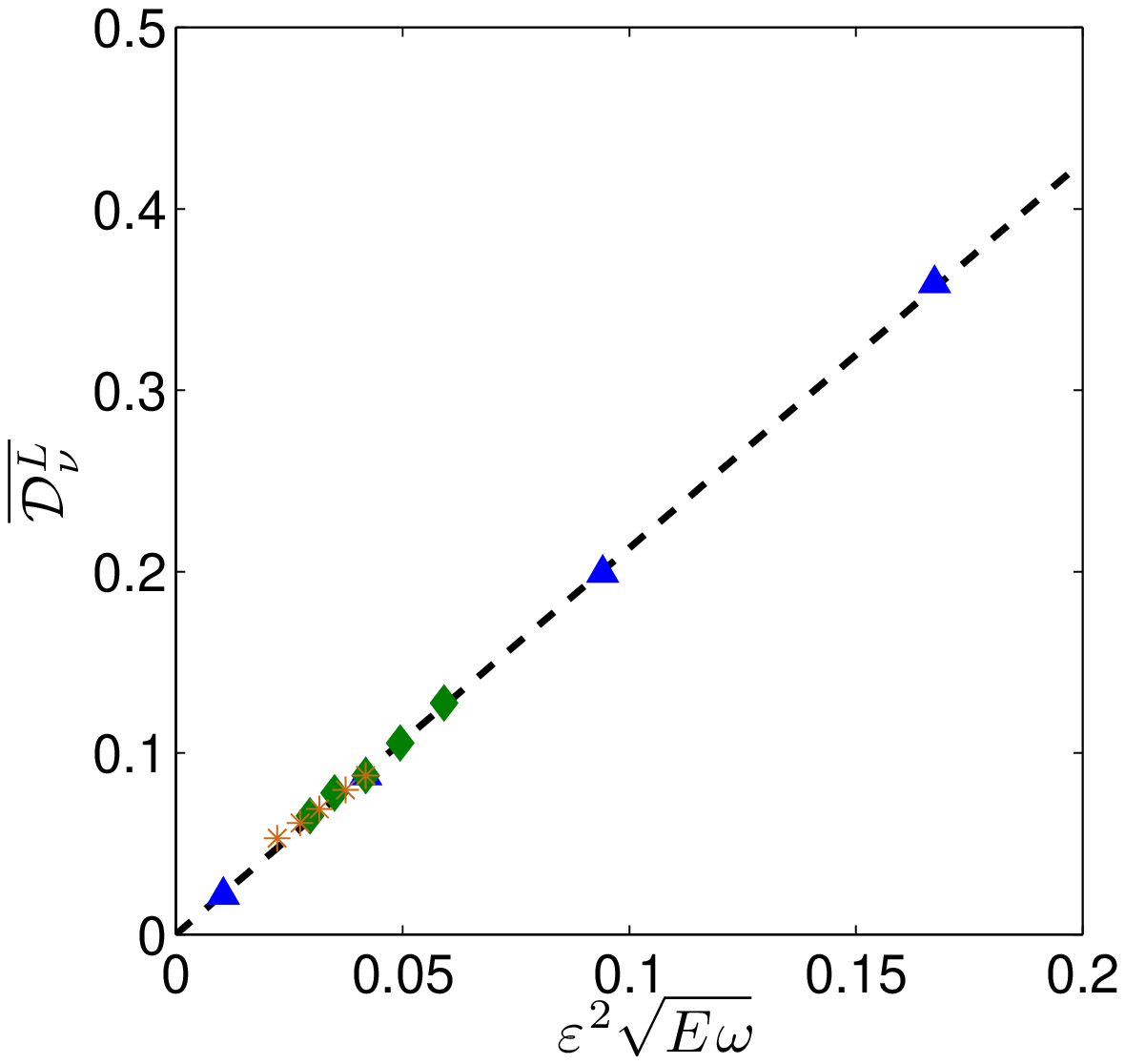}} &
            \setlength{\epsfysize}{5.0cm}
            \subfigure[]{\epsfbox{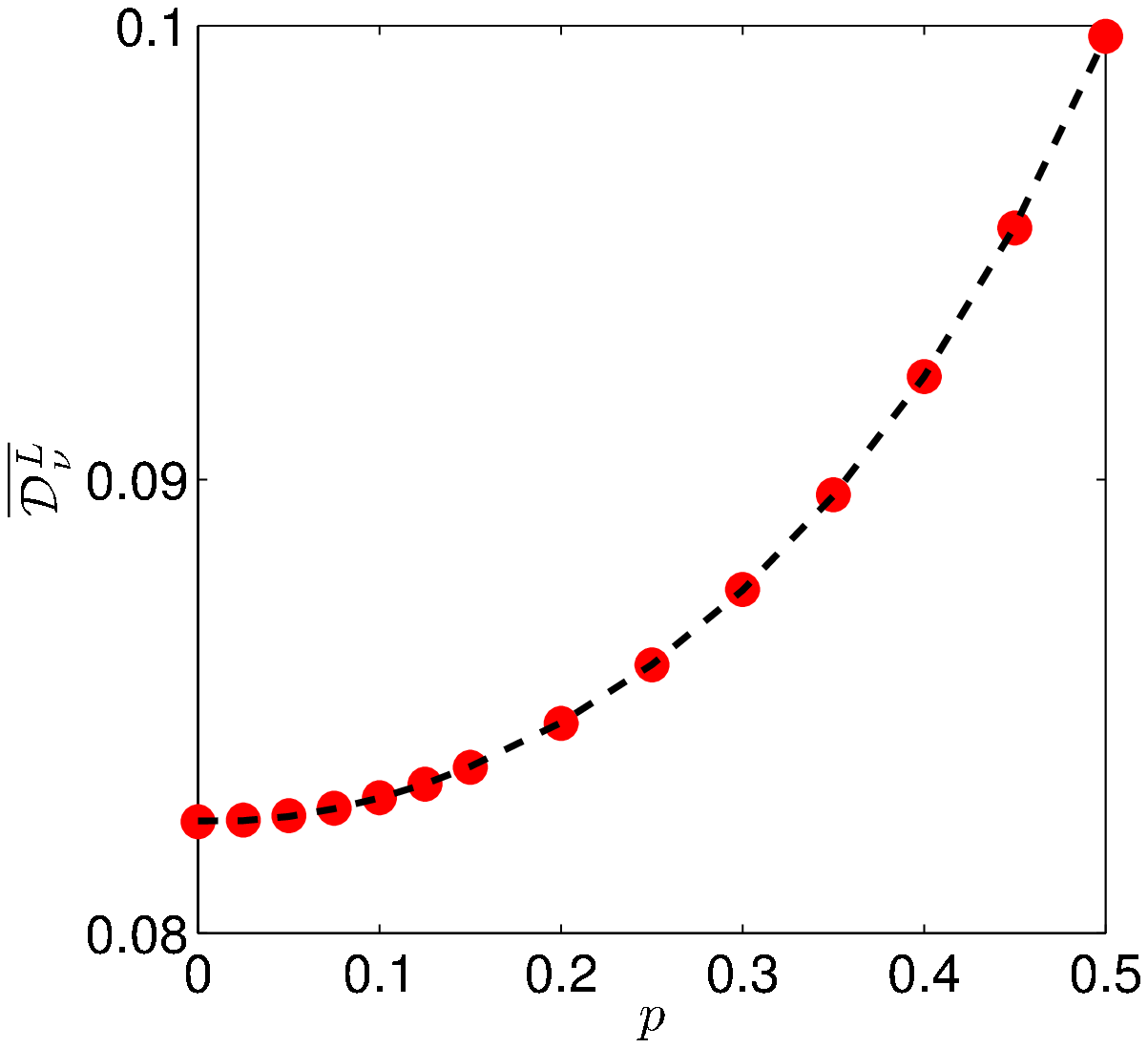}}
       \end{tabular}
       \caption{Viscous dissipation rate  $\overline{\mathcal{D}_{\nu}^{L}}$ of the two-dimensional basic flow for $n=3$. (a) in function of $\varepsilon$, $E$ and $\omega$ with fixed $p=0.3$. Triangles: varying $\varepsilon$, fixed $E=5 \cdot 10^{-4}$ and $\omega=3.5$, diamonds: varying $E$ for fixed $\varepsilon=1$ and $\omega=3.5$, asterisks: varying $\omega$ for fixed $\varepsilon=1$ and $E=5\cdot 10^{-4}$. Dashed line: linear fit to the data points, giving a slope of 2.13. (b) in function of $p$ for fixed $E=5\cdot10^{-4}$, $\varepsilon=1$ and $\omega=3.5$. Circles: numerical data points, dashed line: fourth-order polynomial fit to the data points. }
      \label{fig:dissipation_2D}            
   \end{center}
\end{figure}

\par
In order to verify (\ref{dissipation_poincare}), we have performed extensive 2D numerical simulations of the basic flow in which the four parameters $\varepsilon,p, E$, and $\omega$ are independently varied. The results of this survey are shown in figure \ref{fig:dissipation_2D} and confirm indeed that  $\overline{{\mathcal D}_{\nu}^L}$ scales, to the leading order, as $\overline{{\mathcal D}_{\nu}^L} \sim \sqrt{E \omega}\varepsilon^2$. The slope of the dashed line in this figure is approximately 2.13, which is close to the leading-order coefficient $\pi/\sqrt{2} \approx 2.22$ (difference of $4 \%$). 

In figure \ref{fig:dissipation_2D}(b), we show the dependency of $\overline{{\mathcal D}_{\nu}^{L}}$ on $p$. Performing a fourth-order polynomial fit to these data points, we obtain $\overline{{\mathcal D}_{\nu}^{L}} = 0.08246 - 0.00146p + 0.027p^2 + \mathcal{O}(p^3)$. We see that the prefactor in front of the linear term is almost two orders of magnitude smaller than the ones in front of the constant and quadratic term. The magnitude of this coefficient reduces further when we increase the order of the polynomial fit. This indicates that the first higher-order term in $p$ in (\ref{dissipation_poincare}) is indeed quadratic in $p$. 
The constant factor $0.08246$ can be compared with the prefactor in (\ref{dissipation_poincare}), i.e. $\pi\sqrt{E \omega/2} = 0.092929$ (difference of $11 \%)$.

\end{appendix}

\bibliographystyle{jfm}
\bibliography{convection}
\end{document}